\documentclass[longauth]{aa}
\usepackage{graphicx}	% Including figure files
\usepackage{natbib}
\usepackage{scalerel}
\usepackage{amsmath}	% Advanced maths commands
\usepackage{amssymb}	% Extra maths symbols
\usepackage{txfonts}
\usepackage{xcolor}
\usepackage{multirow}
\newcommand*{\spr}{{\tt SPRITZ}}
\usepackage{euclid}
\newcommand*{\hers}{\textit{Herschel}\,}

\usepackage[pdfencoding=auto,psdextra,draft]{hyperref}
\hypersetup{
    colorlinks=true,
    linkcolor=blue,
    filecolor=magenta,      
    urlcolor=blue,
    citecolor=blue
}
\makeatletter
\renewcommand*\aa@pageof{, page \thepage{} of \pageref*{LastPage}}
\makeatother

\usepackage[utf8]{inputenc}

\usepackage[switch, modulo]{lineno}
\begin{document} 

\title{\Euclid preparation. XLIX.}
\subtitle{Selecting active galactic nuclei using observed colours}

\titlerunning{\Euclid\/:AGN colour selection}
\authorrunning{Euclid Collaboration: L. Bisigello et al.}
		   
%%%% please do not edit the author list -- contact ECEB Bureau for changes
\newcommand{\orcid}[1]{} %% if already defined in aa.cls: comment, or use renewcommand			   
\author{Euclid Collaboration: L.~Bisigello\orcid{0000-0003-0492-4924}\thanks{\email{laura.bisigello@inaf.it}}\inst{\ref{aff1},\ref{aff2}}
\and M.~Massimo\inst{\ref{aff3}}
\and C.~Tortora\orcid{0000-0001-7958-6531}\inst{\ref{aff4}}
\and S.~Fotopoulou\inst{\ref{aff5}}
\and V.~Allevato\orcid{0000-0001-7232-5152}\inst{\ref{aff4}}
\and M.~Bolzonella\orcid{0000-0003-3278-4607}\inst{\ref{aff6}}
\and C.~Gruppioni\orcid{0000-0002-5836-4056}\inst{\ref{aff6}}
\and L.~Pozzetti\orcid{0000-0001-7085-0412}\inst{\ref{aff6}}
\and G.~Rodighiero\orcid{0000-0002-9415-2296}\inst{\ref{aff1},\ref{aff2}}
\and S.~Serjeant\inst{\ref{aff7}}
\and P.~A.~C.~Cunha\orcid{0000-0002-9454-859X}\inst{\ref{aff8},\ref{aff9}}
\and L.~Gabarra\inst{\ref{aff10},\ref{aff1}}
\and A.~Feltre\inst{\ref{aff11}}
\and A.~Humphrey\inst{\ref{aff12},\ref{aff9}}
\and F.~La~Franca\orcid{0000-0002-1239-2721}\inst{\ref{aff13}}
\and H.~Landt\orcid{0000-0001-8391-6900}\inst{\ref{aff14}}
\and F.~Mannucci\orcid{0000-0002-4803-2381}\inst{\ref{aff11}}
\and I.~Prandoni\orcid{0000-0001-9680-7092}\inst{\ref{aff15}}
\and M.~Radovich\orcid{0000-0002-3585-866X}\inst{\ref{aff2}}
\and F.~Ricci\orcid{0000-0001-5742-5980}\inst{\ref{aff13},\ref{aff16}}
\and M.~Salvato\orcid{0000-0001-7116-9303}\inst{\ref{aff17}}
\and F.~Shankar\orcid{0000-0001-8973-5051}\inst{\ref{aff18}}
\and D.~Stern\orcid{0000-0003-2686-9241}\inst{\ref{aff19}}
\and L.~Spinoglio\orcid{0000-0001-8840-1551}\inst{\ref{aff20}}
\and D.~Vergani\orcid{0000-0003-0898-2216}\inst{\ref{aff6}}
\and C.~Vignali\orcid{0000-0002-8853-9611}\inst{\ref{aff21},\ref{aff6}}
\and G.~Zamorani\orcid{0000-0002-2318-301X}\inst{\ref{aff6}}
\and L.~Y.~A.~Yung\orcid{0000-0003-3466-035X}\inst{\ref{aff22}}
\and S.~Charlot\orcid{0000-0003-3458-2275}\inst{\ref{aff23}}
\and N.~Aghanim\inst{\ref{aff24}}
\and A.~Amara\inst{\ref{aff25}}
\and S.~Andreon\orcid{0000-0002-2041-8784}\inst{\ref{aff26}}
\and N.~Auricchio\inst{\ref{aff6}}
\and M.~Baldi\orcid{0000-0003-4145-1943}\inst{\ref{aff3},\ref{aff6},\ref{aff27}}
\and S.~Bardelli\orcid{0000-0002-8900-0298}\inst{\ref{aff6}}
\and P.~Battaglia\orcid{0000-0002-7337-5909}\inst{\ref{aff6}}
\and R.~Bender\orcid{0000-0001-7179-0626}\inst{\ref{aff17},\ref{aff28}}
\and D.~Bonino\inst{\ref{aff29}}
\and E.~Branchini\orcid{0000-0002-0808-6908}\inst{\ref{aff30},\ref{aff31}}
\and S.~Brau-Nogue\inst{\ref{aff32}}
\and M.~Brescia\orcid{0000-0001-9506-5680}\inst{\ref{aff33},\ref{aff4}}
\and S.~Camera\orcid{0000-0003-3399-3574}\inst{\ref{aff34},\ref{aff35},\ref{aff29}}
\and V.~Capobianco\orcid{0000-0002-3309-7692}\inst{\ref{aff29}}
\and C.~Carbone\orcid{0000-0003-0125-3563}\inst{\ref{aff36}}
\and J.~Carretero\orcid{0000-0002-3130-0204}\inst{\ref{aff37},\ref{aff38}}
\and S.~Casas\orcid{0000-0002-4751-5138}\inst{\ref{aff39}}
\and F.~J.~Castander\orcid{0000-0001-7316-4573}\inst{\ref{aff40},\ref{aff41}}
\and M.~Castellano\orcid{0000-0001-9875-8263}\inst{\ref{aff16}}
\and S.~Cavuoti\orcid{0000-0002-3787-4196}\inst{\ref{aff4},\ref{aff42}}
\and A.~Cimatti\inst{\ref{aff43}}
\and G.~Congedo\orcid{0000-0003-2508-0046}\inst{\ref{aff44}}
\and C.~J.~Conselice\inst{\ref{aff45}}
\and L.~Conversi\orcid{0000-0002-6710-8476}\inst{\ref{aff46},\ref{aff47}}
\and Y.~Copin\orcid{0000-0002-5317-7518}\inst{\ref{aff48}}
\and L.~Corcione\orcid{0000-0002-6497-5881}\inst{\ref{aff29}}
\and F.~Courbin\orcid{0000-0003-0758-6510}\inst{\ref{aff49}}
\and H.~M.~Courtois\orcid{0000-0003-0509-1776}\inst{\ref{aff50}}
\and M.~Cropper\inst{\ref{aff51}}
\and A.~Da~Silva\orcid{0000-0002-6385-1609}\inst{\ref{aff52},\ref{aff53}}
\and H.~Degaudenzi\orcid{0000-0002-5887-6799}\inst{\ref{aff54}}
\and A.~M.~Di~Giorgio\orcid{0000-0002-4767-2360}\inst{\ref{aff20}}
\and J.~Dinis\inst{\ref{aff53},\ref{aff52}}
\and X.~Dupac\inst{\ref{aff47}}
\and S.~Dusini\orcid{0000-0002-1128-0664}\inst{\ref{aff10}}
\and A.~Ealet\inst{\ref{aff48}}
\and M.~Farina\orcid{0000-0002-3089-7846}\inst{\ref{aff20}}
\and S.~Farrens\orcid{0000-0002-9594-9387}\inst{\ref{aff55}}
\and S.~Ferriol\inst{\ref{aff48}}
\and M.~Frailis\orcid{0000-0002-7400-2135}\inst{\ref{aff56}}
\and E.~Franceschi\orcid{0000-0002-0585-6591}\inst{\ref{aff6}}
\and P.~Franzetti\inst{\ref{aff36}}
\and M.~Fumana\orcid{0000-0001-6787-5950}\inst{\ref{aff36}}
\and S.~Galeotta\orcid{0000-0002-3748-5115}\inst{\ref{aff56}}
\and B.~Garilli\orcid{0000-0001-7455-8750}\inst{\ref{aff36}}
\and B.~Gillis\orcid{0000-0002-4478-1270}\inst{\ref{aff44}}
\and C.~Giocoli\orcid{0000-0002-9590-7961}\inst{\ref{aff6},\ref{aff27}}
\and B.~R.~Granett\orcid{0000-0003-2694-9284}\inst{\ref{aff26}}
\and A.~Grazian\orcid{0000-0002-5688-0663}\inst{\ref{aff2}}
\and F.~Grupp\inst{\ref{aff17},\ref{aff28}}
\and L.~Guzzo\orcid{0000-0001-8264-5192}\inst{\ref{aff57},\ref{aff26},\ref{aff58}}
\and S.~V.~H.~Haugan\orcid{0000-0001-9648-7260}\inst{\ref{aff59}}
\and W.~Holmes\inst{\ref{aff19}}
\and I.~Hook\orcid{0000-0002-2960-978X}\inst{\ref{aff60}}
\and F.~Hormuth\inst{\ref{aff61}}
\and A.~Hornstrup\orcid{0000-0002-3363-0936}\inst{\ref{aff62},\ref{aff63}}
\and K.~Jahnke\orcid{0000-0003-3804-2137}\inst{\ref{aff64}}
\and E.~Keih\"anen\inst{\ref{aff65}}
\and S.~Kermiche\orcid{0000-0002-0302-5735}\inst{\ref{aff66}}
\and A.~Kiessling\orcid{0000-0002-2590-1273}\inst{\ref{aff19}}
\and M.~Kilbinger\orcid{0000-0001-9513-7138}\inst{\ref{aff67}}
\and T.~Kitching\orcid{0000-0002-4061-4598}\inst{\ref{aff51}}
\and M.~K\"ummel\orcid{0000-0003-2791-2117}\inst{\ref{aff28}}
\and M.~Kunz\orcid{0000-0002-3052-7394}\inst{\ref{aff68}}
\and H.~Kurki-Suonio\orcid{0000-0002-4618-3063}\inst{\ref{aff69},\ref{aff70}}
\and S.~Ligori\orcid{0000-0003-4172-4606}\inst{\ref{aff29}}
\and P.~B.~Lilje\orcid{0000-0003-4324-7794}\inst{\ref{aff59}}
\and V.~Lindholm\orcid{0000-0003-2317-5471}\inst{\ref{aff69},\ref{aff70}}
\and I.~Lloro\inst{\ref{aff71}}
\and E.~Maiorano\orcid{0000-0003-2593-4355}\inst{\ref{aff6}}
\and O.~Mansutti\orcid{0000-0001-5758-4658}\inst{\ref{aff56}}
\and O.~Marggraf\orcid{0000-0001-7242-3852}\inst{\ref{aff72}}
\and K.~Markovic\orcid{0000-0001-6764-073X}\inst{\ref{aff19}}
\and N.~Martinet\orcid{0000-0003-2786-7790}\inst{\ref{aff73}}
\and F.~Marulli\orcid{0000-0002-8850-0303}\inst{\ref{aff21},\ref{aff6},\ref{aff27}}
\and R.~Massey\orcid{0000-0002-6085-3780}\inst{\ref{aff74}}
\and S.~Maurogordato\inst{\ref{aff75}}
\and E.~Medinaceli\orcid{0000-0002-4040-7783}\inst{\ref{aff6}}
\and S.~Mei\orcid{0000-0002-2849-559X}\inst{\ref{aff76}}
\and Y.~Mellier\inst{\ref{aff77},\ref{aff78}}
\and M.~Meneghetti\orcid{0000-0003-1225-7084}\inst{\ref{aff6},\ref{aff27}}
\and E.~Merlin\orcid{0000-0001-6870-8900}\inst{\ref{aff16}}
\and G.~Meylan\inst{\ref{aff49}}
\and M.~Moresco\orcid{0000-0002-7616-7136}\inst{\ref{aff21},\ref{aff6}}
\and L.~Moscardini\orcid{0000-0002-3473-6716}\inst{\ref{aff21},\ref{aff6},\ref{aff27}}
\and E.~Munari\orcid{0000-0002-1751-5946}\inst{\ref{aff56}}
\and S.-M.~Niemi\inst{\ref{aff79}}
\and C.~Padilla\orcid{0000-0001-7951-0166}\inst{\ref{aff37}}
\and S.~Paltani\inst{\ref{aff54}}
\and F.~Pasian\inst{\ref{aff56}}
\and K.~Pedersen\inst{\ref{aff80}}
\and W.~J.~Percival\orcid{0000-0002-0644-5727}\inst{\ref{aff81},\ref{aff82},\ref{aff83}}
\and V.~Pettorino\inst{\ref{aff84}}
\and G.~Polenta\orcid{0000-0003-4067-9196}\inst{\ref{aff85}}
\and M.~Poncet\inst{\ref{aff86}}
\and F.~Raison\orcid{0000-0002-7819-6918}\inst{\ref{aff17}}
\and R.~Rebolo\inst{\ref{aff87},\ref{aff88}}
\and A.~Renzi\orcid{0000-0001-9856-1970}\inst{\ref{aff1},\ref{aff10}}
\and J.~Rhodes\inst{\ref{aff19}}
\and G.~Riccio\inst{\ref{aff4}}
\and E.~Romelli\orcid{0000-0003-3069-9222}\inst{\ref{aff56}}
\and M.~Roncarelli\orcid{0000-0001-9587-7822}\inst{\ref{aff6}}
\and E.~Rossetti\inst{\ref{aff3}}
\and R.~Saglia\orcid{0000-0003-0378-7032}\inst{\ref{aff28},\ref{aff17}}
\and D.~Sapone\orcid{0000-0001-7089-4503}\inst{\ref{aff89}}
\and B.~Sartoris\inst{\ref{aff28},\ref{aff56}}
\and M.~Schirmer\orcid{0000-0003-2568-9994}\inst{\ref{aff64}}
\and P.~Schneider\orcid{0000-0001-8561-2679}\inst{\ref{aff72}}
\and T.~Schrabback\orcid{0000-0002-6987-7834}\inst{\ref{aff90},\ref{aff72}}
\and A.~Secroun\orcid{0000-0003-0505-3710}\inst{\ref{aff66}}
\and G.~Seidel\orcid{0000-0003-2907-353X}\inst{\ref{aff64}}
\and S.~Serrano\orcid{0000-0002-0211-2861}\inst{\ref{aff41},\ref{aff40},\ref{aff91}}
\and C.~Sirignano\orcid{0000-0002-0995-7146}\inst{\ref{aff1},\ref{aff10}}
\and G.~Sirri\orcid{0000-0003-2626-2853}\inst{\ref{aff27}}
\and L.~Stanco\orcid{0000-0002-9706-5104}\inst{\ref{aff10}}
\and C.~Surace\orcid{0000-0003-2592-0113}\inst{\ref{aff73}}
\and P.~Tallada-Cresp\'{i}\orcid{0000-0002-1336-8328}\inst{\ref{aff92},\ref{aff38}}
\and A.~N.~Taylor\inst{\ref{aff44}}
\and I.~Tereno\inst{\ref{aff52},\ref{aff93}}
\and R.~Toledo-Moreo\orcid{0000-0002-2997-4859}\inst{\ref{aff94}}
\and F.~Torradeflot\orcid{0000-0003-1160-1517}\inst{\ref{aff38},\ref{aff92}}
\and I.~Tutusaus\orcid{0000-0002-3199-0399}\inst{\ref{aff32}}
\and E.~A.~Valentijn\inst{\ref{aff95}}
\and L.~Valenziano\orcid{0000-0002-1170-0104}\inst{\ref{aff6},\ref{aff96}}
\and T.~Vassallo\orcid{0000-0001-6512-6358}\inst{\ref{aff28},\ref{aff56}}
\and Y.~Wang\orcid{0000-0002-4749-2984}\inst{\ref{aff97}}
\and J.~Zoubian\inst{\ref{aff66}}
\and E.~Zucca\orcid{0000-0002-5845-8132}\inst{\ref{aff6}}
\and A.~Biviano\orcid{0000-0002-0857-0732}\inst{\ref{aff56},\ref{aff98}}
\and E.~Bozzo\orcid{0000-0002-8201-1525}\inst{\ref{aff54}}
\and C.~Colodro-Conde\inst{\ref{aff87}}
\and D.~Di~Ferdinando\inst{\ref{aff27}}
\and G.~Fabbian\orcid{0000-0002-3255-4695}\inst{\ref{aff99},\ref{aff100}}
\and J.~Graci\'{a}-Carpio\inst{\ref{aff17}}
\and S.~Marcin\inst{\ref{aff101}}
\and N.~Mauri\orcid{0000-0001-8196-1548}\inst{\ref{aff43},\ref{aff27}}
\and Z.~Sakr\orcid{0000-0002-4823-3757}\inst{\ref{aff102},\ref{aff32},\ref{aff103}}
\and V.~Scottez\inst{\ref{aff77},\ref{aff104}}
\and M.~Tenti\orcid{0000-0002-4254-5901}\inst{\ref{aff27}}
\and Y.~Akrami\orcid{0000-0002-2407-7956}\inst{\ref{aff105},\ref{aff106}}
\and C.~Baccigalupi\orcid{0000-0002-8211-1630}\inst{\ref{aff107},\ref{aff56},\ref{aff108},\ref{aff98}}
\and M.~Ballardini\inst{\ref{aff109},\ref{aff110},\ref{aff6}}
\and M.~Bethermin\inst{\ref{aff111},\ref{aff73}}
\and A.~Blanchard\orcid{0000-0001-8555-9003}\inst{\ref{aff32}}
\and S.~Borgani\orcid{0000-0001-6151-6439}\inst{\ref{aff56},\ref{aff112},\ref{aff108},\ref{aff98}}
\and A.~S.~Borlaff\orcid{0000-0003-3249-4431}\inst{\ref{aff113},\ref{aff114},\ref{aff115}}
\and S.~Bruton\inst{\ref{aff116}}
\and C.~Burigana\orcid{0000-0002-3005-5796}\inst{\ref{aff15},\ref{aff96}}
\and R.~Cabanac\orcid{0000-0001-6679-2600}\inst{\ref{aff32}}
\and A.~Calabro\orcid{0000-0003-2536-1614}\inst{\ref{aff16}}
\and A.~Cappi\inst{\ref{aff6},\ref{aff75}}
\and C.~S.~Carvalho\inst{\ref{aff93}}
\and G.~Castignani\orcid{0000-0001-6831-0687}\inst{\ref{aff21},\ref{aff6}}
\and T.~Castro\orcid{0000-0002-6292-3228}\inst{\ref{aff56},\ref{aff108},\ref{aff98}}
\and K.~C.~Chambers\orcid{0000-0001-6965-7789}\inst{\ref{aff117}}
\and A.~R.~Cooray\orcid{0000-0002-3892-0190}\inst{\ref{aff118}}
\and J.~Coupon\inst{\ref{aff54}}
\and O.~Cucciati\orcid{0000-0002-9336-7551}\inst{\ref{aff6}}
\and S.~Davini\inst{\ref{aff31}}
\and G.~De~Lucia\orcid{0000-0002-6220-9104}\inst{\ref{aff56}}
\and G.~Desprez\inst{\ref{aff119}}
\and A.~D\'iaz-S\'anchez\orcid{0000-0003-0748-4768}\inst{\ref{aff120}}
\and S.~Di~Domizio\orcid{0000-0003-2863-5895}\inst{\ref{aff121}}
\and H.~Dole\orcid{0000-0002-9767-3839}\inst{\ref{aff24}}
\and J.~A.~Escartin~Vigo\inst{\ref{aff17}}
\and S.~Escoffier\orcid{0000-0002-2847-7498}\inst{\ref{aff66}}
\and I.~Ferrero\inst{\ref{aff59}}
\and F.~Finelli\orcid{0000-0002-6694-3269}\inst{\ref{aff6},\ref{aff96}}
\and K.~Ganga\orcid{0000-0001-8159-8208}\inst{\ref{aff76}}
\and J.~Garc\'ia-Bellido\orcid{0000-0002-9370-8360}\inst{\ref{aff105}}
\and F.~Giacomini\orcid{0000-0002-3129-2814}\inst{\ref{aff27}}
\and G.~Gozaliasl\orcid{0000-0002-0236-919X}\inst{\ref{aff122},\ref{aff69}}
\and A.~Gregorio\orcid{0000-0003-4028-8785}\inst{\ref{aff112},\ref{aff56},\ref{aff108}}
\and H.~Hildebrandt\orcid{0000-0002-9814-3338}\inst{\ref{aff123}}
\and A.~Jiminez~Mu{\~n}oz\inst{\ref{aff124}}
\and J.~J.~E.~Kajava\orcid{0000-0002-3010-8333}\inst{\ref{aff125},\ref{aff126}}
\and V.~Kansal\inst{\ref{aff47},\ref{aff127},\ref{aff128}}
\and D.~Karagiannis\orcid{0000-0002-4927-0816}\inst{\ref{aff129},\ref{aff130}}
\and C.~C.~Kirkpatrick\inst{\ref{aff65}}
\and L.~Legrand\orcid{0000-0003-0610-5252}\inst{\ref{aff68}}
\and A.~Loureiro\orcid{0000-0002-4371-0876}\inst{\ref{aff131},\ref{aff132}}
\and J.~Macias-Perez\inst{\ref{aff124}}
\and G.~Maggio\orcid{0000-0003-4020-4836}\inst{\ref{aff56}}
\and M.~Magliocchetti\orcid{0000-0001-9158-4838}\inst{\ref{aff20}}
\and G.~Mainetti\inst{\ref{aff133}}
\and R.~Maoli\orcid{0000-0002-6065-3025}\inst{\ref{aff134},\ref{aff16}}
\and M.~Martinelli\orcid{0000-0002-6943-7732}\inst{\ref{aff16},\ref{aff135}}
\and C.~J.~A.~P.~Martins\orcid{0000-0002-4886-9261}\inst{\ref{aff136},\ref{aff9}}
\and S.~Matthew\inst{\ref{aff44}}
\and L.~Maurin\orcid{0000-0002-8406-0857}\inst{\ref{aff24}}
\and R.~B.~Metcalf\orcid{0000-0003-3167-2574}\inst{\ref{aff21},\ref{aff6}}
\and M.~Migliaccio\inst{\ref{aff137},\ref{aff138}}
\and P.~Monaco\orcid{0000-0003-2083-7564}\inst{\ref{aff112},\ref{aff56},\ref{aff108},\ref{aff98}}
\and G.~Morgante\inst{\ref{aff6}}
\and S.~Nadathur\orcid{0000-0001-9070-3102}\inst{\ref{aff25}}
\and L.~Patrizii\inst{\ref{aff27}}
\and V.~Popa\inst{\ref{aff139}}
\and C.~Porciani\inst{\ref{aff72}}
\and D.~Potter\orcid{0000-0002-0757-5195}\inst{\ref{aff140}}
\and M.~P\"{o}ntinen\orcid{0000-0001-5442-2530}\inst{\ref{aff69}}
\and P.-F.~Rocci\inst{\ref{aff75}}
\and A.~G.~S\'anchez\orcid{0000-0003-1198-831X}\inst{\ref{aff17}}
\and A.~Schneider\orcid{0000-0001-7055-8104}\inst{\ref{aff140}}
\and M.~Sereno\orcid{0000-0003-0302-0325}\inst{\ref{aff6},\ref{aff27}}
\and P.~Simon\inst{\ref{aff72}}
\and J.~Stadel\orcid{0000-0001-7565-8622}\inst{\ref{aff140}}
\and S.~A.~Stanford\orcid{0000-0003-0122-0841}\inst{\ref{aff141}}
\and J.~Steinwagner\inst{\ref{aff17}}
\and G.~Testera\inst{\ref{aff31}}
\and M.~Tewes\orcid{0000-0002-1155-8689}\inst{\ref{aff72}}
\and R.~Teyssier\orcid{0000-0001-7689-0933}\inst{\ref{aff142}}
\and S.~Toft\orcid{0000-0003-3631-7176}\inst{\ref{aff63},\ref{aff143},\ref{aff144}}
\and S.~Tosi\orcid{0000-0002-7275-9193}\inst{\ref{aff30},\ref{aff31},\ref{aff26}}
\and A.~Troja\orcid{0000-0003-0239-4595}\inst{\ref{aff1},\ref{aff10}}
\and M.~Tucci\inst{\ref{aff54}}
\and J.~Valiviita\orcid{0000-0001-6225-3693}\inst{\ref{aff69},\ref{aff70}}
\and M.~Viel\orcid{0000-0002-2642-5707}\inst{\ref{aff98},\ref{aff56},\ref{aff107},\ref{aff108}}
\and I.~A.~Zinchenko\inst{\ref{aff28}}}
										   
%%%% please do not edit the affiliation list -- contact ECEB Bureau for changes
\institute{Dipartimento di Fisica e Astronomia "G. Galilei", Universit\`a di Padova, Via Marzolo 8, 35131 Padova, Italy\label{aff1}
\and
INAF-Osservatorio Astronomico di Padova, Via dell'Osservatorio 5, 35122 Padova, Italy\label{aff2}
\and
Dipartimento di Fisica e Astronomia, Universit\`a di Bologna, Via Gobetti 93/2, 40129 Bologna, Italy\label{aff3}
\and
INAF-Osservatorio Astronomico di Capodimonte, Via Moiariello 16, 80131 Napoli, Italy\label{aff4}
\and
School of Physics, HH Wills Physics Laboratory, University of Bristol, Tyndall Avenue, Bristol, BS8 1TL, UK\label{aff5}
\and
INAF-Osservatorio di Astrofisica e Scienza dello Spazio di Bologna, Via Piero Gobetti 93/3, 40129 Bologna, Italy\label{aff6}
\and
School of Physical Sciences, The Open University, Milton Keynes, MK7 6AA, UK\label{aff7}
\and
Faculdade de Ci\^encias da Universidade do Porto, Rua do Campo de Alegre, 4150-007 Porto, Portugal\label{aff8}
\and
Instituto de Astrof\'isica e Ci\^encias do Espa\c{c}o, Universidade do Porto, CAUP, Rua das Estrelas, PT4150-762 Porto, Portugal\label{aff9}
\and
INFN-Padova, Via Marzolo 8, 35131 Padova, Italy\label{aff10}
\and
INAF-Osservatorio Astrofisico di Arcetri, Largo E. Fermi 5, 50125, Firenze, Italy\label{aff11}
\and
DTx -- Digital Transformation CoLAB, Building 1, Azur\'em Campus, University of Minho, 4800-058 Guimar\~aes, Portugal\label{aff12}
\and
Department of Mathematics and Physics, Roma Tre University, Via della Vasca Navale 84, 00146 Rome, Italy\label{aff13}
\and
Department of Physics, Centre for Extragalactic Astronomy, Durham University, South Road, DH1 3LE, UK\label{aff14}
\and
INAF, Istituto di Radioastronomia, Via Piero Gobetti 101, 40129 Bologna, Italy\label{aff15}
\and
INAF-Osservatorio Astronomico di Roma, Via Frascati 33, 00078 Monteporzio Catone, Italy\label{aff16}
\and
Max Planck Institute for Extraterrestrial Physics, Giessenbachstr. 1, 85748 Garching, Germany\label{aff17}
\and
Department of Physics and Astronomy, University of Southampton, Southampton, SO17 1BJ, UK\label{aff18}
\and
Jet Propulsion Laboratory, California Institute of Technology, 4800 Oak Grove Drive, Pasadena, CA, 91109, USA\label{aff19}
\and
INAF-Istituto di Astrofisica e Planetologia Spaziali, via del Fosso del Cavaliere, 100, 00100 Roma, Italy\label{aff20}
\and
Dipartimento di Fisica e Astronomia "Augusto Righi" - Alma Mater Studiorum Universit\`a di Bologna, via Piero Gobetti 93/2, 40129 Bologna, Italy\label{aff21}
\and
NASA Goddard Space Flight Center, Greenbelt, MD 20771, USA\label{aff22}
\and
Sorbonne Universit{\'e}s, UPMC Univ Paris 6 et CNRS, UMR 7095, Institut d'Astrophysique de Paris, 98 bis bd Arago, 75014 Paris, France\label{aff23}
\and
Universit\'e Paris-Saclay, CNRS, Institut d'astrophysique spatiale, 91405, Orsay, France\label{aff24}
\and
Institute of Cosmology and Gravitation, University of Portsmouth, Portsmouth PO1 3FX, UK\label{aff25}
\and
INAF-Osservatorio Astronomico di Brera, Via Brera 28, 20122 Milano, Italy\label{aff26}
\and
INFN-Sezione di Bologna, Viale Berti Pichat 6/2, 40127 Bologna, Italy\label{aff27}
\and
Universit\"ats-Sternwarte M\"unchen, Fakult\"at f\"ur Physik, Ludwig-Maximilians-Universit\"at M\"unchen, Scheinerstrasse 1, 81679 M\"unchen, Germany\label{aff28}
\and
INAF-Osservatorio Astrofisico di Torino, Via Osservatorio 20, 10025 Pino Torinese (TO), Italy\label{aff29}
\and
Dipartimento di Fisica, Universit\`a di Genova, Via Dodecaneso 33, 16146, Genova, Italy\label{aff30}
\and
INFN-Sezione di Genova, Via Dodecaneso 33, 16146, Genova, Italy\label{aff31}
\and
Institut de Recherche en Astrophysique et Plan\'etologie (IRAP), Universit\'e de Toulouse, CNRS, UPS, CNES, 14 Av. Edouard Belin, 31400 Toulouse, France\label{aff32}
\and
Department of Physics "E. Pancini", University Federico II, Via Cinthia 6, 80126, Napoli, Italy\label{aff33}
\and
Dipartimento di Fisica, Universit\`a degli Studi di Torino, Via P. Giuria 1, 10125 Torino, Italy\label{aff34}
\and
INFN-Sezione di Torino, Via P. Giuria 1, 10125 Torino, Italy\label{aff35}
\and
INAF-IASF Milano, Via Alfonso Corti 12, 20133 Milano, Italy\label{aff36}
\and
Institut de F\'{i}sica d'Altes Energies (IFAE), The Barcelona Institute of Science and Technology, Campus UAB, 08193 Bellaterra (Barcelona), Spain\label{aff37}
\and
Port d'Informaci\'{o} Cient\'{i}fica, Campus UAB, C. Albareda s/n, 08193 Bellaterra (Barcelona), Spain\label{aff38}
\and
Institute for Theoretical Particle Physics and Cosmology (TTK), RWTH Aachen University, 52056 Aachen, Germany\label{aff39}
\and
Institute of Space Sciences (ICE, CSIC), Campus UAB, Carrer de Can Magrans, s/n, 08193 Barcelona, Spain\label{aff40}
\and
Institut d'Estudis Espacials de Catalunya (IEEC),  Edifici RDIT, Campus UPC, 08860 Castelldefels, Barcelona, Spain\label{aff41}
\and
INFN section of Naples, Via Cinthia 6, 80126, Napoli, Italy\label{aff42}
\and
Dipartimento di Fisica e Astronomia "Augusto Righi" - Alma Mater Studiorum Universit\`a di Bologna, Viale Berti Pichat 6/2, 40127 Bologna, Italy\label{aff43}
\and
Institute for Astronomy, University of Edinburgh, Royal Observatory, Blackford Hill, Edinburgh EH9 3HJ, UK\label{aff44}
\and
Jodrell Bank Centre for Astrophysics, Department of Physics and Astronomy, University of Manchester, Oxford Road, Manchester M13 9PL, UK\label{aff45}
\and
European Space Agency/ESRIN, Largo Galileo Galilei 1, 00044 Frascati, Roma, Italy\label{aff46}
\and
ESAC/ESA, Camino Bajo del Castillo, s/n., Urb. Villafranca del Castillo, 28692 Villanueva de la Ca\~nada, Madrid, Spain\label{aff47}
\and
Universit\'e Claude Bernard Lyon 1, CNRS/IN2P3, IP2I Lyon, UMR 5822, Villeurbanne, F-69100, France\label{aff48}
\and
Institute of Physics, Laboratory of Astrophysics, Ecole Polytechnique F\'ed\'erale de Lausanne (EPFL), Observatoire de Sauverny, 1290 Versoix, Switzerland\label{aff49}
\and
UCB Lyon 1, CNRS/IN2P3, IUF, IP2I Lyon, 4 rue Enrico Fermi, 69622 Villeurbanne, France\label{aff50}
\and
Mullard Space Science Laboratory, University College London, Holmbury St Mary, Dorking, Surrey RH5 6NT, UK\label{aff51}
\and
Departamento de F\'isica, Faculdade de Ci\^encias, Universidade de Lisboa, Edif\'icio C8, Campo Grande, PT1749-016 Lisboa, Portugal\label{aff52}
\and
Instituto de Astrof\'isica e Ci\^encias do Espa\c{c}o, Faculdade de Ci\^encias, Universidade de Lisboa, Campo Grande, 1749-016 Lisboa, Portugal\label{aff53}
\and
Department of Astronomy, University of Geneva, ch. d'Ecogia 16, 1290 Versoix, Switzerland\label{aff54}
\and
Universit\'e Paris-Saclay, Universit\'e Paris Cit\'e, CEA, CNRS, AIM, 91191, Gif-sur-Yvette, France\label{aff55}
\and
INAF-Osservatorio Astronomico di Trieste, Via G. B. Tiepolo 11, 34143 Trieste, Italy\label{aff56}
\and
Dipartimento di Fisica "Aldo Pontremoli", Universit\`a degli Studi di Milano, Via Celoria 16, 20133 Milano, Italy\label{aff57}
\and
INFN-Sezione di Milano, Via Celoria 16, 20133 Milano, Italy\label{aff58}
\and
Institute of Theoretical Astrophysics, University of Oslo, P.O. Box 1029 Blindern, 0315 Oslo, Norway\label{aff59}
\and
Department of Physics, Lancaster University, Lancaster, LA1 4YB, UK\label{aff60}
\and
von Hoerner \& Sulger GmbH, Schlossplatz 8, 68723 Schwetzingen, Germany\label{aff61}
\and
Technical University of Denmark, Elektrovej 327, 2800 Kgs. Lyngby, Denmark\label{aff62}
\and
Cosmic Dawn Center (DAWN), Denmark\label{aff63}
\and
Max-Planck-Institut f\"ur Astronomie, K\"onigstuhl 17, 69117 Heidelberg, Germany\label{aff64}
\and
Department of Physics and Helsinki Institute of Physics, Gustaf H\"allstr\"omin katu 2, 00014 University of Helsinki, Finland\label{aff65}
\and
Aix-Marseille Universit\'e, CNRS/IN2P3, CPPM, Marseille, France\label{aff66}
\and
AIM, CEA, CNRS, Universit\'{e} Paris-Saclay, Universit\'{e} de Paris, 91191 Gif-sur-Yvette, France\label{aff67}
\and
Universit\'e de Gen\`eve, D\'epartement de Physique Th\'eorique and Centre for Astroparticle Physics, 24 quai Ernest-Ansermet, CH-1211 Gen\`eve 4, Switzerland\label{aff68}
\and
Department of Physics, P.O. Box 64, 00014 University of Helsinki, Finland\label{aff69}
\and
Helsinki Institute of Physics, Gustaf H{\"a}llstr{\"o}min katu 2, University of Helsinki, Helsinki, Finland\label{aff70}
\and
NOVA optical infrared instrumentation group at ASTRON, Oude Hoogeveensedijk 4, 7991PD, Dwingeloo, The Netherlands\label{aff71}
\and
Universit\"at Bonn, Argelander-Institut f\"ur Astronomie, Auf dem H\"ugel 71, 53121 Bonn, Germany\label{aff72}
\and
Aix-Marseille Universit\'e, CNRS, CNES, LAM, Marseille, France\label{aff73}
\and
Department of Physics, Institute for Computational Cosmology, Durham University, South Road, DH1 3LE, UK\label{aff74}
\and
Universit\'e C\^{o}te d'Azur, Observatoire de la C\^{o}te d'Azur, CNRS, Laboratoire Lagrange, Bd de l'Observatoire, CS 34229, 06304 Nice cedex 4, France\label{aff75}
\and
Universit\'e Paris Cit\'e, CNRS, Astroparticule et Cosmologie, 75013 Paris, France\label{aff76}
\and
Institut d'Astrophysique de Paris, 98bis Boulevard Arago, 75014, Paris, France\label{aff77}
\and
Institut d'Astrophysique de Paris, UMR 7095, CNRS, and Sorbonne Universit\'e, 98 bis boulevard Arago, 75014 Paris, France\label{aff78}
\and
European Space Agency/ESTEC, Keplerlaan 1, 2201 AZ Noordwijk, The Netherlands\label{aff79}
\and
Department of Physics and Astronomy, University of Aarhus, Ny Munkegade 120, DK-8000 Aarhus C, Denmark\label{aff80}
\and
Waterloo Centre for Astrophysics, University of Waterloo, Waterloo, Ontario N2L 3G1, Canada\label{aff81}
\and
Department of Physics and Astronomy, University of Waterloo, Waterloo, Ontario N2L 3G1, Canada\label{aff82}
\and
Perimeter Institute for Theoretical Physics, Waterloo, Ontario N2L 2Y5, Canada\label{aff83}
\and
Universit\'e Paris-Saclay, Universit\'e Paris Cit\'e, CEA, CNRS, Astrophysique, Instrumentation et Mod\'elisation Paris-Saclay, 91191 Gif-sur-Yvette, France\label{aff84}
\and
Space Science Data Center, Italian Space Agency, via del Politecnico snc, 00133 Roma, Italy\label{aff85}
\and
Centre National d'Etudes Spatiales -- Centre spatial de Toulouse, 18 avenue Edouard Belin, 31401 Toulouse Cedex 9, France\label{aff86}
\and
Instituto de Astrof\'isica de Canarias, Calle V\'ia L\'actea s/n, 38204, San Crist\'obal de La Laguna, Tenerife, Spain\label{aff87}
\and
Departamento de Astrof\'isica, Universidad de La Laguna, 38206, La Laguna, Tenerife, Spain\label{aff88}
\and
Departamento de F\'isica, FCFM, Universidad de Chile, Blanco Encalada 2008, Santiago, Chile\label{aff89}
\and
Universit\"at Innsbruck, Institut f\"ur Astro- und Teilchenphysik, Technikerstr. 25/8, 6020 Innsbruck, Austria\label{aff90}
\and
Satlantis, University Science Park, Sede Bld 48940, Leioa-Bilbao, Spain\label{aff91}
\and
Centro de Investigaciones Energ\'eticas, Medioambientales y Tecnol\'ogicas (CIEMAT), Avenida Complutense 40, 28040 Madrid, Spain\label{aff92}
\and
Instituto de Astrof\'isica e Ci\^encias do Espa\c{c}o, Faculdade de Ci\^encias, Universidade de Lisboa, Tapada da Ajuda, 1349-018 Lisboa, Portugal\label{aff93}
\and
Universidad Polit\'ecnica de Cartagena, Departamento de Electr\'onica y Tecnolog\'ia de Computadoras,  Plaza del Hospital 1, 30202 Cartagena, Spain\label{aff94}
\and
Kapteyn Astronomical Institute, University of Groningen, PO Box 800, 9700 AV Groningen, The Netherlands\label{aff95}
\and
INFN-Bologna, Via Irnerio 46, 40126 Bologna, Italy\label{aff96}
\and
Infrared Processing and Analysis Center, California Institute of Technology, Pasadena, CA 91125, USA\label{aff97}
\and
IFPU, Institute for Fundamental Physics of the Universe, via Beirut 2, 34151 Trieste, Italy\label{aff98}
\and
Center for Computational Astrophysics, Flatiron Institute, 162 5th Avenue, 10010, New York, NY, USA\label{aff99}
\and
School of Physics and Astronomy, Cardiff University, The Parade, Cardiff, CF24 3AA, UK\label{aff100}
\and
University of Applied Sciences and Arts of Northwestern Switzerland, School of Engineering, 5210 Windisch, Switzerland\label{aff101}
\and
Universit\'e St Joseph; Faculty of Sciences, Beirut, Lebanon\label{aff102}
\and
Institut f\"ur Theoretische Physik, University of Heidelberg, Philosophenweg 16, 69120 Heidelberg, Germany\label{aff103}
\and
Junia, EPA department, 41 Bd Vauban, 59800 Lille, France\label{aff104}
\and
Instituto de F\'isica Te\'orica UAM-CSIC, Campus de Cantoblanco, 28049 Madrid, Spain\label{aff105}
\and
CERCA/ISO, Department of Physics, Case Western Reserve University, 10900 Euclid Avenue, Cleveland, OH 44106, USA\label{aff106}
\and
SISSA, International School for Advanced Studies, Via Bonomea 265, 34136 Trieste TS, Italy\label{aff107}
\and
INFN, Sezione di Trieste, Via Valerio 2, 34127 Trieste TS, Italy\label{aff108}
\and
Dipartimento di Fisica e Scienze della Terra, Universit\`a degli Studi di Ferrara, Via Giuseppe Saragat 1, 44122 Ferrara, Italy\label{aff109}
\and
Istituto Nazionale di Fisica Nucleare, Sezione di Ferrara, Via Giuseppe Saragat 1, 44122 Ferrara, Italy\label{aff110}
\and
Universit\'e de Strasbourg, CNRS, Observatoire astronomique de Strasbourg, UMR 7550, 67000 Strasbourg, France\label{aff111}
\and
Dipartimento di Fisica - Sezione di Astronomia, Universit\`a di Trieste, Via Tiepolo 11, 34131 Trieste, Italy\label{aff112}
\and
NASA Ames Research Center, Moffett Field, CA 94035, USA\label{aff113}
\and
Kavli Institute for Particle Astrophysics \& Cosmology (KIPAC), Stanford University, Stanford, CA 94305, USA\label{aff114}
\and
Bay Area Environmental Research Institute, Moffett Field, California 94035, USA\label{aff115}
\and
Minnesota Institute for Astrophysics, University of Minnesota, 116 Church St SE, Minneapolis, MN 55455, USA\label{aff116}
\and
Institute for Astronomy, University of Hawaii, 2680 Woodlawn Drive, Honolulu, HI 96822, USA\label{aff117}
\and
Department of Physics \& Astronomy, University of California Irvine, Irvine CA 92697, USA\label{aff118}
\and
Department of Astronomy \& Physics and Institute for Computational Astrophysics, Saint Mary's University, 923 Robie Street, Halifax, Nova Scotia, B3H 3C3, Canada\label{aff119}
\and
Departamento F\'isica Aplicada, Universidad Polit\'ecnica de Cartagena, Campus Muralla del Mar, 30202 Cartagena, Murcia, Spain\label{aff120}
\and
Dipartimento di Fisica, Universit\`a degli studi di Genova, and INFN-Sezione di Genova, via Dodecaneso 33, 16146, Genova, Italy\label{aff121}
\and
Department of Computer Science, Aalto University, PO Box 15400, Espoo, FI-00 076, Finland\label{aff122}
\and
Ruhr University Bochum, Faculty of Physics and Astronomy, Astronomical Institute (AIRUB), German Centre for Cosmological Lensing (GCCL), 44780 Bochum, Germany\label{aff123}
\and
Univ. Grenoble Alpes, CNRS, Grenoble INP, LPSC-IN2P3, 53, Avenue des Martyrs, 38000, Grenoble, France\label{aff124}
\and
Department of Physics and Astronomy, Vesilinnantie 5, 20014 University of Turku, Finland\label{aff125}
\and
Serco for European Space Agency (ESA), Camino bajo del Castillo, s/n, Urbanizacion Villafranca del Castillo, Villanueva de la Ca\~nada, 28692 Madrid, Spain\label{aff126}
\and
ARC Centre of Excellence for Dark Matter Particle Physics, Melbourne, Australia\label{aff127}
\and
Centre for Astrophysics \& Supercomputing, Swinburne University of Technology, Victoria 3122, Australia\label{aff128}
\and
School of Physics and Astronomy, Queen Mary University of London, Mile End Road, London E1 4NS, UK\label{aff129}
\and
Department of Physics and Astronomy, University of the Western Cape, Bellville, Cape Town, 7535, South Africa\label{aff130}
\and
Oskar Klein Centre for Cosmoparticle Physics, Department of Physics, Stockholm University, Stockholm, SE-106 91, Sweden\label{aff131}
\and
Astrophysics Group, Blackett Laboratory, Imperial College London, London SW7 2AZ, UK\label{aff132}
\and
Centre de Calcul de l'IN2P3/CNRS, 21 avenue Pierre de Coubertin 69627 Villeurbanne Cedex, France\label{aff133}
\and
Dipartimento di Fisica, Sapienza Universit\`a di Roma, Piazzale Aldo Moro 2, 00185 Roma, Italy\label{aff134}
\and
INFN-Sezione di Roma, Piazzale Aldo Moro, 2 - c/o Dipartimento di Fisica, Edificio G. Marconi, 00185 Roma, Italy\label{aff135}
\and
Centro de Astrof\'{\i}sica da Universidade do Porto, Rua das Estrelas, 4150-762 Porto, Portugal\label{aff136}
\and
Dipartimento di Fisica, Universit\`a di Roma Tor Vergata, Via della Ricerca Scientifica 1, Roma, Italy\label{aff137}
\and
INFN, Sezione di Roma 2, Via della Ricerca Scientifica 1, Roma, Italy\label{aff138}
\and
Institute of Space Science, Str. Atomistilor, nr. 409 M\u{a}gurele, Ilfov, 077125, Romania\label{aff139}
\and
Department of Astrophysics, University of Zurich, Winterthurerstrasse 190, 8057 Zurich, Switzerland\label{aff140}
\and
Department of Physics and Astronomy, University of California, Davis, CA 95616, USA\label{aff141}
\and
Department of Astrophysical Sciences, Peyton Hall, Princeton University, Princeton, NJ 08544, USA\label{aff142}
\and
Niels Bohr Institute, University of Copenhagen, Jagtvej 128, 2200 Copenhagen, Denmark\label{aff143}
\and
Cosmic Dawn Center (DAWN)\label{aff144}}        

   \date{Received ; accepted }

\abstract{
The \Euclid space mission will cover over $14\,000\,\rm deg^{2}$ with two optical and near-infrared spectro-photometric instruments, and is expected to detect around ten million active galactic nuclei (AGN). This unique data set will make a considerable impact on our understanding of galaxy evolution in general, and AGN in particular. For this work we identified the best colour selection criteria for AGN, based only on \Euclid photometry or including ancillary photometric observations, such as the data that will be available with the Rubin Legacy Survey of Space and Time (LSST) and observations already available from \textit{Spitzer}/IRAC. The analysis was performed for unobscured AGN, obscured AGN, and composite (AGN and star-forming) objects. We made use of the spectro-photometric realisations of infrared-selected targets at all-$z$ (\spr{}) to create mock catalogues mimicking both the Euclid Wide Survey (EWS) and the Euclid Deep Survey (EDS). Using these mock catalogues, we estimated the best colour selection, maximising the harmonic mean (F1) of: (a) completeness, that is, the fraction of AGN correctly selected with respect to the total AGN sample; and (b) purity, that is, the fraction of AGN inside the selection with respect to the selected sample. The selection of unobscured AGN in both \Euclid surveys (Wide and Deep) is possible with \Euclid photometry alone with $\rm F1=0.22$--0.23 (Wide and Deep), which can increase to $\rm F1=0.43$--0.38 (Wide and Deep) if we limit out study to objects at $z>0.7$. Such a selection is improved once the Rubin/LSST filters, that is, a combination of the $u$, $g$, $r$, or $z$ filters, are considered, reaching an F1 score of 0.84 and 0.86 for the EDS and EWS, respectively. The combination of a \Euclid colour with the $[3.6]-[4.5]$ colour, which is possible only in the EDS, results in an F1 score of 0.59, improving the results using only \Euclid filters, but worse than the selection combining \Euclid and LSST colours. The selection of composite ($f_{{\rm AGN}}=0.05$--0.65 at 8--40$\,\micron$) and obscured AGN is challenging, with $\rm F1\leq0.3$ even when including Rubin/LSST or IRAC filters. This is unsurprising since it is driven by the similarities between the broad-band spectral energy distribution of these AGN and star-forming galaxies in the wavelength range 0.3--5\,$\micron$.
} 


   \maketitle
%
%%%%%%%%%%%%%%%%%%%%%%%%%%%%%%%%%%%%%%%%

%%%%%%%%%%%%%%%%% BODY OF PAPER %%%%%%%%%%%%%%%%%%

\section{Introduction}

Dynamical signatures of supermassive black holes (SMBHs) appear to be ubiquitous at the centres of local galaxies observed with a high sensitivity. The discoveries of scaling relations between the masses of SMBHs and different galactic properties point to a close co-evolution between the two systems \citep[e.g.][]{Magorrian1998,Silk1998,Gebhardt2000,Ferrarese2002,Mullaney2012,Delvecchio2022}, but the physical driver of such an interconnection is still among the most debated topics in extragalactic astrophysics. 
%\par 

Accreting SMBHs shining as active galactic nuclei (AGN) have also been observed in large numbers at different redshifts and environments and thus represent key probes for developing a more comprehensive knowledge of SMBHs growth and demography \citep[see, e.g.][]{Georgakakis2019,Aird2019,Allevato2021,Delvecchio2022}. Recently, the \textit{James Webb} Space Telescope observations uncovered a population of faint AGN at $z\geq5$ \citep[e.g. ][]{Kocevski2023,Barro2023,Labbe2023}, indicating that AGN in the primeval Universe could be more numerous than previously thought. However, the study of AGN is hampered by their relatively low numbers compared to the total number of galaxies (i.e. the low duty cycle). 
%\par

In general, AGN selection is based on emission-line diagnostics \citep[e.g.][]{BPT1981,Feltre2016}, X-ray emission \citep[e.g.][]{Luo2017}, radio observations \citep[e.g.][]{Smolcic2017}, photometric variability \citep[e.g.][]{Butler2011,MacLeod2011,Peters2015}, optical to mid-infrared colour criteria \citep[e.g.][]{Richards2002,Stern2005,Wu2010,Donley2012,Kirkpatrick2013,Wang2016}, or optical-to-infrared spectral decomposition \citep[e.g.][]{Berta2013,Delvecchio2014}. More recently, a new promising alternative consists of the use of various machine-learning algorithms, which can also combine multi-wavelength data sets \citep[e.g.][]{Cavuoti2014,Brescia2015,Jin2019,Khramtsov2019,Shu2019,SanchezSaez2019,Decicco2021,Cunha2022,Doorenbos2022}. However, each of these methods generally traces different physical processes and, unfortunately, none of them is capable of providing a complete census of the AGN population \citep{Delvecchio2017,Lyu2022}.
%\par

In the coming years, the \Euclid space mission \citep{Laureijs2011,Mellier2024} is expected to provide a significant boost to our understanding of galaxy evolution in general, and the role of AGN in particular. Briefly, \Euclid will undertake two surveys: the ${>}\,14\,000\,\rm deg^{2}$ Euclid Wide Survey \citep[EWS;][]{Scaramella2021} and the $53\,\rm deg^{2}$ Euclid Deep Survey (EDS). These surveys will entail broadband optical imaging with the visible imager \citep[\IE\ covering $\lambda=0.53$--0.92$\,\micron$;][]{Cropper2024} and multi-band near-infrared spectroscopy and imaging with the Near-Infrared Spectrometer and Photometer \citep[NISP;][]{Jahnke2024}. The latter will use three filters (\YE, \JE, and \HE) covering from 0.95$\,\micron$ to 2.02$\,\micron$. These data sets are expected to detect about ten million AGN observed with the \Euclid imager (Selwood et al. in preparation) and several hundred thousand with the spectrometer, boosting the number of known AGN enormously. In particular, if we consider the red grism covering between 1.25 and 1.86$\,\micron$, \Euclid is expected to reach a $3.5\,\sigma$ detection limit for emission lines of $(2.5\pm0.6)\times10^{-16} \rm \,erg\,s^{-1}\,cm^{-2}$ in the EWS and $(6.9\pm2.8)\times10^{-17} \rm \,erg\,s^{-1}\,cm^{-2}$ in the EDS \citep{Gabarra2023}. For AGN with fainter nebular emission lines and at redshifts where an emission line diagnostic cannot be used, given the wavelength coverage of NISP, which also include a blue grism (0.92--1.30 $\micron$) that will observe the EDS, the classification of AGN observed with \Euclid will rely on photometric data only. It is therefore of primary importance to identify a robust selection method that takes advantage of all the available photometric data. 
%\par

In this paper we take a first step, using simulated data, towards identifying AGN in the \Euclid surveys by exploring different colour-colour selections based only on \Euclid observed colours or by combining \Euclid with ancillary optical and near-infrared observations. We chose colour-colour selections because they can be quickly applied even to a very large data set; as will be the case with \Euclid, they can be easily compared to previous (or future) selections available in the literature, and they can be directly interpreted by comparing the results with spectral energy distribution (SED) tracks and known features in the templates. Moreover, in this work we do not consider the additional information that could be derived from morphology, which could improve the selection of unobscured AGN, which are generally point-like. We leave more complex methods, such as machine-learning algorithms, as well as a morphological analysis, to future works. We present the simulated \Euclid catalogues in Sect.~\ref{sec:sim}, we explain how we tested different colour-colour selections in Sect.~\ref{sec:col}, and we summarise our main findings in Sect.~\ref{sec:conclusions}. Throughout the paper, we consider a $\Lambda$CDM cosmology with $H_0=70\,{\rm km}\,{\rm s}^{-1}\,{\rm Mpc}^{-1} $, $\Omega_{\rm m}=0.27$, $\Omega_\Lambda=0.73$, and a Chabrier initial mass function  \citep[IMF,][]{Chabrier2003}. All magnitudes are in the AB system \citep{Oke1983}.
%%%%%%%%%%%%%%%%%%%%%%%%%%%%%%%%%%%%%%
\section{Simulated data}\label{sec:sim}
\subsection{\spr}\label{sec:spritz}
The \Euclid simulated catalogues were derived from the spectro-photometric realisations of infrared-selected targets at all-$z$ \citep[\spr{}\footnote{\href{http://spritz.oas.inaf.it/}{http://spritz.oas.inaf.it/}} v1.13;][]{Bisigello2021,Bisigello2022}. Briefly, the simulation starts from a series of observed luminosity functions and galaxy stellar mass functions to derive the number of galaxies expected at different redshifts and infrared (IR) luminosities. In particular, the simulation includes galaxies with IR luminosity from $\logten(L_{\rm IR} L_{\odot}^{-1})=5$ to 15 and out to $z=10$. We did not include galaxies at higher redshifts as we expect them to be faint or completely undetected in the \Euclid filters due to absorption by the intergalactic medium and because the simulation has not been extensively tested at higher redshifts. 

The galaxy populations included in the simulation can be broadly divided in four categories, as follows. 
\begin{itemize}
    \item Star formation dominated systems correspond to galaxies currently forming stars with no evident sign of AGN activity. Among these, spirals and starbursts (SBs) are derived from the observed \hers{} IR luminosity functions \citep{Gruppioni2013}. Spirals have specific star formation rates (sSFR) $\logten({\rm sSFR}/{\rm yr^{-1}})=-10.4$ to $-8.9$, while SBs have $\logten({\rm sSFR}/{\rm yr^{-1}})=-8.8$ to $-8.1$. The simulation also includes dwarf irregulars (Irr), derived from their observed galaxy stellar mass function \citep{Huertas-Company2016,Moffett2016}, which has a characteristic stellar mass (i.e. mass at the knee of the mass function) $\logten(M_{*}/M_{\odot})\leq11$.  Both the infrared luminosity functions and the Irr galaxy stellar mass function have been observed out to $z\simeq3$. The  number of Irrs at $z\geq3$ is obtained by extrapolating the evolution of the number density observed at lower redshifts, while the characteristic stellar mass is kept constant. Similarly, the number of spirals and SBs at $z>3$ is obtained by keeping the characteristic luminosity (i.e. luminosity at the knee of the luminosity function) constant and decreasing the number density as $(1+z)^{-1}$, since this extrapolation gives results consistent with observations at $z\sim6$ \citep{Gruppioni2020}.
    \item AGN-dominated systems are defined here to be galaxies whose mid-infrared emission is dominated by AGN activity. They are split into two populations, AGN1 and AGN2, depending on their optical extinction. Their number densities have been derived by \citet{Bisigello2021} starting from the observed AGN IR and UV observed luminosity functions \citep{Gruppioni2013,Croom2009,McGreer2013,Ross2013,Akiyama2018,Schindler2019}, available, at least partially, out to $z=5$. The characteristic luminosity and the number density of AGN-dominated systems is extrapolated at higher redshfits using the evolution observed at $z\leq5$.  
    \item Composite systems are galaxies whose energetics are dominated by star formation, but which have a faint AGN component. In particular, star-forming AGN (SF-AGN) are galaxies with an intrinsically faint AGN (i.e. $L_{\rm BOL}\leq10^{13}\,L_{\odot}$), while SB-AGN have a bright, but extremely obscured (i.e. $\logten(N_{\rm H}/{\rm cm}^{-2})=23.5-24.5$) AGN. They are derived starting from the observed \hers luminosity function \citep{Gruppioni2013}, with some updates reported here in Appendix~\ref{sec:newspritz}. As for spirals and SBs, the IR luminosity functions of composite systems are extrapolated at $z>3$, keeping the characteristic luminosity constant and decreasing the number density as $(1+z)^{-1}$.
    \item Passive galaxies are elliptical galaxies (Ell) derived from the observed $K$-band luminosity functions \citep{Arnouts2007,Cirasuolo2007,Beare2019}. Some of these galaxies may host an obscured AGN (Ell-AGN, see Appendix~\ref{sec:newspritz}). At $z > 2$ the luminoity function is extrapolated, starting from the observed K-band luminosity function by \citet{Cirasuolo2007}, by keeping the characteristic luminosity constant and decreasing the number density as $(1+z)^{-1}$.
\end{itemize}
Based on these galaxy populations, we assign to each simulated galaxy a set of SED models \citep{Polletta2007,Rieke2009,Gruppioni2010,Bianchi2018} to extract the photometric fluxes expected in different filters. The probability of each template varies with redshift and IR luminosity, following \hers{} observations, when available.  A large set of physical properties, such as stellar mass, star formation rate (SFR), accretion luminosity, and metallicity are then assigned considering available theoretical or empirical relations or by fitting the empirical SED assigned to each simulated galaxy using 
%{\sc sed3fit} 
the software
{\tt SED3FIT} 
\citep[for AGN;][]{Berta2013} or 
%{\sc Magphys} 
the multi-wavelength analysis of galaxy physical properties
{\tt MAGPHYS}
\citep[for non-active objects;][]{daCunha2008}. 
%\par 

The presence of strong emission lines inside broad-band filters can create a boost to the observed fluxes \citep[e.g.][]{Stark2013,Bisigello2017}. Therefore, before deriving the \Euclid, Rubin/LSST, and \textit{Spitzer} expected fluxes, we first used the best-fit templates previously mentioned to recover the stellar continuum and remove any nebular emission line already included in the empirical templates. This is performed in order to include line ratios that are not fixed for each template but may change with other physical properties, such as gas metallicity. Then, we incorporated optical emission lines due to star formation (i.e. hydrogen lines, $[\ion{\rm O}{II}]$3727, $[\ion{\rm Ne}{III}]$3869, $[\ion{\rm N}{II}]$6548,6584, $[\ion{\rm O}{III}]$4959,5007, $[\ion{\rm S}{II}]$6717,6731, and $[\ion{\rm S}{III}]$9069,9532]) using a set of empirical relations starting from the SFR \citep{Kennicutt1998,Pettini2004,Kewley2013,Jones2015,Dopita2016,Kashino2019,Proxauf2014,Mingozzi2020}. Given its complex nature, the Ly$\alpha$ line was not included in the simulation, but this only impacts the photometry of the most star-forming galaxies at $z>5$. 
%\par

We also included the contribution to the emission lines mentioned above, from the narrow-line gas emitting regions, by considering the theoretical predictions from the photo-ionisation models developed by \citet{Feltre2016} using \texttt{CLOUDY} \citep[version c13.3;][]{Ferland2013}. The emission coming from the broad-line regions was not modelled and therefore emission lines of AGN1 should be considered as lower limits, as they contain only the narrow-line component. The impact of this underestimation depends on the redshift and filters considered and thus we discuss it later along with the results (Sect.~\ref{sec:col}). We refer to section 2.3 of \citet{Bisigello2021} for more details regarding emission lines. 
%\par

Overall, \spr{} is consistent with a large set of observations out to $z\simeq6$, including luminosity functions and number counts from X-rays to radio, AGN diagnostic diagrams \citep[e.g.][and Appendix~\ref{sec:colsec_spritz}]{BPT1981}, the global galaxy stellar mass function, and the SFR versus stellar mass plane. We refer to \citet{Bisigello2021,Bisigello2022} for more comparisons with observations, while in Appendix~\ref{sec:colsec_spritz} we verify the accuracy of the AGN near-IR colours in the simulation. 

\subsubsection{\spr{} simulated \Euclid catalogues} \label{sec:cat}
With \spr\ we simulated two different catalogues, mimicking the EWS and the EDS. We perturbed the flux densities associated with each simulated galaxy, considering a Gaussian function with a standard deviation equal to the expected photometric uncertainties. For the \Euclid filters, the standard deviation associated with each flux density is the sum, in quadrature, of the background flux density error ($\sigma_{\rm bkg}$) and the photon noise ($\sigma_{\rm noise}$).\footnote{The entire procedure described here is taken from 
 \href{https://github.com/jcoupon/euclid_phz_testing}{https://github.com/jcoupon/euclid\_phz\_testing}} The latter is derived as
\begin{equation}
    \sigma_{\rm noise}=\frac{f_{5\,\sigma}}{5}\frac{r}{r_{\rm ref}}\;,
\end{equation}
where $f_{5\,\sigma}$ is the expected survey depth at $5\,\sigma$, $r$ is the radius of each simulated galaxy based on an assumed galaxy mass-size relation \citep{Wel2014}, and $r_{\rm ref}$ is the median effective radius of galaxies with flux densities equal to $f_{5\,\sigma}$. In our simulated catalogues this is equal to \ang{;;0.25} for the EWS and \ang{;;0.16} for the EDS. In \spr, we assumed two different mass-size relations for star-forming and passive galaxies, but we did not force the radius of AGN1 to be equal to the filter point-spread-function, which is the case for unresolved galaxies, but it depends on the contribution of the AGN in the different filters. This has a negligible impact of the final S/N, with an overestimation below $30\%$ for all filters and the larger difference happening for the most massive, hence the brightest, objects.
The background noise was derived as
\begin{equation}
    \sigma_{\rm bkg}=\sigma_{\rm noise}\sqrt{\frac{f}{f_{\rm sky}\,\pi\, r^{2}}}\;,
\end{equation}
where $f_{\rm sky}$ is the reference sky surface background and corresponds to 22.33, 22.10, 22.11, and 22.28 AB ${\rm mag}\,{\rm arcsec}^{-2}$ in the \IE, \YE, \JE, and \HE\ filters, respectively. We do not include Galactic extinction in the simulation, assuming observed magnitudes will be corrected accordingly.
The observational depths expected in the EWS are presented in \citet{Scaramella2021} and reported in Table~\ref{tab:depths}, while we expect the EDS to be 2 magnitudes deeper. 
%\par

\subsubsection{\spr{} simulated LSST catalogues} \label{sec:cat_lsst}
It is necessary to take into account that the depths of the optical ancillary data will evolve with time. In particular, in the north, the Ultraviolet Near-Infrared Optical Northern Survey \citep[UNIONS;][]{Ibata2017} is gathering observations in the $u$, $g$, $r$, $i$, and $z$ filters to reach final $5\,\sigma$ depths of 24.35, 25.25, 24.85, 24.35, and 24.15 mag, respectively. Among these, only the depth of the $i$ band is expected to evolve with time, from 23.95 to 24.35 from the first to the fifth year of \Euclid observations (i.e. first and third data release, DR). In the south, the first year of \Euclid observations\footnote{Assuming the Rubin/LSST survey starts in spring 2024.} will rely on the public Dark Energy Survey \citep[DES;][]{DES2}, which includes $g$, $r$, $i$, and $z$ filters at $5\,\sigma$ depths of 25.25, 24.85, 23.95, and 24.15 mag, respectively. The absence of a $u$ filter precludes the use of some of the colour selections considered in the next sections. From the second year of \Euclid observations, Rubin/LSST data should be available, reaching depths of 25.90, 27.10, 27.20, 26.50, and 25.80 in the $u$, $g$, $r$, $i$, and $z$ filters at the fifth year of \Euclid observations \citep{Guy2022}. For the first Rubin/LSST year of observations we assumed $5\,\sigma$ depths of 24.45, 25.65, 25.75, 25.05, and 24.35 in the $u$, $g$, $r$, $i$, and $z$ filters \citep{Brandt2018}. All the considered observational depths are summarised in Table~\ref{tab:depths} and  we assume that the Euclid Deep fields will have Rubin/LSST data 2 magnitudes deeper than the depths available in the EWS. The main results presented in this work are derived considering the depths of the \Euclid DR3 combined with Rubin/LSST data available in the south, but we analyse the evolution of our selection criteria with time/depth in Sect.~\ref{sec:time}. The photometric noise for the ancillary filters was derived by directly scaling the expected survey depths. For information on the methods used to combine \Euclid and optical ancillary data, we refer to \citet{Mellier2024} and \citet{Guy2022}.
%\par

\subsubsection{\spr{} simulated IRAC catalogues} \label{sec:cat_irac}
As presented in detail in \citet{Moneti2022}, the EDF is also covered by \textit{Spitzer} observations. However, while there is complete coverage from the two filters at the shortest wavelengths, i.e. [3.6] and [4.5], only $21.8\,\rm deg^{2}$ of the field  are covered with the [5.6] and [8.0] filters. Therefore, the selections by \citet{Stern2005} and \citet{Lacy2004}, which make use of all four filters, cannot be applied to the majority of the objects. Instead it would be possible to apply a single IRAC colour selection, such as $[3.6]-[4.5]>0.16$, as presented by \citet{Stern2012}. Therefore, we also tested if the additional information given by the $[3.6]-[4.5]$ colour could be used to improve the selection based on the \Euclid filters. The observational depths of the IRAC bands vary a lot around the field, given that observations are taken from different legacy programmes. The depths of the \textit{Spitzer} observations are not uniform, given that it is a collection of different surveys. In particular, the 5$\sigma$ depths are 23.9 at 3.6$\,\micron$ and 23.8 at 4.5$\,\micron$ in the EDF-South (EDF-S), while they are 24.8 at 3.6$\,\micron$ and 24.7 at 4.5$\,\micron$ in the EDF-North (EDF-N) and EDF-Fornax (EDF-F; EC: McPartland in prep.) fields. In this work, selections are derived considering the deepest fields, but we verify their effectiveness also in the EDF-S.

\begin{figure}
    \centering
    \includegraphics[width=\linewidth, keepaspectratio]{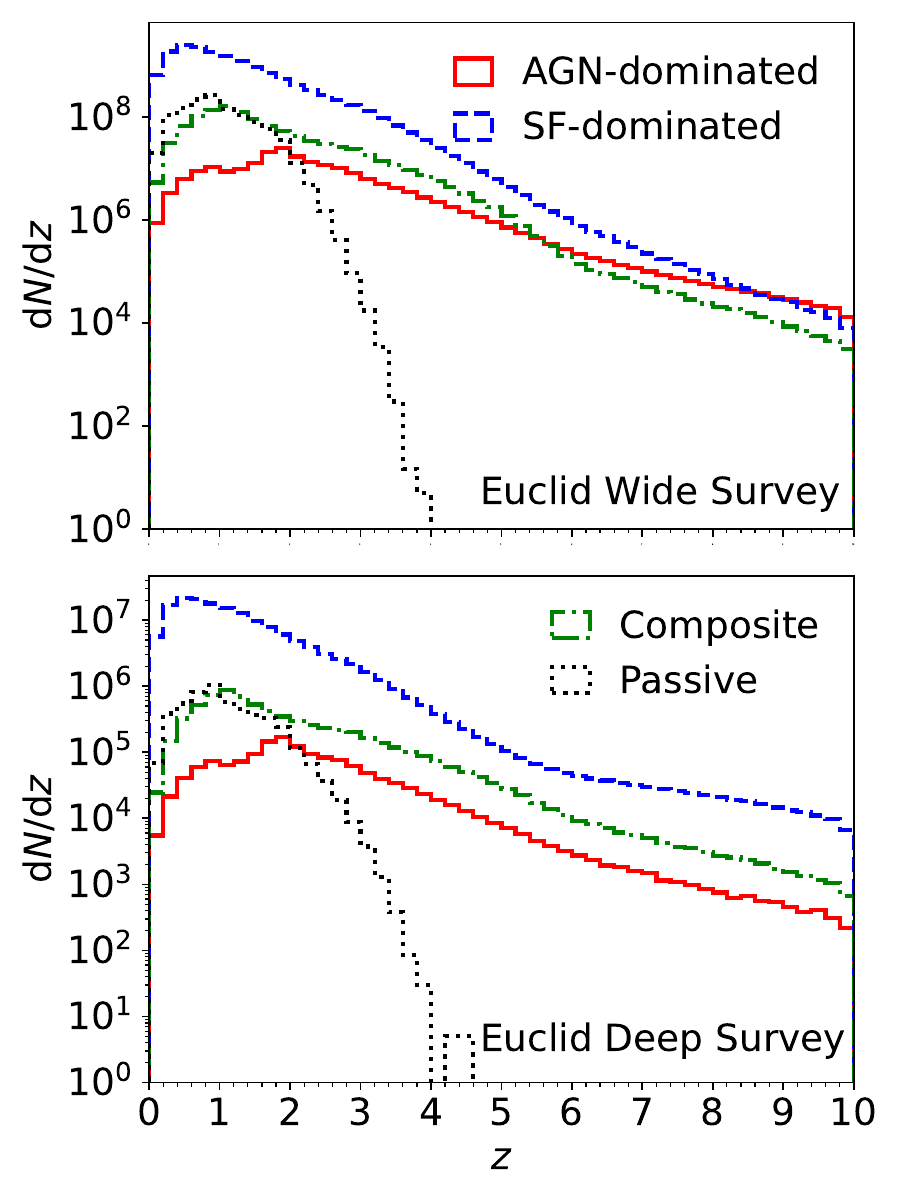}
    \caption{Redshift distribution of AGN-dominated, SF-dominated, composite systems, and passive galaxies, observed (${\rm S/N}>3$) in at least one \Euclid filter, as derived in the simulated \Euclid catalogues obtained from \spr{} for the EWS (top) and the EDS (bottom).  We use a redshift bin width of 0.2 for the histograms.}
    \label{fig:zdist_SPRITZ}
\end{figure}

The EDS and EWS simulated catalogues contain only galaxies detected (signal-to-noise ratio ${\rm S/N}>3$) in at least one \Euclid filter, scaling the noise levels starting from those reported in Table~\ref{tab:depths}. The redshift distributions of the  galaxy populations in the two samples are shown in Fig.~\ref{fig:zdist_SPRITZ}. When testing each colour selection, we included only the sub-sample of galaxies with ${\rm S/N}>3$ in the four filters considered in the selection. In Appendix~\ref{sec:Selection} we investigate the biases introduced by this additional requirement.

\begin{table}[htbp!]
    \caption{Observational depths.}
    \centering
    \begin{tabular}{cccccc}
        \hline\hline
        \noalign{\vskip 1pt}
        & \multicolumn{3}{c}{South} & \multicolumn{2}{c}{North} \\
        Filter & DR1 & DR2 & DR3  & DR1 & DR3 \\
        \hline
        \noalign{\vskip 1pt}
        \rule{0pt}{\dimexpr.7\normalbaselineskip+1mm}
        \IE\ & 26.20 & 26.20 & 26.20 & 26.20 & 26.20\\
        \YE\ & 24.30 & 24.30 & 24.30 & 24.30 & 24.30\\
        \JE\ & 24.50 & 24.50 & 24.50 & 24.50 & 24.50\\
        \HE\ & 24.40 & 24.40 & 24.40 & 24.40 & 24.40\\
        $u$ & -- & 24.45 & 25.90 & 24.35 & 24.35\\
        $g$ & 25.25 & 25.65 & 27.10 & 25.25 & 25.25\\
        $r$ & 24.85 & 25.75 & 27.20 & 24.85 & 24.85\\
        $i$ & 23.95 & 25.05 & 26.50 & 23.95 & 24.35\\
        $z$ & 24.15 & 24.35 & 25.80 & 24.15 & 24.15\\
        \hline
    \end{tabular}
    \label{tab:depths}
    \tablefoot{The top table shows the observational depths ($5\,\sigma$) expected for different \Euclid data releases (DRs). To create the EWS mock catalogue we consider as reference the depths of DR3 in the south. The optical depths refer to UNIONS in the north and to DES (DR1) and Rubin/LSST (DR2-DR3) in the south (see Sec. \ref{sec:cat_lsst}). The observational depths of the EDS are 2 magnitudes deeper.}
\end{table}

\subsection{\Euclid mocks from observed galaxies}
To validate results obtained with the \Euclid mock catalogues described above, we obtained additional mock catalogues starting from observed galaxies using a complementary approach. In particular, we took advantage of the publicly available Kilo-Degree Survey \citep[KiDS;][]{deJong2013} Bright ExtraGalactic Objects \citep[KiDS-BEXGO;][]{Khramtsov2019}, which is based on KiDS Data Release 4 \citep{Kuijken2019} and the VISTA (Visible and Infrared Survey Telescope for Astronomy) Kilo-degree Infrared Galaxy \citep[VIKING;][]{Edge2013} survey. The KiDS-BEXGO catalogue was created using a tree-based machine-learning algorithm and contains $2\times10^5$ quasi-stellar objects (QSOs) and $4\times10^{6}$ galaxies with magnitude $r<22$. The catalogue contains  9-band Gaussian-aperture and point-spread-function \citep[{\tt GAaP};][]{Kuijken2015} magnitudes ($u$, $g$, $r$, $i$, $Z$, $Y$, $J$, $H$, and $K_{\rm s}$) in AB units and has been corrected for Galactic extinction (using the \citealt{SF11} prescription). To select objects with the most reliable photometry, we conservatively removed masked regions (e.g. with star halos). Then, we selected galaxies and QSOs considering a probability above 0.8 to be one or the other, as reported in the KiDS-BEXGO catalogue. The catalogue was constructed and tested on QSOs and does not include composite systems.
%\par

\citet{euclid-nisp2022} presented relations to convert the fluxes from several near-IR filters to the NISP only filters. However, such transformations are not available for the \IE\ filter and were derived for a general galaxy population without AGN. Following the method exploited in \cite{Ginolfi+20}, we therefore used a principal component analysis (PCA) approach to link the available KiDS+VIKING filters to the \Euclid ones, using two different template sets for QSOs and galaxies. PCA is a parameter transformation technique that diagonalizes the covariance matrix of a set of variables (magnitudes in our case). As a result, PCA produces the linear combinations of magnitudes, known as eigenvectors (also called principal components), which define the orientations of a hyper-plane. These eigenvectors minimise the covariance and are, by definition, mutually orthogonal. We applied the PCA to sets of four magnitudes, safely assuming that the correlations among the different magnitudes can be considered approximately linear. Among the four principal components the one with the least variance is considered. Therefore this component can be set to zero and inverted to give a prediction of the most dominant magnitude, which depends on the others. The accuracy of the prediction was then calculated by fitting the residuals to a Gaussian function.

In particular, for QSOs we used as reference a set of 28 empirical and synthetic QSO templates at $z<6$ taken from the {\tt LePhare} software \citep{Arnouts1999,Ilbert2006}.\footnote{These templates include the SEDs named Sey2, QSO1, BQSO1, TQSO1, QSO2, Torus, Mrk231,  I19254, and N6240 from \citet{Polletta2007} and the templates with 90, 80, and 70\% AGN contribution from \citet{Salvato2009}.}
We derived the following relations linking KiDS and VIKING filters to \Euclid ones for QSOs,
\begin{align}
\IE&=0.058+0.354\,r+0.451\,i+0.195\,Z\;, \nonumber \\ 
\YE&=0.011+0.657\,Y+0.377\,J-0.035\,H\;, \label{eq:pca_qso} \\ 
\JE&=0.026-0.056\,Y+0.748\,J+0.307\,H\;, \nonumber \\ 
\HE&=-0.019+0.0133\,Y-0.233\,J+ 1.219\,H\;, \nonumber
\end{align}
where $Z$, $Y$, $J$, and $H$ are the magnitudes of the VIKING filters, and $r$ and $i$ are the magnitudes of the KiDS filters. 
With this method, the \IE, \YE, \JE, and \HE\ magnitudes are derived with a scatter of 0.05, 0.02, 0.03, and 0.05 mag, respectively. 
%\par

To create mock \Euclid and Rubin/LSST magnitudes for galaxies, we adopted \citet{Bruzual2003} synthetic models, redshifted to $0 < z < 3$, assuming a \citet{Chabrier2003} IMF, stellar metallicities in the range $0.2$--$2.5 \, Z_{\odot}$, an exponentially declining star formation history with time duration $\tau$ ranging from 0.1 to 30\,Gyr, and galaxy ages up to 13.5\,Gyr. Internal extinction is accounted for by using the \citet{Calzetti2000} extinction curve with $E(B-V) = 0$, 0.1, 0.2, 0.3, 0.4, and 0.5, and emission lines are added, using the prescription provided in the {\tt LePhare} software \citep{Arnouts1999,Ilbert2006}. 
Using the PCA analysis we found that for galaxies, the $Y$, $J$, and $H$ VIKING bands are good proxies for the \Euclid \YE, \JE, and \HE\ bands. On the other hand, for \IE\ we derived the following relation:
\begin{equation}
    \IE=0.149+0.369\,r + 0.475\,i + 0.154\,Z\;,
\end{equation}
with a scatter of 0.03 mag.

We used these relations to derive the \Euclid magnitudes for all objects (QSOs and galaxies) in the KiDS-BEXGO catalogue, which is generally limited to $z<1$. We did not include any additional photometric error, in addition to the KiDS-BEXGO uncertainties, as this survey is shallower than the depth expected for both \Euclid surveys. These two Euclidised catalogues have the advantage of being based on real observed galaxies, but the derived colours may be intrinsically correlated, given the PCA analysis.

%%%%%%%%%%%%%%%%%%%%%%%%%%%%%%%%%%%%%%%%%
\section{AGN colour selections }\label{sec:col}
\subsection{Considered colour selections}
\begin{figure}
    \centering
    \includegraphics[width=\linewidth, keepaspectratio]{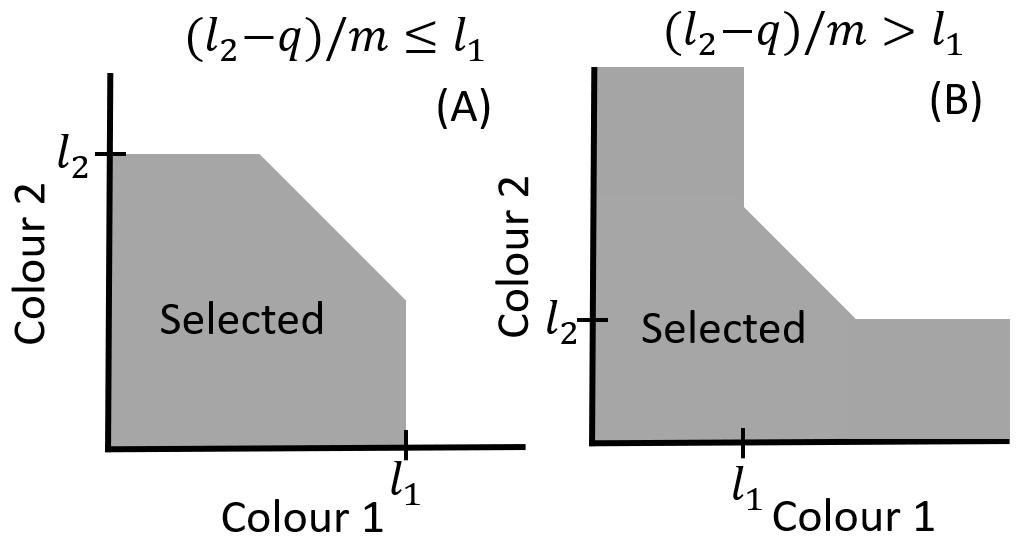}
    \caption{Sketch of the two selection criteria considered in this work, referred to as type~A (left) and type~B (right). The grey areas indicate the selected regions. The diagonal separation is defined by a line with slope $m$ and vertical intercept $q$.}
    \label{fig:Sel}
\end{figure}

Given the wavelength coverage of \Euclid filters, we expect AGN1 
(i.e. unobscured AGN-dominated systems) to be the easiest AGN population to identify, since the optical continuum fluxes of AGN2 (i.e. obscured AGN-dominated systems) and composite systems are dominated or highly contaminated by the stellar emission, while nebular emission lines coming from the AGN component may contribute too little to the overall broad-band fluxes. Indeed, we take the \YE\ filter as an extreme example, as it is the narrowest among the \Euclid filters and therefore the more sensitive to nebular emission lines. In this case, to produce a boost with respect to the continuum in the observed flux 3 times above the noise, a line needs to have an equivalent width above $1800\,\AA$, $1000\,\AA$, and $500\,\AA$ at $z=0$, $z=0.7$, and $z=2.4$, respectively. The same line needs to be brighter to produce a similar boost with respect to the continuum in the other \Euclid filters, since they are wider than the \YE\ filter. For this reason, in the next sections we first investigated a possible AGN1 colour selection and, then, a selection for all AGN included in \spr. These colour selections were analysed considering only the four \Euclid filters, including both \Euclid and Rubin/LSST filters, or both \Euclid and \textit{Spitzer} 3.6- and 4.5-$\micron$ filters. In the following tests we separately considered the depths of the EDS and EWS, and we assumed that stars have been previously selected and removed from the sample. 
%\par

We tested two different selection criteria (Fig.~\ref{fig:Sel}), which, in the colour-colour plane, resemble in shape other colour selections present in the literature that focused on AGN \citep[e.g.][]{Lacy2004,Stern2005}. The considered selections are specific for the filters used in this work and correspond to

\begin{align}\label{eq:sel}
(c_{1}<l_{1})\, \land\,(c_{2}<l_{2})\,\land\,(c_{2}<m\,c_{1}+q), \;\text{if } (l_2-q)&/m\leq l_{1}\\
&\rm{(type\,A)},\nonumber \\
(c_{1}<l_{1})\, \lor\,(c_{2}<l_{2})\,\lor\,(c_{2}<m\,c_{1}+q), \;\text{if } (l_2-q)&/m> l_{1} \\
&\rm{(type\,B)},\nonumber
\end{align}

where $c_{1}$ and $c_{2}$ indicate two different colours, $l_{1}\in[-1,3)$ and $l_{2}\in[-1,3)$ indicate the threshold in colour of the selection, while $m\in[-4,0)$ and $q\in[0,4)$ indicate the slope and intercept of the additional diagonal selection. We highlight that the ranges spanned by the four parameters also include criteria in which the additional diagonal selection is negligible or absent, that is, $(l_2-q)/m=l_{1}$. For all four variables we considered steps of 0.1. We also tested a finer grid with steps of 0.01, but the improvement in the selection was negligible (i.e. less than 0.1$\%$ in the metrics explained below when selecting AGN1 in the EDS). Throughout the text, we refer to the first selection, which includes the logical AND operator  ($\land$), as type~A and the second selection, which includes the logical OR operator  ($\lor$), as type~B. 
%\par

\subsubsection{Evaluation metrics}
To evaluate the quality of each colour-colour selection, we define several useful metrics. The first is the purity of the selection, defined as
\begin{equation}
    P:=\frac{N_{\rm TP}}{N_{\rm TP}+N_{\rm FP}}\;,
\end{equation}
where $N_{\rm TP}$ is the number of true positives, that is, AGN correctly recovered, and $N_{\rm FP}$ is the number of false positive, that is, non-active galaxies wrongly identified as AGN.
We also derived the completeness of the selection, which is defined as
\begin{equation}
    C:=\frac{N_{\rm TP}}{N_{\rm TP}+N_{\rm FN}}\;,
\end{equation}
where $N_{\rm FN}$ is the number of false negative, that is, AGN not recovered by the selection.
Then, we calculated the F1 score \citep{Dice1945,Sorensen1948}, which is the harmonic mean of the purity and completeness, as a useful metric to quantify the selection quality,
\begin{equation}
    {\rm F1}:=2\, \frac{P\, C}{P+C}\;.
\end{equation}
All three quantities, that is, purity, completeness, and F1 score, range from a minimum of 0, indicating a poor selection, to a maximum of 1 for an optimal selection. In this work we define the best selection criterion as the one that maximises the F1 score, but we also report the best criteria with $P>0.9$ or $C>0.9$, for the science cases that prefer purity over completeness or the opposite. We chose the harmonic mean, instead of, for example, the arithmetic or quadratic means, because it under-weights extreme uninformative cases, for example, selecting the entire galaxy sample may result in $C=1$ and $P=0.01$, since AGN are a minority, and this corresponds to a low $\rm F1=0.18$ but to a high arithmetic mean of around 0.5. In the next section, when we select AGN1, we consider all the other objects, including other AGN, as contaminants. On the other hand, when selecting all AGN, we consider only inactive galaxies as contaminants. 
The F1 scores derived for all colour combinations are reported in Appendix~\ref{sec:tables}, while in the next sections we limit our analysis to the best colour selections.
\subsubsection{Bootstrap analysis}\label{sec:bootstrap}
In order to assess the uncertainties on the boundaries of each colour selection and the corresponding metrics, we performed the following bootstrap analysis. Each time we derive an optimise colour selection for a specific survey depth or filter combination we repeated the analysis for that specific colour, but for all parameters defining a selection with F1 score larger than 80$\%$ of the maximum F1 value. We did not consider the entire parameter space for computational reasons. We then randomised the fluxes considering the respective flux uncertainties, the expected sky background for \Euclid filters (see Sect.~\ref{sec:cat}), and the scatter for the PCA analysis for KiDS objects. We performed the bootstrap analysis, considering the entire sample in the EDS and only $1\%$ of the objects in EWS, selected randomly. We finally derived a new selection criterion that maximises the F1 score. We repeated the entire procedure 10 times to obtain uncertainties associated to each best selection criteria and the associated F1 score, completeness, and purity.
%%%%%%%%%%%%%%
\subsection{\Euclid-only colours}
We now present the best colour selections using only \Euclid filters. We first focus on the EDS and we then move to the EWS.
\subsubsection{AGN1 in the EDS}
\begin{figure}
    \centering
    \includegraphics[width=\linewidth, keepaspectratio]{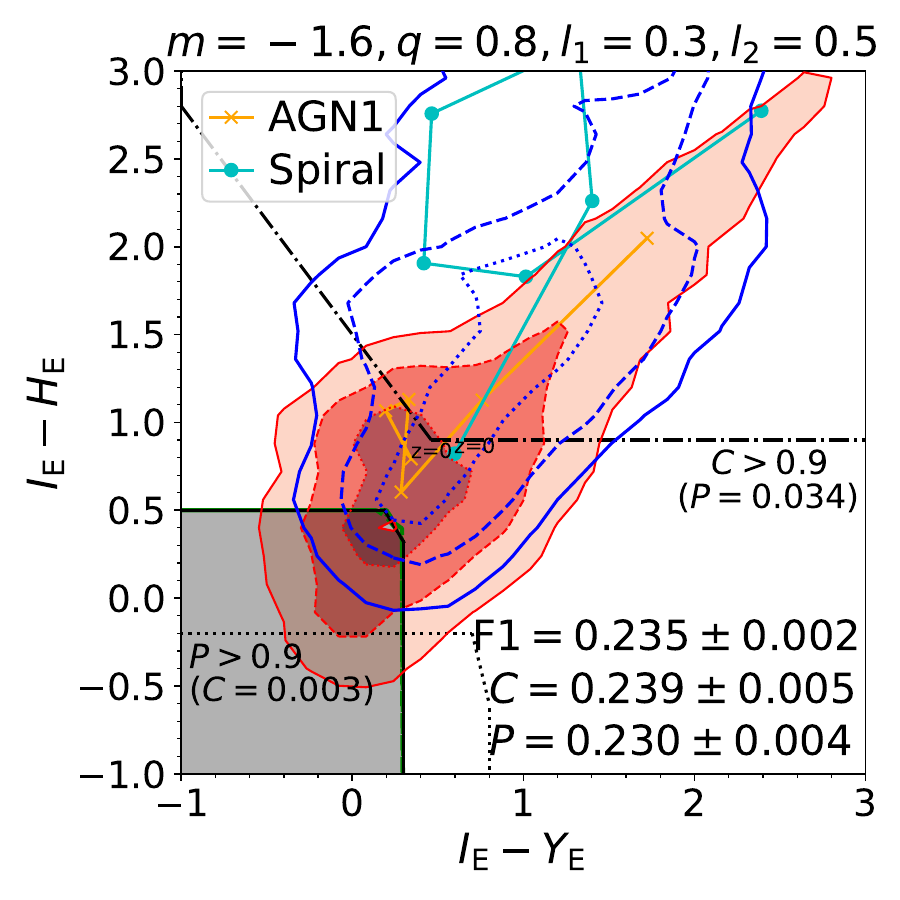}
    \caption{Best selection (type~A) criterion for AGN1 in the EDS using only \Euclid filters (shaded grey area). The shaded red areas correspond to the colour distribution of AGN1, while the blue lines are the contour levels of the remaining galaxies. Levels correspond to 68\%, 95\%, and 99.7\% of the distribution. We also present example SED tracks for one AGN1 (yellow crosses) and one spiral galaxy (cyan circles) from $z=0$ to $z=6$, with steps of $\Delta z=1$. As an indication of how strict the best selection is, the two dashed green lines, which almost completely overlap with the black one, show the extent of all the best selection criteria derived with the bootstrap approach. The dotted black line and the dash-dotted black line correspond to the best selection with $P>0.9$ and $C>0.9$, respectively.}
    \label{fig:Best_Deep_Euclid}
\end{figure}

In Fig.~\ref{fig:Best_Deep_Euclid} we show the best selection criterion, which corresponds to a type~A criterion, recovered for AGN1, derived considering the area and depth expected for the EDS:
\begin{align}\label{eq:AGN1_Deep}
      \left(\IE-\YE<0.3_{-0.0}^{+0.0}\right) \,&\land\,
      \left(\IE-\HE<0.5_{-0.0}^{+0.0}\right) \,\\
      &\land\,
      \left[\IE-\HE<-1.6_{-0.0}^{+0.7}\,(\IE-\YE)+0.8_{-0.1}^{+0.0}\right] \;. \nonumber     
\end{align}
This colour selection corresponds to a completeness $C=0.239\pm0.005$, a purity $P=0.230\pm0.004$, and $\rm F1=0.235\pm0.002$. Uncertainties were estimated by a bootstrap analysis, as described in Sect.~\ref{sec:bootstrap}. 
If we consider a different colour selection that includes all four \Euclid filters, that is, $\IE-\YE$ and $\JE-\HE$, we obtain a smaller F1 score, $\rm F1=0.189$, since the slope between the \JE\ and \HE\ filters is probably less informative than the slope between the \IE\ and \HE\ filters. Similarly, if we use a three-colour criterion combining $\IE-\YE$, $\YE-\JE$, and $\JE-\HE$, the F1 score is $\rm F1=0.239$ ($P=0.228$, $C=0.250$), showing little improvement with respect to the two-colour selection in Eq. \ref{eq:AGN1_Deep} and reducing the AGN1 sample by 2\%, as it requires detection in all the four filters. The statistics of all the other two-colours criteria for AGN1 in the EDS using only \Euclid filters are listed in Table~\ref{tab:Deep_Euclid}.
%\par

\begin{figure}
    \centering
    \includegraphics[width=\linewidth, trim={23 28 0 0},clip, keepaspectratio]{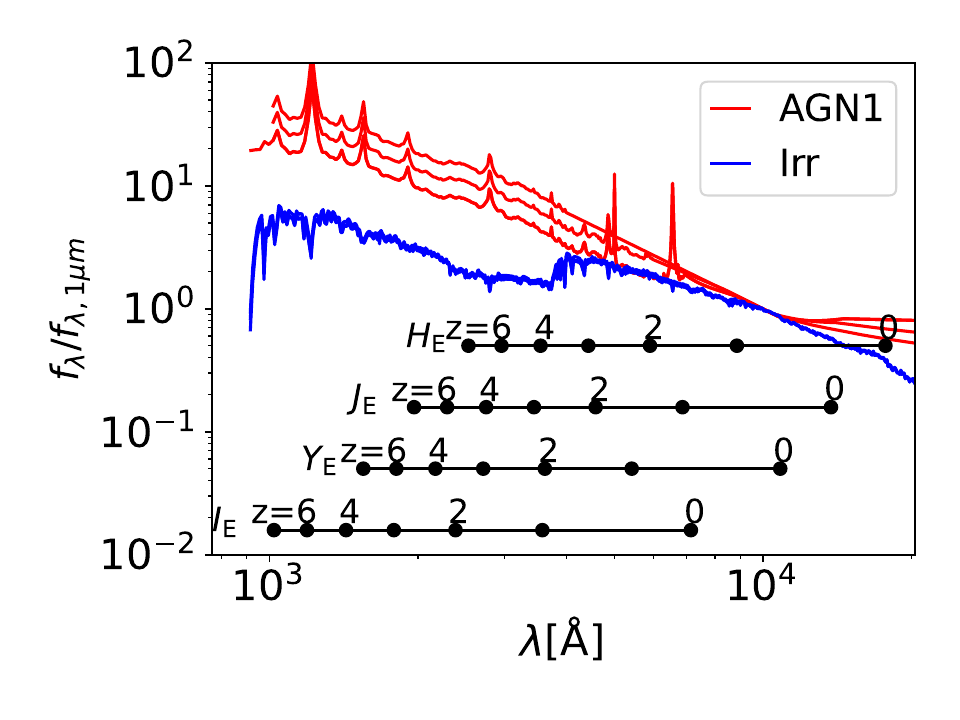}
    \caption{SEDs of AGN1 (solid red lines) and dwarf irregulars (solid blue line), normalised at 1$\,\micron$. We also report the rest-frame wavelength observed by the \Euclid filters at different redshifts (black circles).}
    \label{fig:SED_AGN1Irr}
\end{figure}

In detail, among the galaxies that are wrongly identified as AGN1 by the colour selection in Eq. (\ref{eq:AGN1_Deep}), 99.9\% are dwarf irregular and 0.1\% are spiral galaxies. Dwarf irregulars have a young stellar population with a weak $4000\,\AA$ break. Therefore the part of their SEDs observed by \Euclid mimics AGN1 at $z<1$ (Fig.~\ref{fig:SED_AGN1Irr}). On the one hand, we expect dwarf irregulars to be more difficult to observe with \Euclid at increasing redshift, given that their mean redshift in our mock catalogue is $z=1.1$ and $z=1.3$ in the EWS and EDS, respectively. On the other hand, high-$z$ star-forming galaxies may have an SED similar to dwarf galaxies, given their increasingly smaller stellar masses and higher SFR, but at $z>1$ the $4000\,\AA$ break is inside \Euclid's wavelength coverage and helps to discriminate among these different galaxy populations.

\begin{figure}
    \centering
    \includegraphics[width=\linewidth, keepaspectratio]{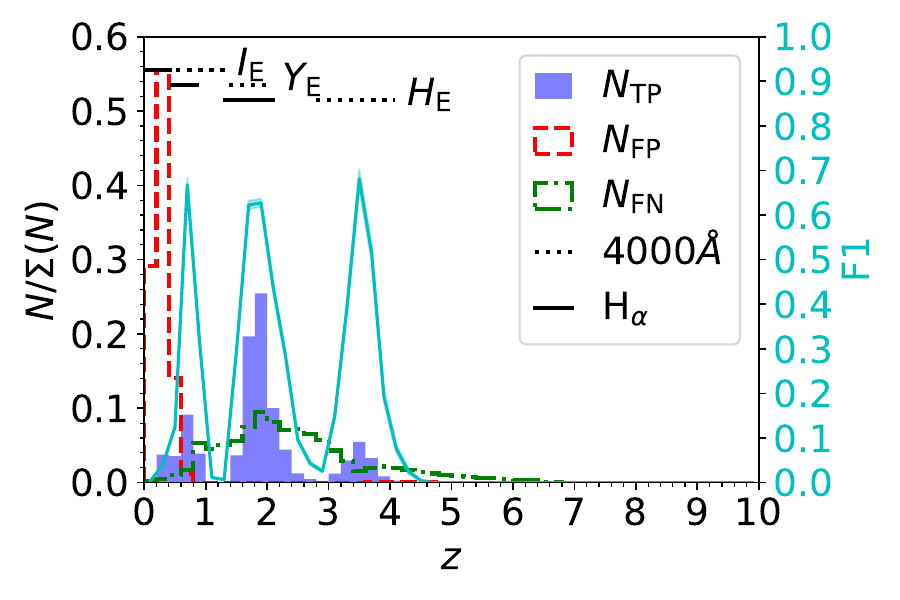}
    \caption{Normalised redshift distributions of the $N_{\rm TP}$ (blue filled histogram), $N_{\rm FP}$ (dashed red histogram), and $N_{\rm FN}$ (dot-dashed green histogram) for the \Euclid-only selection criteria (Eq.~\ref{eq:AGN1_Deep}) of AGN1 in the EDS. The horizontal solid and dotted lines show the redshift range in which H${\alpha}$ and the $4000\,\AA$-break, respectively, are inside the \IE, \YE, or \HE\ filters (\JE\ is not used in Eq.~\ref{eq:AGN1_Deep}). We also report the variation of the F1 score with redshift (cyan solid line and right vertical axis). }
    \label{fig:z_AGN1}
\end{figure}

We further investigate the best colour criterion reported in Eq. (\ref{eq:AGN1_Deep}) by considering how the F1 score varies with redshift as well as analysing the normalised redshift distribution of $N_{\rm FP}$, $N_{\rm FN}$, and $N_{\rm TP}$ (Fig.~\ref{fig:z_AGN1}). For this analysis, the redshift is the true value and it is directly taken from the simulation. $N_{\rm FP}$, $N_{\rm FN}$, and $N_{\rm TP}$ are derived considering the sub-sample of galaxies in each redshift bin ($\Delta z=0.1$) and without considering possible redshift uncertainties, which depends on the method used for the redshift estimation \citep[][Euclid Collaboration: Humphrey
et al. in prep.]{Euclidz2020,Bisigello2023} and goes beyond the scope of this work. From Fig.~\ref{fig:z_AGN1} it is evident that the selection does not perform equally well at all redshifts, with the F1 score showing peaks at $z=0.7$, $z=1.9$, and $z=3.5$ with values at $\rm F1\simeq0.66$. These peaks are due to the presence inside one of the \Euclid  filters of the $4000\,\AA$ break, which is not prominent in AGN1. At $z>6.6$ there are no AGN1 detected in both \Euclid colours, but all AGN1 move out of the colour selection already at $z=4.4$, decreasing the $\rm F1$ score to $0$. 
%\par

Given the variation of the F1 score with redshift, the AGN1 selection criteria presented in Eq. (\ref{eq:AGN1_Deep}) is particularly effective in the redshift range $0.7\leq z\leq4.4$. When limiting only to those redshifts, considering that this redshift selection depends on the estimated redshift uncertainties that are not included and, moreover, on a a galaxy/AGN pre-classification, we derive a completeness $C=0.238\pm0.005$, a purity $P=0.988\pm0.001$, and $\rm F1=0.384\pm0.006$. However, as discussed in Sect.~\ref{sec:spritz}, the hydrogen emission lines of AGN1 may be underestimated in our model. This would make the selection less complete, when the H${\alpha}$ nebular emission line is inside the \YE\ ($z\sim0.7$) or \HE\ filter ($z\sim2$). Indeed, AGN1 would move to the right at $z\sim0.7$, as the \YE\ flux becomes brighter, and to the top of Fig.~\ref{fig:Best_Deep_Euclid} at $z\sim2$, as the \HE\ flux becomes brighter. 
%\par

\begin{figure}
    \centering
    \includegraphics[width=\linewidth, keepaspectratio]{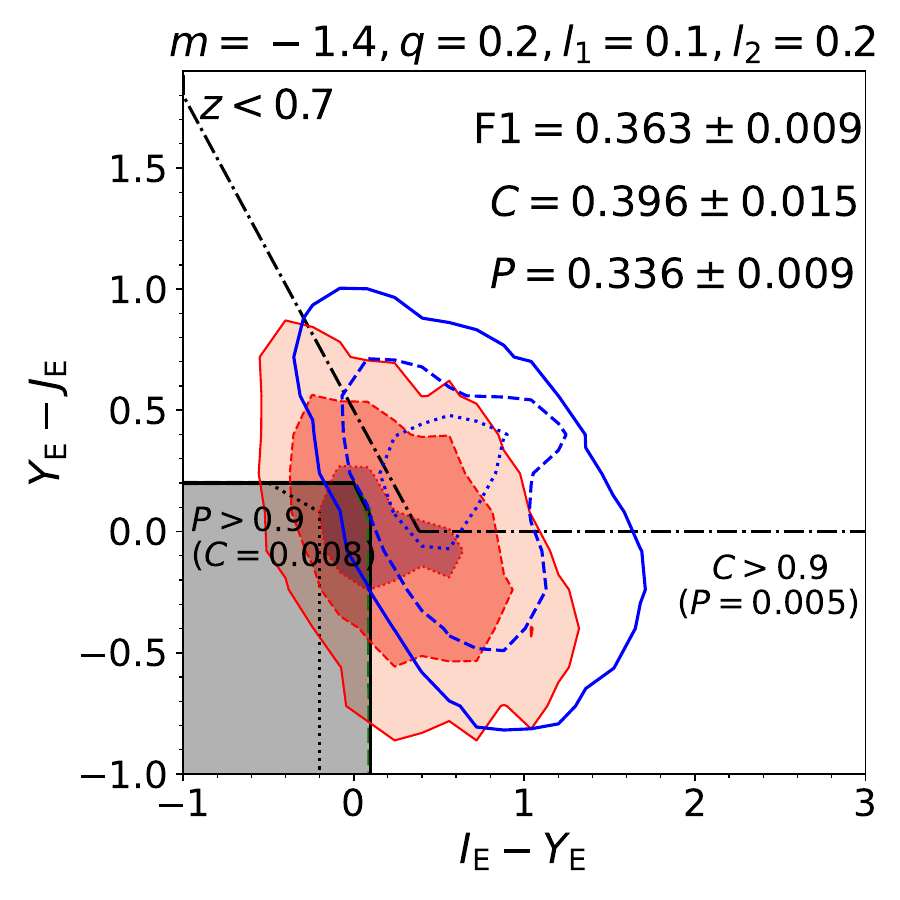}
    \includegraphics[width=\linewidth,keepaspectratio]{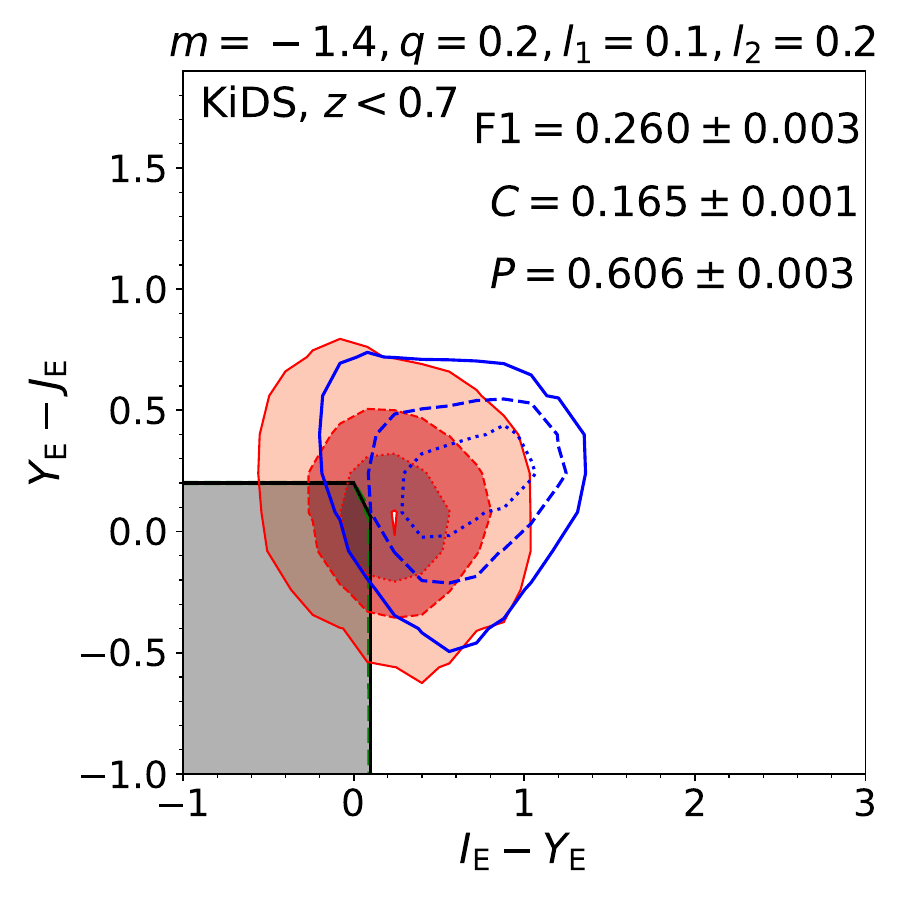}
    \caption{Best selection (type~A) criterion for AGN1 at $z<0.7$ in \spr{} (top panel) and KiDS (bottom panel) in the EDS using \Euclid filters (shaded grey area). The shaded red areas correspond to the colour distribution of AGN1, while the blue lines are the contour levels of the remaining galaxies. Levels correspond to 68\%, 95\%, and 99.7\% of the distribution. As an indication of how strict the best selection is, the hatched green area limited by the two dashed green lines shows the extent of all the best selection criteria derived with the bootstrap approach. The dotted black line and the dash-dotted black line correspond to the best selections, with $P>0.9$ and $C>0.9$, respectively.}
    \label{fig:Best_Deep_Euclid_z1}
\end{figure}

We repeated the analysis only with AGN1 at $z<0.7$. The best selection uses the  $\IE-\YE$ and $\YE-\JE$ colours with $m=-1.4$, $q=0.2$, $l_1=0.1$, and $l_2=0.2$, but the diagonal cut is negligible. This colour selection produces an F1 score of $\rm F1=0.316$ (Fig.~\ref{fig:Best_Deep_Euclid_z1}, top panel). This selection is more pure than the one derived for the full redshift range, but it is less complete. %\par

\begin{figure}
    \centering
    \includegraphics[width=\linewidth, keepaspectratio]{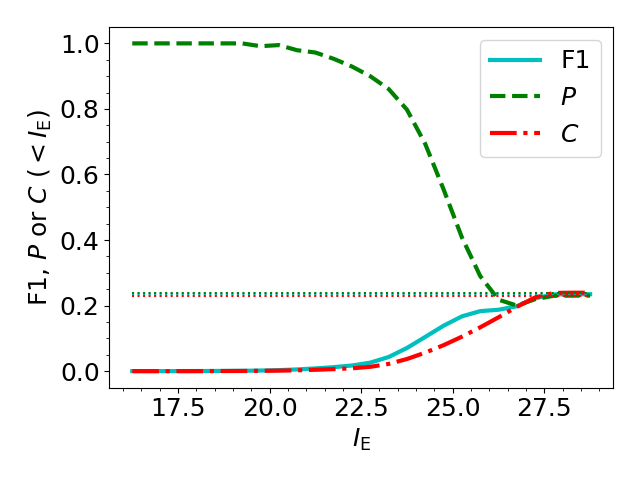}
    \caption{Variation of the F1 score (solid cyan line), completeness (dash-dotted red line), and purity (dashed green line) with \IE\ magnitude limit. Horizontal dotted lines show the values using Eq. (\ref{eq:AGN1_Deep}) without applying any additional magnitude cut.}
    \label{fig:VIScut}
\end{figure}

Alternatively, instead of applying a redshift selection, which has a precision varying with the redshift uncertainties, we could include an additional magnitude limit, given that dwarf galaxies tend to be fainter than AGN1 at similar redshifts. However, this does not show any major improvement over the selection reported in Eq. (\ref{eq:AGN1_Deep}), since an increase in purity due to the removal of dwarf galaxies is counterbalanced by a decrease in completeness given by the removal of AGN1 located at higher redshifts (on average $z=2.5$). This is illustrated in Fig.~\ref{fig:VIScut} using different \IE\ limits. 
%\par

An additional constraint could be given by morphological information, which could help to distinguish point-like objects, such as AGN1, from star-forming galaxies. For example, the \IE\ filter is expected to resolve galaxies down to $10^8\,M_{\odot}$ at $z=1$, considering the stellar mass-size relation given by \citet{Wel2014}. However, the details of such analysis, which cannot be performed on integrated fluxes alone, is left to future work. 
%\par 

Using the same colours we search for a selection with a high completeness, that is, $C>0.9$, but we obtain a low purity $P=0.034$, since the number of contaminants quickly exceed the number of AGN1. This selection is shown in Fig.~\ref{fig:Best_Deep_Euclid} as a black, dash-dotted line and is defined by
$\IE-\HE<0.9$ or $\IE-\HE<-1.3\,(\IE-\YE)+1.5$, with no additional $\IE-\YE$ cut. At the same time, looking for a selection with high purity, that is, $P>0.9$ (black dotted line in Fig.~\ref{fig:Best_Deep_Euclid}), results in a sample that is highly incomplete, $C<0.01$. This selection is defined by $\IE-\HE<-0.2$, with no additional $\IE-\YE$ or diagonal cuts. 
%\par

\subsubsection{Comparison with KiDS}
The majority ($93\%$) of objects in the KiDS-BEXGO catalogue are at $z<0.7$, with a median redshift of $z=0.38$. Therefore, we decided to limit the comparison between these observed QSOs and the \spr{} simulated AGN1 to $z<0.7$. Figure~\ref{fig:Best_Deep_Euclid_z1} presents the best selection criterion derived for AGN1 in the EDS, limited to galaxies at $z<0.7$. The selection is performed using \spr{}, but we also show the colours of objects in the KiDS-BEXGO catalogues. In the rest of this section we discuss the possible reasons behind the differences in the colour distributions of the two catalogues and the corresponding F1 scores. 

First, it is necessary to remember that we used a PCA analysis to estimate the \Euclid fluxes in the KiDS-BEXGO catalogues. As a consequence, there is correlation between the derived \Euclid colours, that is, \YE\ and \JE\ are both derived from a combination of the $Y$, $J$, and $H$ VIKING filters (see Eq.~\ref{eq:pca_qso}). 

\begin{figure}
    \centering
    \includegraphics[width=\linewidth, keepaspectratio]{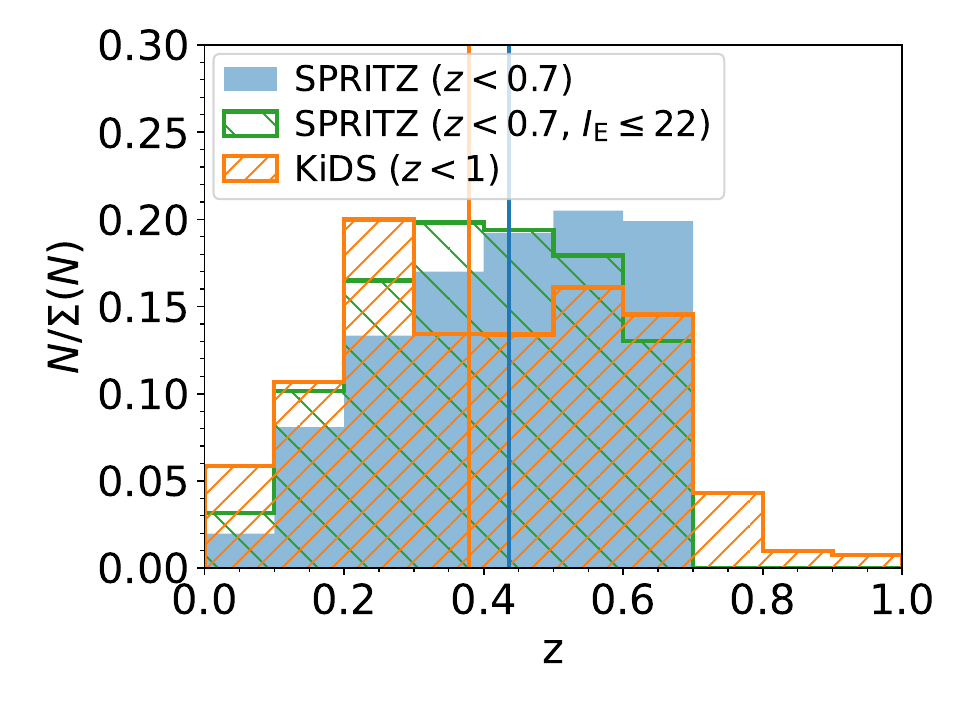}
    \includegraphics[width=\linewidth, keepaspectratio]{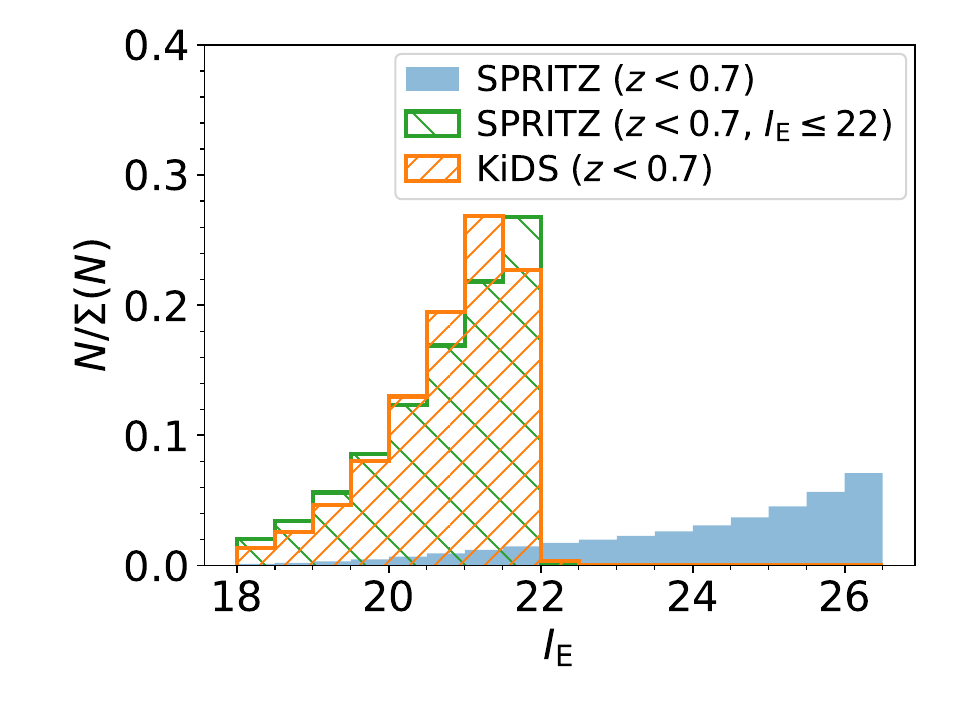}
    \caption{Normalised distributions of the redshift (top panel) and the \IE\ magnitude (bottom panel) for the simulated catalogues from \spr{} (filled blue and hatched green histograms) and KiDS (orange hatched histogram). Vertical lines in the top panel show the average redshift of the two samples at $z<0.7$.}
    \label{fig:zcomp}
\end{figure}

Second, as visible in the bottom panel of Fig.~\ref{fig:zcomp}, the KiDS-BEXGO catalogue is limited to $\IE\leq 22$, not reaching the bulk of the objects that are expected to be observed by \Euclid. Even when limiting the \spr{} catalogue to all objects at $z<0.7$ and $\IE\leq 22$, the redshift distribution is not the same as the KiDS-BEXGO one (top panel of Fig.~\ref{fig:zcomp}). Differences may arise from the selection of the KiDS-BEXGO catalogue, since sources were required to be detected in all nine KiDS and VIKING filters, or on an overestimation of bright galaxies in \spr{} for $z=0.3$--0.5.

%\par
Finally, if we limit the \spr{} sample to objects at $z<0.7$ and $\IE\leq 22$ and we weight the remaining galaxies in order to have the same redshift distribution as the KiDS-BEXGO samples, the F1 score moves to $0.253\pm0.006$, which is close to the value recovered for the KiDS-derived samples ($\rm F1=0.260\pm0.003$). The completeness is however still higher than the KiDS-BEXGO samples, i.e. $C=0.289\pm0.005$, while the purity remains lower, i.e. $P=0.225\pm0.007$. Remaining differences may arise from an underestimation of the broad component of hydrogen nebular emission lines in AGN1, since H${\alpha}$ is in the \IE\ filter at $z<0.4$, causing a possible underestimation of the $\IE-\YE$ and $\IE-\JE$ colours of AGN1. Overall, the similarity between the results of the two approaches, once redshift and magnitude differences are considered, provides reassurance as to the validity of the results presented here.

\subsubsection{All AGN in the EDS}
In this section we investigate colour selections to identify all AGN types, including dust-obscured AGN2 and composite systems, instead of limiting the selection to AGN1. The statistics of all the tested colour criteria to identify allf AGN types in the EDS with \Euclid filters are listed in Table~\ref{tab:Deep_Euclid}. 
%\par 

The selection of all AGN types is challenging; all the colour criteria based only on two or three \Euclid filters provide F1 scores between 0.115 and 0.124. However, these selections include all objects, independently on their nature. To better understand the meaning of such low F1 scores, we randomly assigned each object to the AGN category, assuming we have no knowledge on the expected number of AGN, and we derive the F1 score associated with this random selection. This extreme and unrealistic case corresponds to $\rm F1=0.116\,$--$\,0.124$ ($C=0.500$ and $P=0.066\,$--$\,0.071$), depending on the \Euclid filters considered. These results are similar to the best F1 scores, which are therefore not reliable selection criteria. At the same time, if we limit our analysis to criteria corresponding to a large completeness $C>0.9$ or a large purity $P>0.9$, we obtain an extremely low purity ($P<0.1$) or an extremely low completeness ($C\sim10^{-6}$).

To improve the identification of these objects, it is therefore preferable to rely on spectroscopic data, as provided by \Euclid, or the use of more complex identification methods, such as machine-learning techniques \citep[e.g.][]{Signor24}. Ancillary X-ray observations may also help, but not for the identification of the most obscured or intrinsically faint AGN \citep[][Euclid Collaboration: Selwood in prep.]{Barchiesi2021}. We subsequently discuss the impact of optical or near-IR ancillary observations on the selection of these AGN.
%\par

\subsubsection{AGN1 in the EWS}
We now investigate AGN selection in the EWS. The statistics for all tested selection criteria for AGN1 in the EWS are listed in Table~\ref{tab:Wide_Euclid}. The best selection criterion in the EWS can be different than the one in the EDS, because the two surveys include different AGN populations, given the different areas and observational depths. In particular, the brightest and rarest objects that will be present in the EWS have a low probability to be observed as part of the EDS, while the faintest and most numerous sources that will be observed in the EDS will not be detectable in the EWS. At the same time, the same object will have larger observational uncertainties in the EWS than in the EDS.
%\par

\begin{figure}
    \centering
    \includegraphics[width=\linewidth, keepaspectratio]{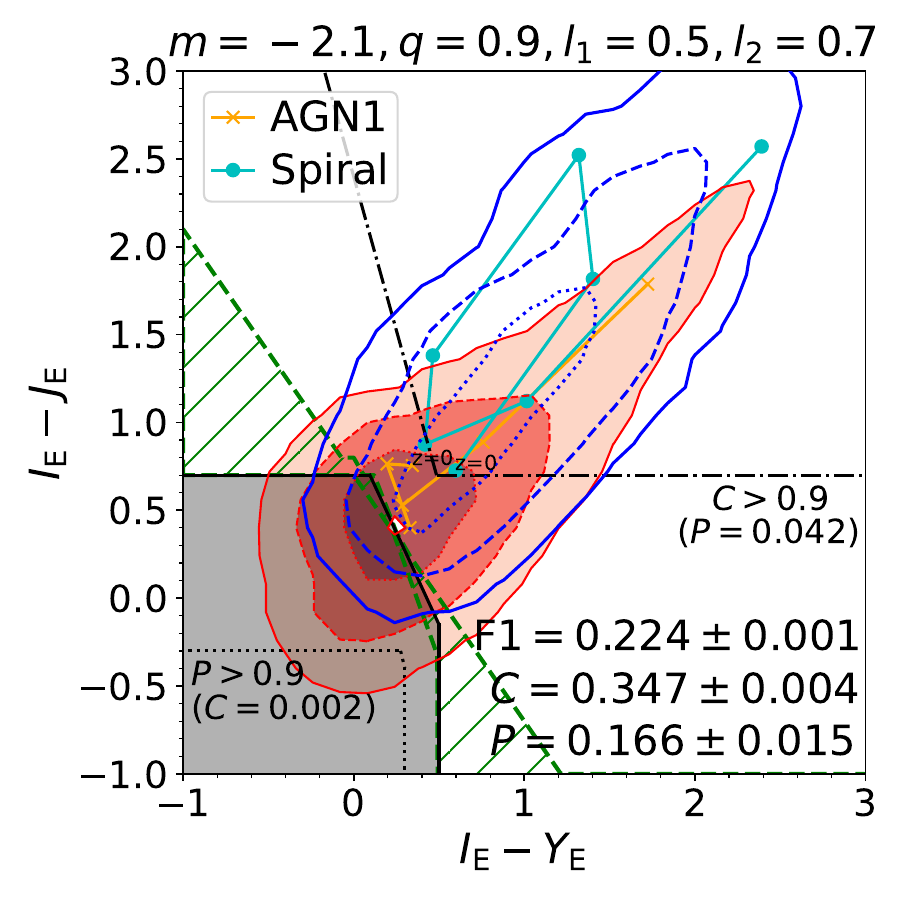}
    \caption{Best selection (type~A) criterion for AGN1 in the EWS using \Euclid filters (shaded grey area). The red shaded areas correspond to the colour distribution of AGN1, while the blue lines are the contour levels of the remaining galaxies. Levels correspond to 68\%, 95\%, and 99.7\% of the distribution. We also present example SED tracks for one AGN1 (yellow crosses) and one spiral galaxy (cyan circles) from $z=0$ to $z=6$, with steps of $\Delta z=1$. As an indication of how strict the best selection is, the hatched green area limited by the two dashed green lines shows the extent of all the best selection criteria derived with the bootstrap approach. The dotted black line and the dash-dotted black line correspond to the best selections with $P>0.9$ and $C>0.9$, respectively.}
    \label{fig:Best_Wide_Euclid}
\end{figure}

The best selection criterion using only two \Euclid filters, as shown in Fig.~\ref{fig:Best_Wide_Euclid}, is
\begin{align}\label{eq:AGN1_Wide}
      \left(\IE-\YE<0.5_{-0.0}^{+0.1}\right)\,&\land\,
      \left(\IE-\JE<0.7^{+0.1}_{-0.0}\right)\,\\
      &\land\,
      \left[\IE-\JE<-2.1_{-0.4}^{+0.5}\,(\IE-\YE)+0.9^{+0.1}_{-0.1}\right] \;.\nonumber    
\end{align}
This selection provides an F1 score of $0.224\pm0.001$, derived from a low purity ($P=0.166\pm0.015$) and a low completeness ($C=0.347\pm0.004$). However, as visible from the bootstrap analysis, the vertical and horizontal selection has negligible effect, since it is mainly important to apply a diagonal separation. Comparable results (i.e. a difference in F1 of 0.002, Table~\ref{tab:Wide_Euclid}) are obtained using $\IE-\YE$ and $\IE-\HE$ colours, which are the colours used in the EDS selection criteria. If we consider exactly the same colour selection derived for the EDS, we obtain $\rm F1=0.211$, with a purity $P=0.193$ and a completeness $C=0.233$. The difference in the completeness is probably due to a slightly redder $\IE-\HE$ colour distribution for AGN1. Indeed, the median $\IE-\HE$ colour for AGN1 is 0.04 redder in the EWS than in the EDS, while it remains similar for the contaminants. A decrease in purity is instead driven by an extension of the blue tail of the colour distribution in the EWS with respect with the EDS. This happens for both AGN1 and contaminants, but it is more prominent for the latter. Indeed, the number density of the contaminants matches the AGN1 one (i.e. $P=0.5$) at $\IE-\HE=0.23$ in the EWS and at $\IE-\HE=0.40$ in the EDS.
%\par

\begin{figure}
    \centering
    \includegraphics[width=\linewidth, keepaspectratio]{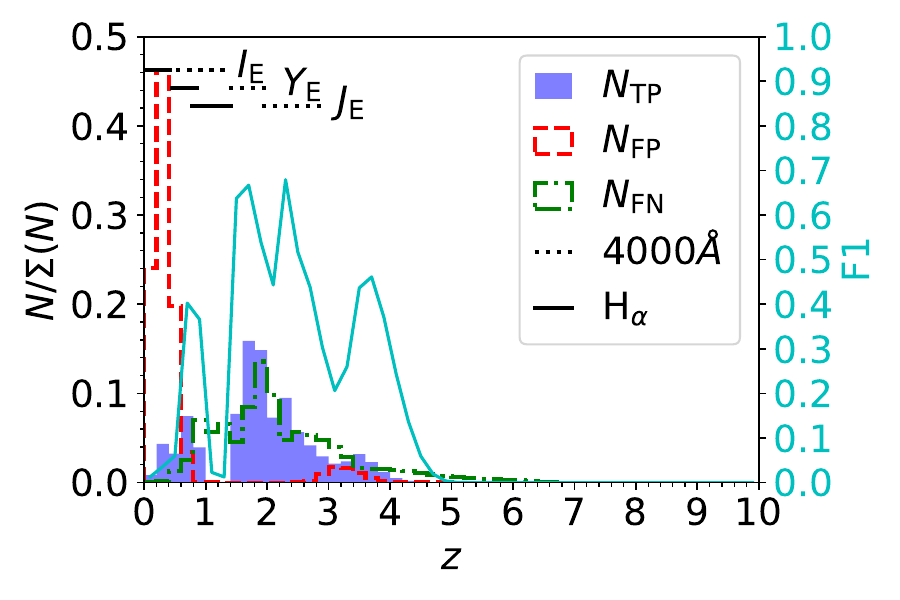}
    \caption{Normalised redshift distribution of the $N_{\rm TP}$ (filled blue histogram), $N_{\rm FP}$ (dashed red histogram), and $N_{\rm FN}$ (dot-dashed green histogram) for the \Euclid-only selection criteria (Eq.~\ref{eq:AGN1_Wide}) of AGN1 in the EWS. The horizontal solid and dotted lines show the redshift range in which H${\alpha}$ and the $4000\,\AA$-break, respectively, are inside the \IE, \YE, or \JE\ filters (\HE\ is not used in Eq.~\ref{eq:AGN1_Wide}). We also report the variation of the F1 score with redshift (solid cyan line and right vertical axis).}
    \label{fig:z_WAGNAll}
\end{figure}

\begin{figure}
    \centering
    \includegraphics[width=\linewidth, keepaspectratio]{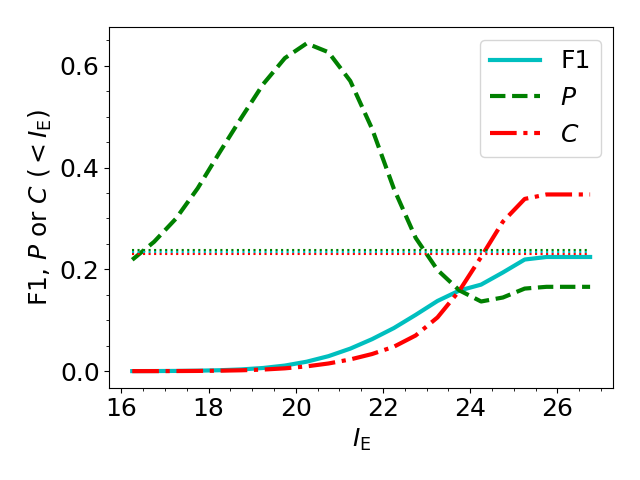}
    \caption{Variation of the F1 score (solid cyan line), completeness (dash-dotted red line), and purity (dashed green line) with \IE\ magnitude limit for AGN1 in the EWS. Horizontal dotted lines show the values using Eq. \ref{eq:AGN1_Wide} without applying any additional magnitude cut. The decrease of the purity for $\IE<20$ is due to objects at $z<0.05$.}
    \label{fig:Viscut_wide}
\end{figure}

As for the EDS, the best selection criterion is affected by strong contamination by dwarf irregular galaxies ($98\%$ of the false positives), particularly at $z<0.7$ (Fig.~\ref{fig:z_WAGNAll}). Limiting the sample above this redshift improves the selection to $\rm F1=0.445$, $P=0.706$, and $C=0.325$. In general, the F1 score varies with redshift; it reaches a maximum of 0.679 at $z=2.3$, but AGN1 move completely out of the selection at $z=5$. The proposed colour selection is therefore is particularly effective at $0.7\leq z<5$. Moreover, we verify that applying a magnitude cut combined with the colour selection proposed in Eq. \ref{eq:AGN1_Wide} does not improve the F1 score (Fig. \ref{fig:Viscut_wide}), as an improvement in completeness is counterbalanced by a reduction in completeness.
%\par

Next, if we consider the same colours, but we search for a selection with a high completeness, that is, $C>0.9$, we obtain an extremely low purity $P\leq0.015$. At the same time, the highly pure selection defined results in a very incomplete sample with $C<0.002$.
%\par

\begin{figure}
    \centering
    \includegraphics[width=\linewidth, keepaspectratio]{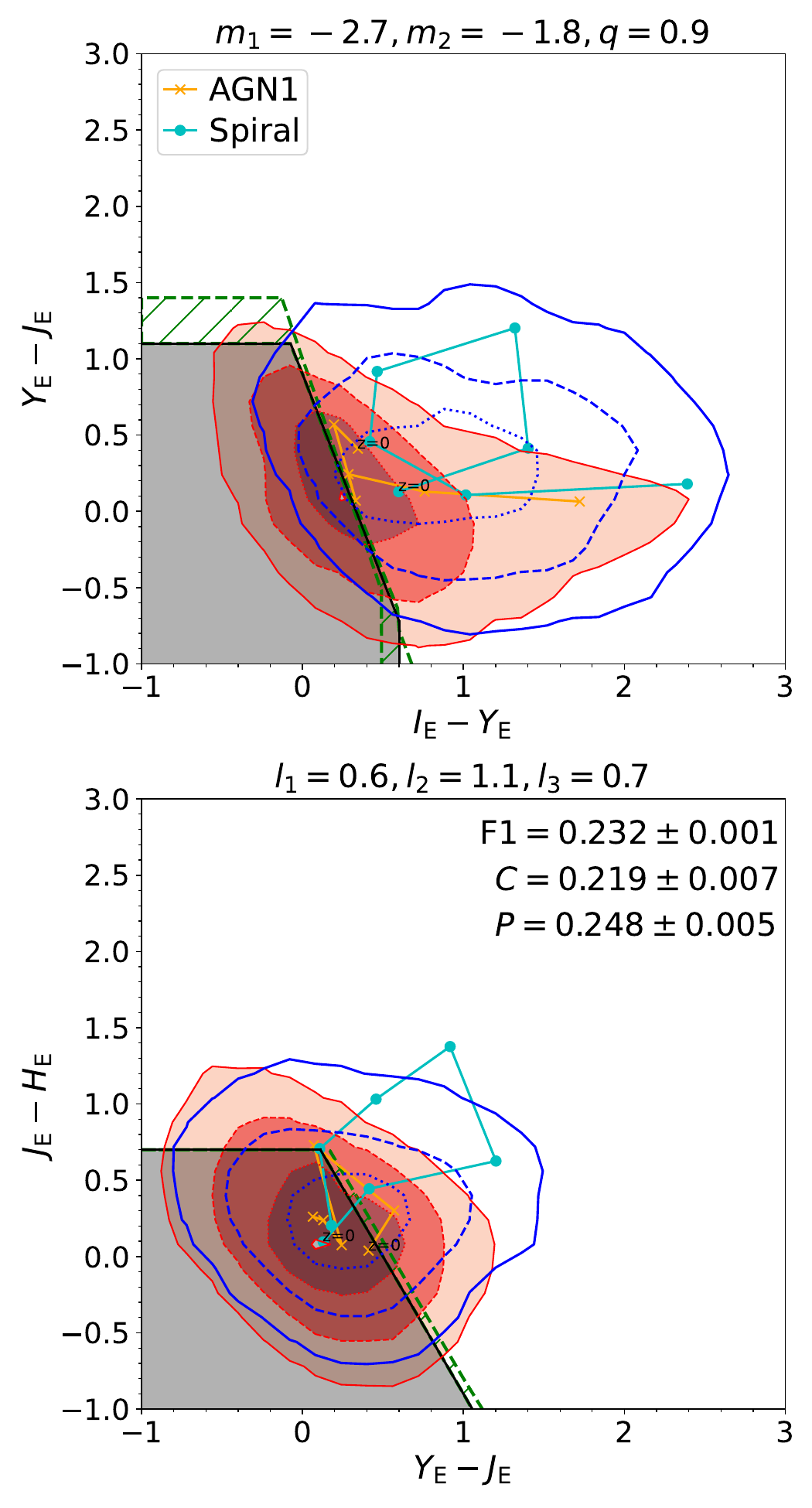}
    \caption{Two-dimensional projections of the best three-colour selection criterion for AGN1 in the EWS using \Euclid filters (shaded grey area). The shaded red areas correspond to the colour distribution of AGN1, while the blue lines are the contour levels of the remaining galaxies. Levels correspond to 68\%, 95\%, and 99.7\% of the distribution. We also present example SED tracks for one AGN1 (yellow crosses) and one spiral galaxy (cyan circles) from $z=0$ to $z=6$, with steps of $\Delta z=1$. As an indication of how strict the best selection is, the hatched green area limited by the two dashed green lines shows the extent of all the best selection criteria derived with the bootstrap approach. }% The black dotted line and the black dash-dotted line correspond to the best selections with $P>0.9$ and $C>0.9$, respectively.}
    \label{fig:Best_Wide_Euclid_3c}
\end{figure}

Considering a three-colour selection based on all \Euclid filters bring a light improvement in the selection, increasing the F1 score to $\rm F1=0.232\pm0.001$, which corresponds to a lower completeness $C=0.219\pm0.007$, but a higher purity $P=0.248\pm0.005$ than the two-colour selection. The projections of these three-colour selection criterion is shown in Fig.~\ref{fig:Best_Wide_Euclid_3c}) and corresponds to:
\begin{align}\label{eq:AGN1_Wide_3c}
      \left(\IE-\YE<0.6_{-0.1}^{+0.1}\right)\,&\land\,
      \left(\YE-\JE<1.1^{+0.3}_{-0.0}\right)\,\land\,
      \left(\JE-\HE<0.7^{+0.0}_{-0.0}\right)\,\nonumber \\
      &\land\,
      \left[\YE-\JE<-2.7_{-0.4}^{+0.1}\,(\IE-\YE)+0.9^{+0.1}_{-0.0}\right]\, \nonumber \\
      &\land\,
      \left[\JE-\HE<-1.8_{-0.0}^{+0.1}\,(\YE-\JE)+0.9^{+0.1}_{-0.0}\right]\;.    
\end{align}
This criterion has a $\IE-\YE$ colour cut consistent with the best two-colour criterion in Eq. \ref{eq:AGN1_Wide}. We also note that requiring a detection on all four \Euclid filters, i.e. adding the $\HE$ with respect to the two-colour criterion, results on a decrease of $4\%$ in the number of detected AGN1.

\subsubsection{All AGN-types in the EWS}

In this section, we present the search for a selection criterion in the EWS for all AGN, including AGN2 and composite systems. This task is challenging, as it was for the EDS. Colour selections solely based on two or three \Euclid filters correspond to F1 scores between 0.153 and 0.156 (see Table~\ref{tab:Wide_Euclid} for the full list). The optimal selection criterion
does not serve our purposes and provides no discriminatory power because it includes all available galaxies. To understand better how poor these selections are, randomly assigning each object in the full catalogue to the AGN category would corresponds to $\rm F1=0.153$--0.156 ($C=0.500$ and $P=0.090$--0.092), depending on the colours considered. Given the limited utility of such selections, we do not analyse them further.
%\par

\subsection{\Euclid and LSST colours}
In this section we investigate if the colour selections based only on \Euclid observations could be improved once additional optical ancillary data are available, such as $u$, $g$, $r$, $i$, and $z$ imaging by Rubin/LSST. These additional data could help break degeneracies between the different galaxy templates, improving the selection of AGN. 

\subsubsection{AGN1 in the EDS}
\begin{figure}
    \centering
    \includegraphics[width=\linewidth, keepaspectratio]{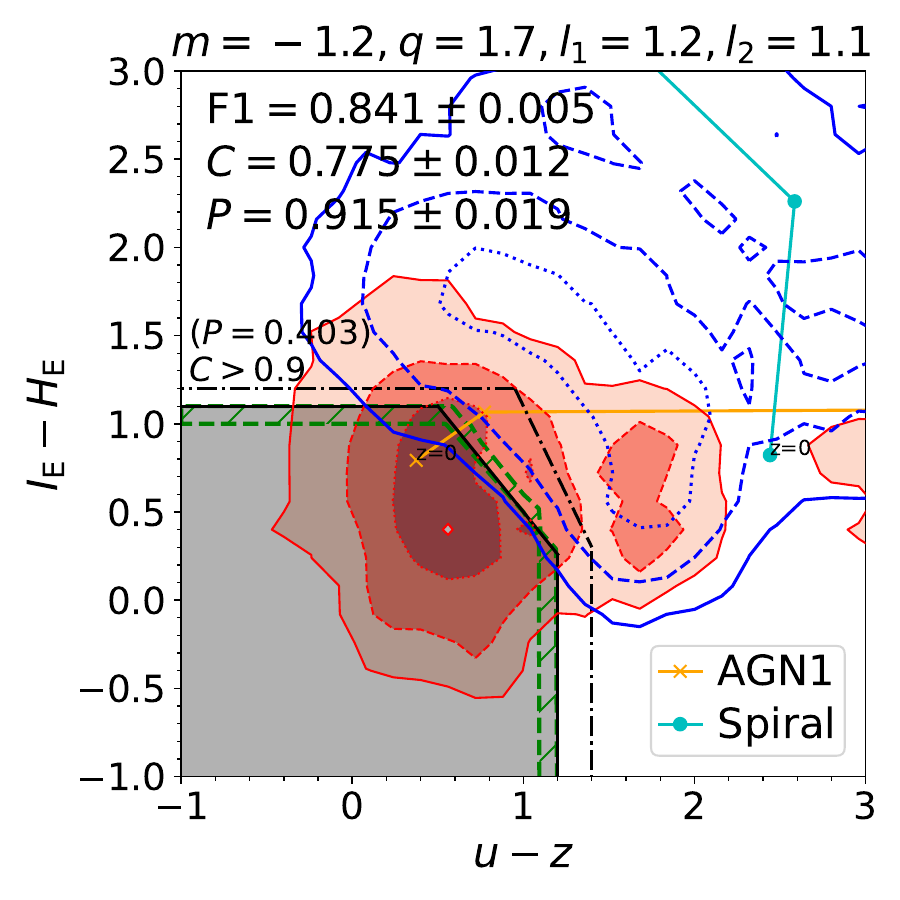}
    \caption{Best selection (type~A) criterion for AGN1 in the EDS using \Euclid and Rubin/LSST filters (shaded grey area). The shaded red areas correspond to the colour distribution of AGN1, while the blue lines are the contour levels of the remaining galaxies. Levels correspond to 68\%, 95\%, and 99.7\% of the distribution. We also present example SED tracks for one AGN1 (yellow crosses) and one spiral galaxy (cyan circles) from $z=0$ to $z=6$, with steps of $\Delta z=1$. As an indication of how strict the best selection is, the two dashed green lines show the extent of all the best selection criteria derived with the bootstrap approach. The dash-dotted black line corresponds to the best selection with $C > 0.9$.}
    \label{fig:Deep_LSST}
\end{figure}

As a first step, we concentrate on the selection of AGN1 in the EDS. The colour selection that maximises the F1 score is shown in Fig.~\ref{fig:Deep_LSST} and corresponds to
\begin{align}
\label{eq:AGN1_Deep_LSST}
      \left(\IE-\HE<1.1_{-0.1}^{+0.0}\right)\,&\land\,
      \left(u-z<1.2_{-0.1}^{+0.0}\right)\, \\
      &\land\,
      \left[\IE-\HE<-1.2_{-0.1}^{+0.1}\,(u-z)+1.7_{-0.0}^{+0.1}\right]\;.\nonumber
\end{align}
This selection has an F1 score of $0.841\pm0.005$ and identifies AGN1 with a purity $P=0.915\pm0.019$ and a completeness $C=0.775\pm0.012$. The selection is effective out to $z=2.1$, where the F1 score remains above 0.5. Limiting the sample to this redshift produce only minor changes, i.e. $\rm F1=0.858\pm0.002$, $P=0.898\pm0.014$, and $C=0.820\pm0.013$, since the number of objects detected in all four filters and inside the selection at higher redshifts is limited.

Using the same colours and no redshift cut, but looking for a selection with a completeness $C>0.9$, we obtain $P=0.403$ and $\rm F1=0.556$. This selection, which is shown as a black dash-dotted line in Fig.~\ref{fig:Deep_LSST}, is defined by $u-z<1.4$, $\IE-\HE<1.2$, and $\IE-\HE<-2.0\,(u-z)+3.1$.

We note that underestimation of the broad component of the H${\alpha}$ in \spr{}, as mentioned in Sect.~\ref{sec:cat}, would lead to an overestimation of $\IE-\HE$ colour at $z<0.4$. It would also result in an underestimation of the $u-z$ colour in the range $z=0.25$--0.40. These changes would make the colour selection of AGN1 less complete at $z=0.25$--0.4, but more complete at $z<0.25$. The statistics of all the other colour criteria for AGN1 in the EDS using \Euclid and Rubin/LSST filters are listed in Table~\ref{tab:Deep_Euclid_LSST}.

\subsubsection{All AGN in the EDS}
\begin{figure}
    \centering
    \includegraphics[width=\linewidth, keepaspectratio]{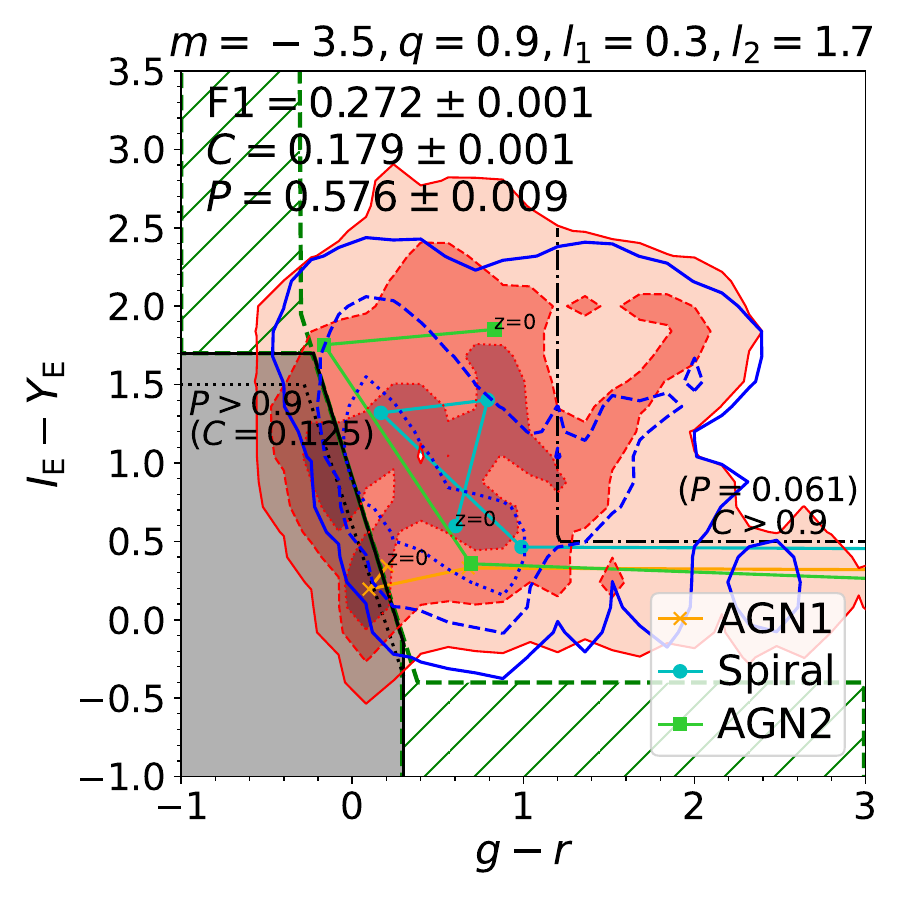}
    \caption{Best selection (type~A) criterion for all AGN in the EDS using \Euclid and Rubin/LSST filters (shaded grey area). The shaded red areas correspond to the colour distribution of all AGN, while the blue lines are the contour levels of the remaining galaxies. Levels correspond to 68\%, 95\%, and 99.7\% of the distribution. We also present example SED tracks for one AGN1 (yellow crosses), for one AGN2 (green squares), and one spiral galaxy (cyan circles) from $z=0$ to $z=6$, with steps of $\Delta z=1$. As an indication of how strict the best selection is, the hatched green area limited by the two dashed green lines show the extent of all the best selection criteria derived with the bootstrap approach. The dotted black line and the dash-dotted black line correspond to the best selection with $P > 0.9$ and $C > 0.9$, respectively.}
    \label{fig:Best_Deep_LSST_All}
\end{figure}

The selection of all AGN types is improved by the inclusion of Rubin/LSST filters. The best selection criterion (type A) corresponds to
\begin{align}\label{eq:AGNAll_Deep_LSST}
      \left(\IE-\YE<1.7_{-0.0}^{+\infty}\right) \,&\land\,
      \left(g-r<0.3_{-0.0}^{+\infty}\right) \, \\
      &\land\,
      \left[\IE-\YE<-3.5_{-0.1}^{+0.0}\,(g-r)+0.9_{-0.0}^{+0.0}\right] \;, \nonumber
\end{align}
where the horizontal and vertical cuts have only minor importance.
Such a colour selection results in $\rm F1=0.272\pm0.001$, $P=0.576\pm0.009$, and $C=0.179\pm0.001$ (Fig.~\ref{fig:Best_Deep_LSST_All}). The statistics of this selection are better than those derived using only \Euclid filters, but 60\% or less than the ones obtained using using Rubin/LSST and \Euclid filters for AGN1. The majority of false negatives correspond to composite systems, that is, 65\% of SF-AGN and 23\% of SB-AGN, since their broad-band photometry is dominated by the stellar continuum. At the same time, the majority of false positives are dwarf irregulars ($88\%$) and spirals (12$\%$), with an optical continuum that is, by definition, similar to the one for AGN2 and  SF-AGN. The colour criterion overall selects $8\%$ of SF-AGN, $21\%$ of SB-AGN, $16\%$ of AGN2, and $58\%$ of AGN1, showing that the latter are the AGN population that is most easily separable from inactive galaxies.

\begin{figure}
    \centering
    \includegraphics[width=\linewidth, keepaspectratio]{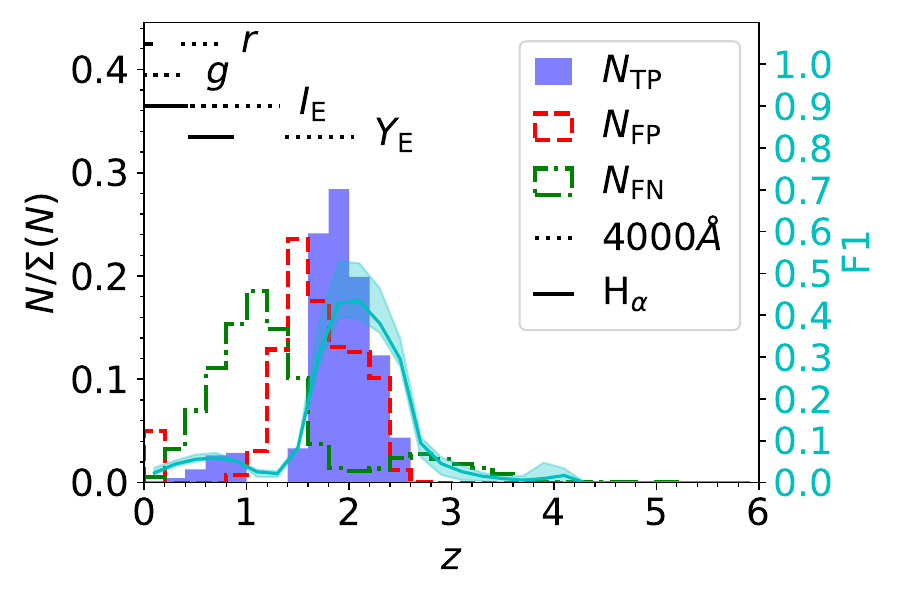}
    \caption{Normalised redshift distribution of the $N_{\rm TP}$ (filled blue histogram), $N_{\rm FP}$ (dashed red histogram), and $N_{\rm FN}$ (dot-dashed green histogram) for the \Euclid and Rubin/LSST selection criteria of all AGN in Eq. (\ref{eq:AGNAll_Deep_LSST}). The horizontal solid and dotted lines show the redshift range in which H${\alpha}$ and the $4000\,\AA$-break, respectively, are inside the \IE\, \YE, $r$, or $g$ filters. We also report the variation of the F1 score with redshift (solid cyan line and right vertical axis).}
    \label{fig:zdeepall_LSST}
\end{figure}

As visible in Fig.~\ref{fig:zdeepall_LSST}, the selection criterion is mainly effective for $z=1.7$--$2.5$, where F1 remains equal or larger than 0.5. Limiting the analysis to this redshift interval, we obtained $\rm F1=0.739\pm0.003$ with a purity $P=0.765\pm0.006$ and completeness $C=0.715\pm0.001$. The selection includes $78\%$ of SF-AGN, $59\%$ of SB-AGN, $88\%$ of AGN1, and $48\%$ of AGN2. The F1 score does not improve, particularly if we focus only on AGN-dominated systems instead of applying a redshift selection, yielding $\rm F1=0.338$. 

If we consider the same $g-r$ and $\IE-\YE$ colours, but we search for a selection with a high completeness, that is, $C>0.9$, we obtain an extremely low purity $P\leq0.06$, since it is necessary to select almost all available sources. At the same time, a very pure selection (i.e. $P>0.9$) corresponds to a low completeness $C\leq0.125$. 

\subsubsection{AGN1 in the EWS}
\begin{figure}
    \centering
    \includegraphics[width=\linewidth, keepaspectratio]{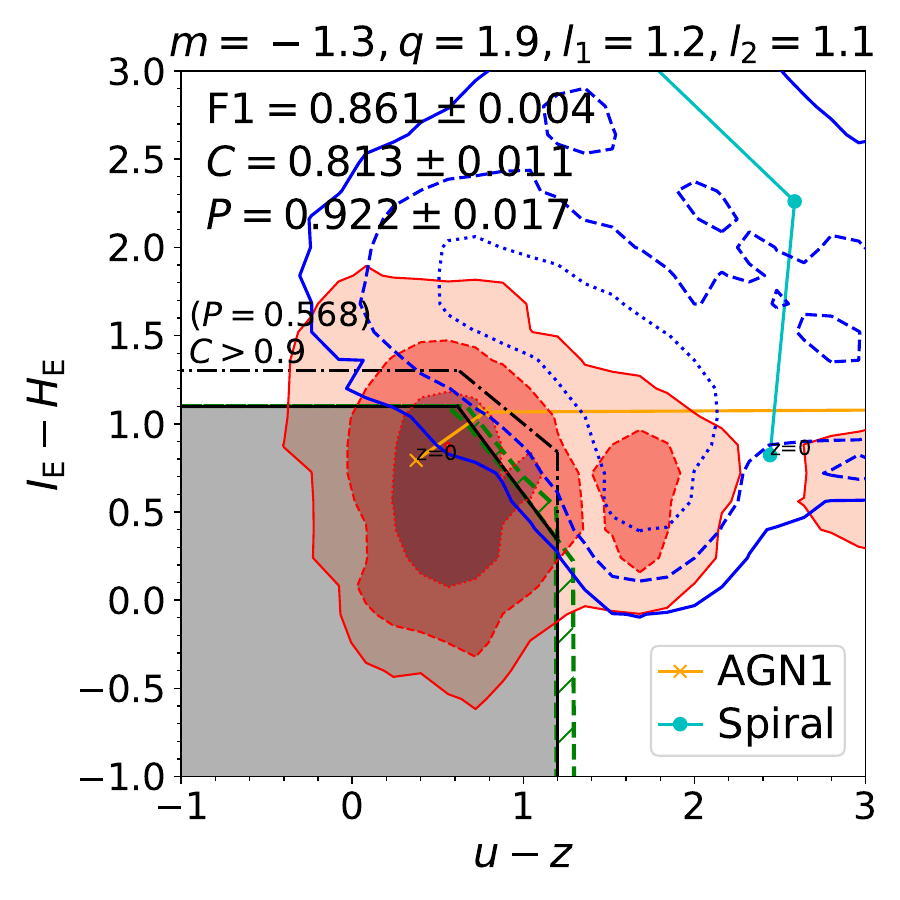}
    \caption{Best selection (type~A) criterion for AGN1 in the EWS using \Euclid and Rubin/LSST filters (shaded grey area). The shaded red areas correspond to the colour distribution of AGN1, while the blue lines are the contour levels of the remaining galaxies. Levels correspond to 68\%, 95\%, and 99.7\% of the distribution. We also present example SED tracks for one AGN1 (yellow crosses) and one spiral galaxy (cyan circles) from $z=0$ to $z=6$, with steps of $\Delta z=1$. As an indication of how strict the best selection is, the two dashed green lines show the extent of all the best selection criteria derived with the bootstrap approach. The dash-dotted black line correspond to the best selection with $C > 0.9$.}
    \label{fig:WAGN1_LSST}
\end{figure}

As for the EDS, the addition of the Rubin/LSST filters improves the selection of AGN1. The best selection criterion for AGN1 in the EWS is based on the $u-z$ and $\IE-\HE$ colours, as for the EDS (Fig.~\ref{fig:WAGN1_LSST}):
\begin{align}\label{eq:AGN1_Wide_LSST}
      \left(\IE-\HE<1.1_{-0.0}^{+0.0}\right)\,&\land\,
      \left(u-z<1.2_{-0.0}^{+0.0}\right)  \, \\
      &\land\,
      \left[\IE-\HE<-1.3_{-0.0}^{+0.2}\,(u-z)+1.9_{-0.1}^{+0.0}\right] \;. \nonumber
\end{align}
This selection is consistent with the one for the EDS within the confidence interval. This selection performs similarly than the one in the EDS, with the maximum F1 score that is $\rm F1=0.861\pm0.004$, in comparison with $\rm F1=0.841\pm0.005$ in the EDS, with $C=0.813\pm0.011$ and $P=0.922\pm0.017$. As for the EDS, the F1 score remains above 0.5 at $z\leq2.1$, but limiting the selection to this redshift results on marginal improvement, that is, $\rm F1=0.874\pm0.001$, $P=0.900\pm0.011$, and $C=0.850\pm0.010$.

%\par

Finally, using the same colours but imposing $C>0.9$, which could be preferred by science cases that value completeness over purity, we found a selection identified by $\IE-\HE<1.3$, $u-z<1.2$, and $\IE-\HE<-0.8\,(u-z)+1.8$. This criterion corresponds to $P=0.586$ and $\rm F1=0.710$.  

\subsubsection{All AGN in the EWS}
\begin{figure}
    \centering
    \includegraphics[width=\linewidth, keepaspectratio]{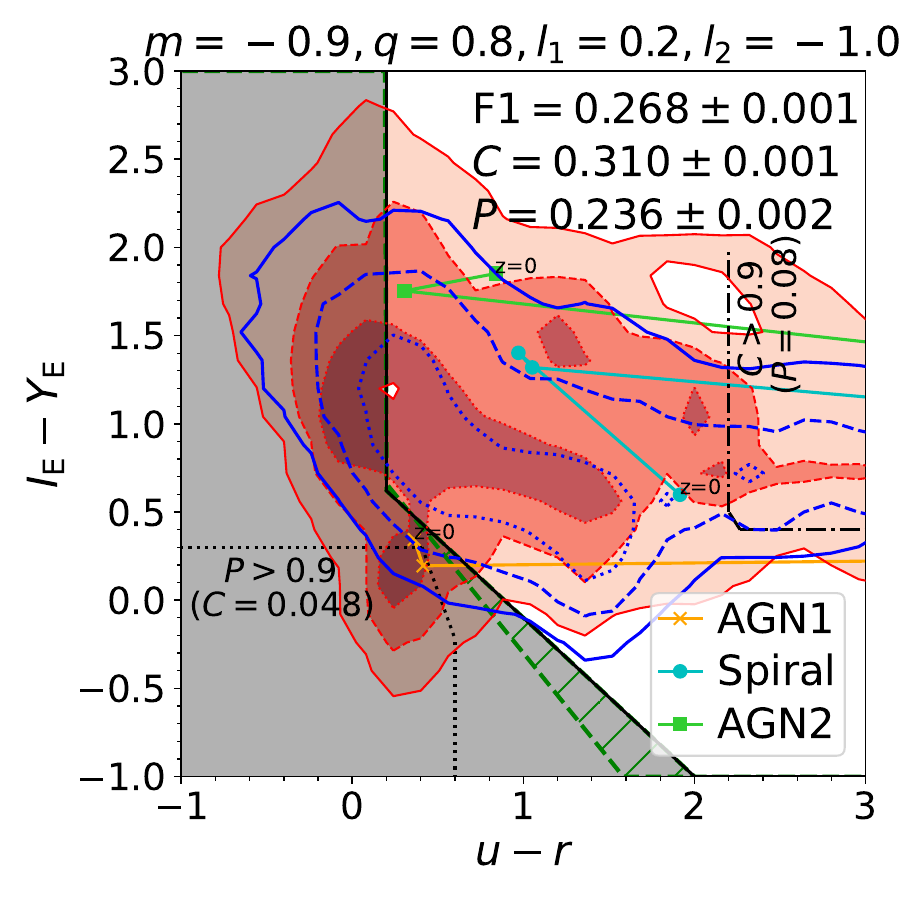}
    \caption{Best selection (type~B) criterion for all AGN in the EWS using \Euclid and Rubin/LSST filters (shaded rey area). The shaded red areas correspond to the colour distribution of all AGN, while the blue lines are the contour levels of the remaining galaxies. Levels correspond to 68\%, 95\%, and 99.7\% of the distribution. We also present example SED tracks for one AGN1 (yellow crosses), for one AGN2 (green squares), and one spiral galaxy (cyan circles) from $z=0$ to $z=6$, with steps of $\Delta z=1$. As an indication of how strict the best selection is, the hatched green area limited by the two dashed green lines show the extent of all the best selection criteria derived with the bootstrap approach. The dotted black line and the dash-dotted black line correspond to the best selection with $P > 0.9$ and $C > 0.9$, respectively.}
    \label{fig:WAGNAll_LSST}
\end{figure}

As for the EDS, adding the Rubin/LSST filters allows for finding a selection (type~B) for all AGN, corresponding to
\begin{equation}\label{eq:AGNAll_Wide_LSST}
      \left[\IE-\YE<-0.9_{-0.1}^{+0.0}\,(u-r)+0.8_{-0.0}^{+0.0}\right]\,\land\,
      \left(u-r<0.2_{-0.0}^{+0.0}\right)      \;.
\end{equation}
This selection produces $C=0.310\pm0.001$, $P=0.236\pm0.002$, and $\rm F1=0.268\pm0.001$ (Fig.~\ref{fig:WAGNAll_LSST}). The performance is slightly better that for the EDS, although it is based on a different set of Rubin/LSST filters and it is well below the results for AGN1. Considering the $g-r$ colour and the same criteria as for the EDS would slightly decrease the F1 score to 0.112, since the purity increases from 0.236 to 0.467, but the completeness decreases for 0.310 to 0.064, as a result of both AGN and galaxies being on average redder in $g-r$. 

In the best selection criterion, the majority (80$\%$) of false negatives, that is, AGN not correctly identified, are SF-AGN, that is, composite galaxies hosting an intrinsically faint AGN. However, even if we focus on AGN-dominated systems, that is, AGN1 and AGN2, the completeness reaches $C=0.623$, but the overall F1 score is very low, that is, $\rm F1=0.109$, because of a very low purity in the sample selection ($P<0.1$). The F1 score of the best selection varies with redshift, but it is above 0.5 on a narrow redshift range (i.e. $z=1.8$--2.0). Moreover, the number of detected sources quickly drops at $z>2.3$. Overall, limiting the selection to $z=1.7$--2.3, similar to the EDS, increases the F1 score to $0.589\pm0.003$, because of an increase of both the completeness ($C=0.609\pm0.004$) and the purity ($0.571\pm0.002$).

Finally, if we look for a selection criterion based on the same $\IE-\YE$ and $u-r$ colours, but with $C>0.9$, we obtain a quite low purity $P=0.08$, since AGN and non-active galaxies share almost the same region in colour space. For the same reason, a pure selection criterion, described by $u-r<0.3$, $\IE-\YE<0.4$, and $\IE-\YE<-2.7\,(u-r)+1.4$, results in $P=0.901$, but $C=0.048$ and it mainly selects AGN1. 

\subsubsection{Evolution with the survey depth}\label{sec:time}
As mentioned in Sect.~\ref{sec:cat_lsst}, ancillary data in the south will become deeper over time. Having deeper observations implies a reduction of the photometric uncertainties, but also the inclusion of fainter objects, which may be at higher redshift, intrinsically fainter, or more dust obscured. Therefore, we explore the impact of such evolution on our colour selections. %\par

\begin{figure}
    \centering
    \includegraphics[width=\linewidth, keepaspectratio]{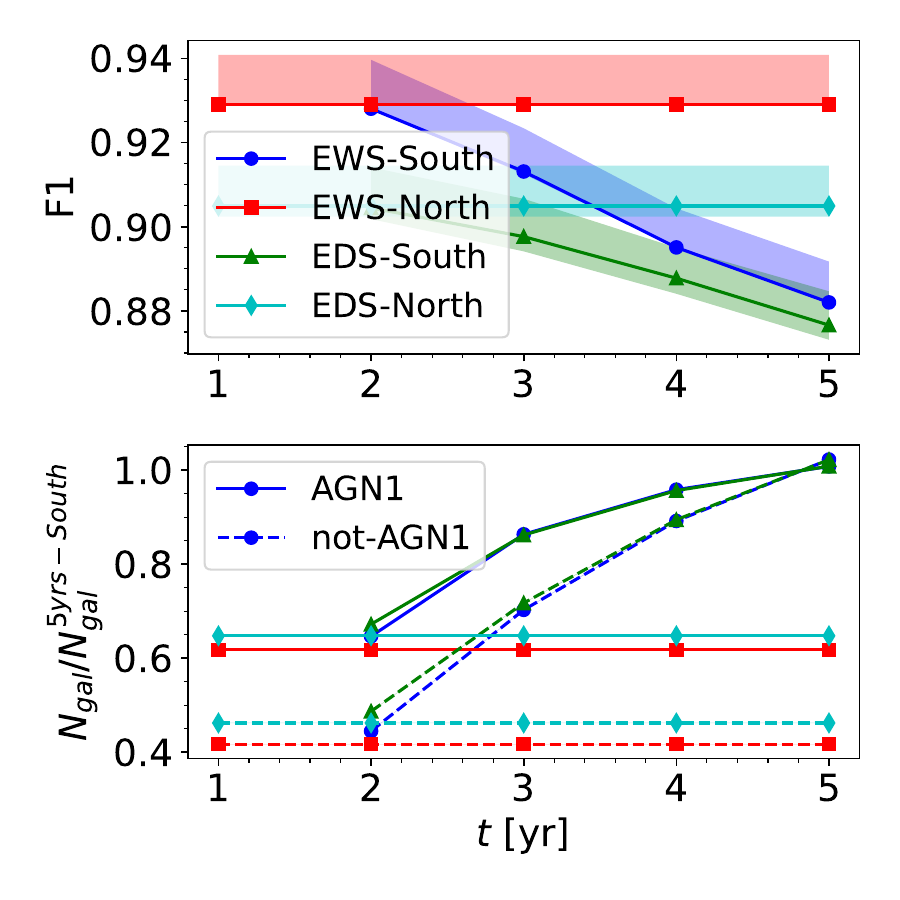}
    \caption{Effects of the increasing depths of the optical ancillary data on the AGN selection. \textit{Top:} variation of the F1 score with time due to the increasing depths of optical ancillary data. The depths of the ancillary data in the north remains constant with time, but we include them for comparison. The F1 score here corresponds to AGN1 derived using \Euclid and Rubin/LSST filters. \textit{Bottom:} variation of the number of AGN1 (solid lines) or non-AGN1 (i.e. all other AGN, SF, and passive galaxies, dashed lines) with time normalised to the number of AGN1 or non-AGN1 observed by the fifth year in the south. In the south no $u$-band filter observations will be available in the first year, so the analysis starts from the second year. Some of the F1 scores decrease with time because shallower $u$ band observations correspond to smaller samples and the brightest AGN are the easiest ones to identify.}
    \label{fig:AGN1_time}
\end{figure}

\begin{figure}
    \centering
    \includegraphics[width=\linewidth, keepaspectratio]{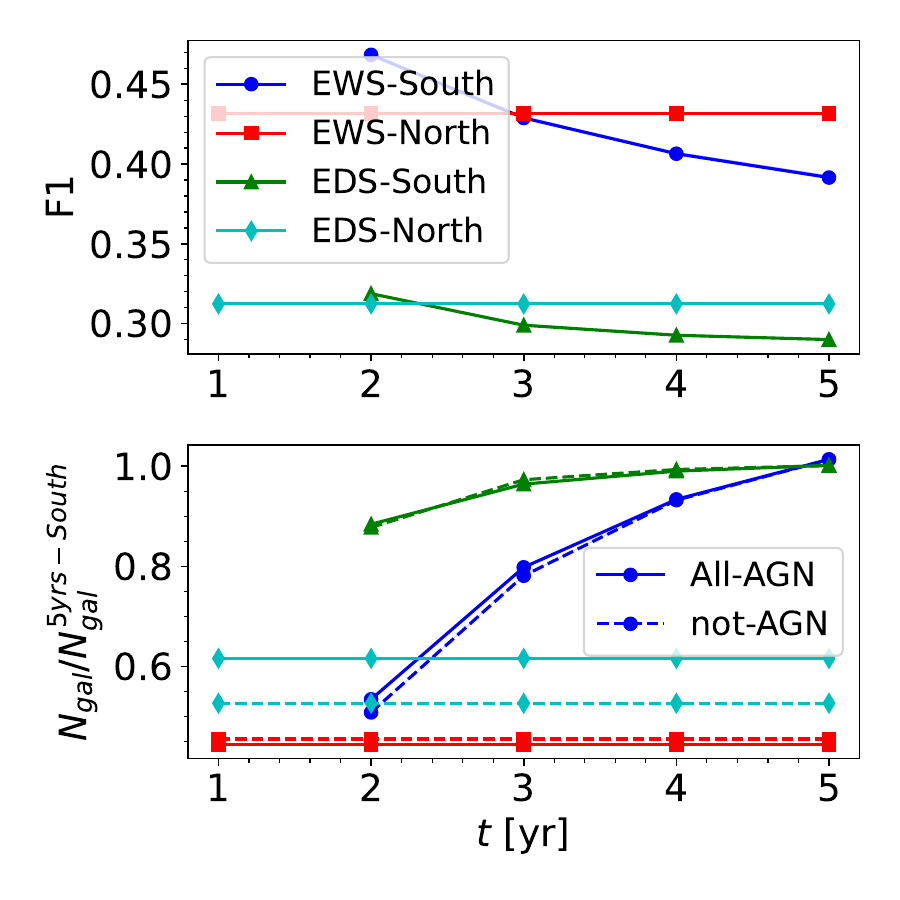}
    \caption{Same as Fig.~\ref{fig:AGN1_time}, but considering all AGN.}
    \label{fig:AGNAll_time}
\end{figure}

In Fig.~\ref{fig:AGN1_time} and \ref{fig:AGNAll_time} (top panels) we report how the F1 score will change with time, adopting the colour selections using ancillary data. For comparison, we also include the F1 scores in the north, which depths remain constant with time. Encouragingly, our selection criteria are quite stable for AGN1 and all AGN, with differences in the F1 score less than 0.07 and 0.08, respectively. In particular, the F1 score generally decreases with time. This counter-intuitive result is due to the increasing depth of both \Euclid and ancillary data,
%fact that at the beginning of the \Euclid surveys there will be shallower $u$ band observations, 
the sample of galaxies initially available will be smaller than the one at the end of the mission (bottom panels in Fig.~\ref{fig:AGN1_time} and \ref{fig:AGNAll_time}) and it will mainly include the brightest AGN, which are the easiest ones to identify. Therefore, if preferred, an additional selection in $u$-band magnitude could improve the colour criteria presented before, although decreasing the overall completeness of the selected sample.

\subsection{\Euclid and IRAC colours}

As previously mentioned, the EDS will have ancillary data available at 3.6 and 4.5\,$\micron$ with \textit{Spitzer}-IRAC. As a first step, the detection in both IRAC filters is alone a good discriminator. Indeed, depending on the observational depths, only $22$--$34\%$ of non-active galaxies are expected to have ${\rm S/N}>3$ in both IRAC filters, compared with $28$--$42\%$ of AGN1 and $54$--$72\%$ of all AGN. This is encouraging for the use of IRAC filters to select AGN, but also highlights that, unfortunately, these existing observations are not as deep as the \Euclid data in order to observe all AGN.

\begin{figure}
    \centering
    \includegraphics[width=\linewidth, keepaspectratio]{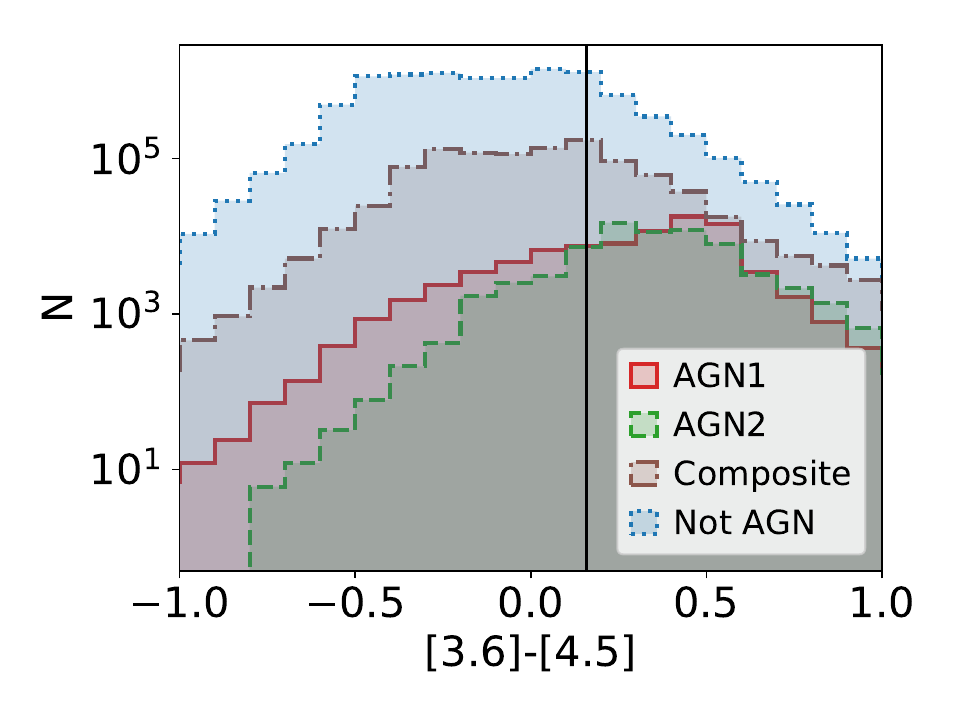}
    \caption{$[3.6]-[4.5]$ colour distribution for objects in our mock catalogue in the EDF, divided by different AGN populations and objects without an AGN. We include only objects with ${\rm S/N}>3$ in both IRAC filters. The vertical black line corresponds to the selection $\rm [3.6]-[4.5]>0.16$ \citep{Stern2012}.}
    \label{fig:irac_only}
\end{figure}

In addition, these colours have been previously used to select AGN1 using the selection $[3.6]-[4.5]>0.16$ \citep{Stern2012}. Using our mock observations, this selection corresponds to $\rm F1=0.039\pm0.006$ and $\rm F1=0.207\pm0.018$ for AGN1 and all AGN in the EDS, respectively. In both cases, the low F1 score is due to a low purity ($P=0.020\pm0.003$ and $P=0.145\pm0.018$) with a large completeness ($C=0.676\pm0.018$ and  $C=0.365\pm0.004$). As visible in Fig.~\ref{fig:irac_only}, even after imposing a detection in both IRAC filters, the criterion is effective at selecting AGN-dominated systems, which, however, are outnumbered by the red tail of the colour distribution of objects without an AGN, in particular SBs. As suggested by other colour criteria \citep[e.g.][]{Lacy2004,Donley2012,Stern2005} it is necessary to have information in the other two IRAC filters (i.e, [5.6] and [8.0]) in order to effectively remove contaminants. 

The two longest wavelength IRAC bands are unfortunately available only for a part of the field. Therefore, in the next section we analyse whether combining the $[3.6]-[4.5]$ colour with a \Euclid one could improve over the IRAC-only or \Euclid-only colour selections.

\subsubsection{AGN1 in the EDS}
\begin{figure}
    \centering
    \includegraphics[width=\linewidth, keepaspectratio]{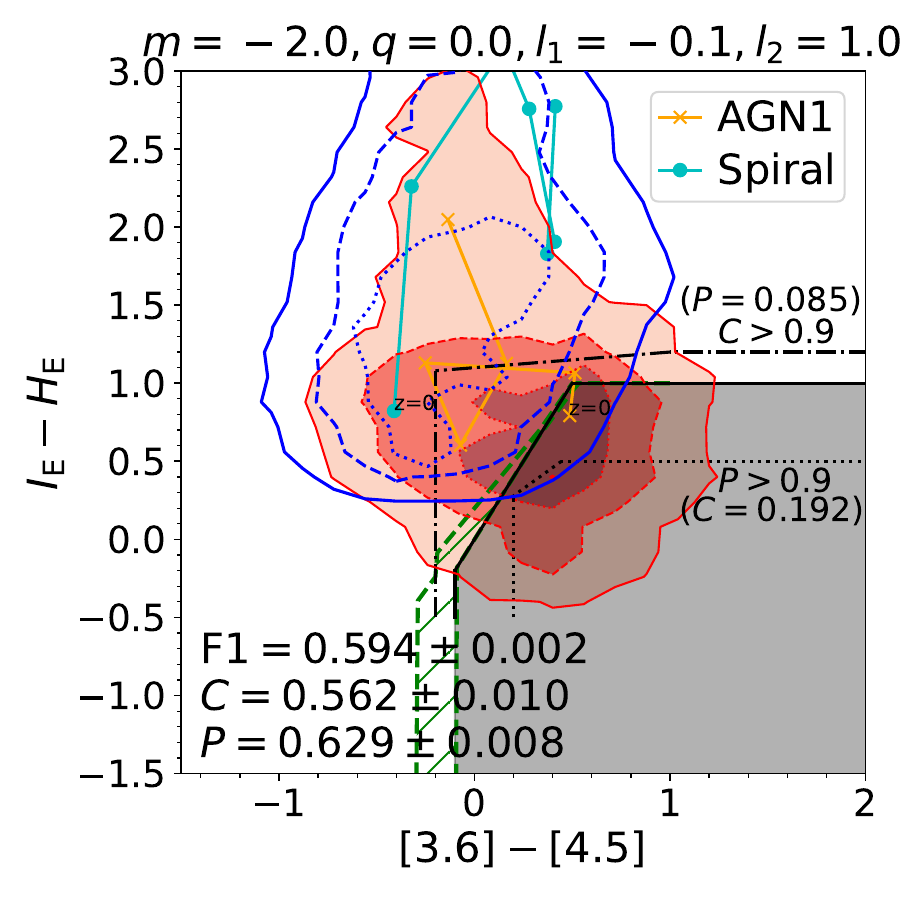}
    \caption{Best selection (type~A) criterion for AGN1 in the EDS using \Euclid and IRAC filters (shaded grey area). The shaded red areas correspond to the colour distribution of AGN1, while the blue lines are the contour levels of the remaining galaxies. Levels correspond to 68\%, 95\%, and 99.7\% of the distribution. We also present example SED tracks for one AGN1 (yellow crosses) and one spiral galaxy (cyan circles) from $z=0$ to $z=6$, with steps of $\Delta z=1$. As an indication of how strict the best selection is, the hatched green area limited by the two dashed green line shows the extent of all the best selection criteria derived with the bootstrap approach. The dotted black line and the dash-dotted black line correspond to the best selection with $P > 0.9$ and $C > 0.9$, respectively.}
    \label{fig:irac_AGN1}
\end{figure}

In Fig.~\ref{fig:irac_AGN1} we show the best selection criterion using \Euclid together with the 3.6- and 4.5-$\micron$ IRAC bands. The selection (type~A) corresponds to
\begin{align}\label{eq:AGN1_IRAC}
      \left(\IE-\HE<1.0_{-0.0}^{+0.0}\right)\,&\land\,
      \left([3.6]-[4.5]>-0.1_{-0.2}^{+0.0}\right)\, \\
      &\land\,
      \left\{\IE-\HE<2.0_{-0.3}^{+0.0}\,([3.6]-[4.5]) + 0.0_{-0.0}^{+0.1}\right\}\;.\nonumber
\end{align}
This selection is an improvement with respect to the \Euclid-only criterion, as it results in $\rm F1=0.591\pm0.002$, $C=0.562\pm0.010$ and $P=0.624\pm0.008$. The selection is particularly effective between $z=0.7$ and $2.4$, where the F1 score remains above 0.5. Indeed, if we limit the selection to only this redshift interval, the overall statistics improve to $F1=0.859\pm0.005$, $P=0.982\pm0.002$ and $C=0.764\pm0.008$, resulting on the best selection for AGN1 in a specific redshift range.

False positives are as follows: $75\%$ are dwarf irregulars; $13\%$ are star-forming galaxies; and $11\%$ are dust-obscured AGN, that is, AGN2 and SB-AGN. As partially visible in Fig.~\ref{fig:irac_AGN1}, the tracks of both star-forming galaxies, and dwarf irregulars (not shown), are outside the AGN1 selection criterion, but photometric uncertainties in the IRAC bands scatter some of these objects inside the selection. Indeed, the median ${\rm S/N}$  for false positive dwarfs galaxies is 4.9 and 3.3 in the 3.6- and 4.5-$\micron$ filters. For the same reason, 
if we increase the ${\rm S/N}$ threshold for a detection in all bands from 3 to 5 or 10, the F1 score increases to $0.711\pm0.0024$ and $0.765\pm0.0024$, respectively. This is driven by an increase in the purity to $0.885\pm0.008$ for ${\rm S/N}>5$ and $0.951\pm0.008$ for ${\rm S/N}>10$, while the completeness remains around 0.6. At the same time, when increasing the ${\rm S/N}$ threshold, dust-obscured AGN 
account for 47$\%$ and 98$\%$ of the false positives, showing that the increase in purity is mainly driven by the removal of non-active galaxies. However, it is necessary to consider that increasing the ${\rm S/N}$ threshold has also an impact on the number of selected (true and false positives) sources, as they decrease from ($1371\pm40$)\,sources\,deg$^{-2}$ for ${\rm S/N}>3$ to ($811\pm17$)\,sources\,deg$^{-2}$ for ${\rm S/N}>5$ and ($575\pm7$)\,sources\,deg$^{-2}$ for ${\rm S/N}>10$.

Leaving the threshold to ${\rm S/N}=3$, but looking for a selection with $C>0.9$, we obtain a selection described by $[3.6]-[4.5]>-0.2$ and $\IE-\HE<1.2$, which, however, corresponds to a low purity, i.e. $P=0.085$.  A selection with a high purity $P>0.9$ corresponds to $C=0.192$ and is described by $[3.6]-[4.5]>0.2$, $\IE-\HE<0.5$ and $[3.6]-[4.5]>0.9\,(\IE-\HE)+0.1$.

\subsubsection{All AGN in the EDS}
We now move to analysing the selection of all AGN using both \Euclid and the 3.6- and 4.5-$\micron$ \textit{Spitzer}/IRAC bands. As reported in Table~\ref{tab:Deep_Euclid_LSST}, the F1 scores obtained combining both facilities varies between 0.207 and 0.222, which is similar to the F1 score obtained using only 3.6- and 4.5-$\micron$ \textit{Spitzer}/IRAC bands (i.e. $0.207\pm0.018$). Indeed, the majority of these selections include no colour cut in the \Euclid filters or only a very marginal one, since composite systems and optically obscured AGN have colours similar to non-active galaxies in the optical. Differences in the F1 score arises from selection effects, since we request a detection in the four filters used in each colour-colour criterion (Fig.~\ref{fig:fraction_IRAC}).\\

%%%%%%%%%%%%%%%%%%%%%%%%%%%%%%%%%%%%%
\section{Conclusions}\label{sec:conclusions}

\begin{table*}[hbpt!]
\centering
    \caption{Summary of the most promising selection criteria.} 
 \resizebox{\textwidth}{!}{
 \addtolength{\tabcolsep}{-0.4em}
	\begin{tabular}{ccccccccccccc}
 \hline\hline\noalign{\vskip 1pt}
        Survey & AGN type & Colours & $z$ cut & $l_{1}$ & $l_{2}$ & $m$ & $q$ & $C$ & $P$ & F1 & $N$ [deg$^{-2}$] & Fig. \\
        \hline\noalign{\vskip 1pt}
        \rule{0pt}{\dimexpr.7\normalbaselineskip+1mm}
        EDS & AGN1 & $\IE-\YE$, $\IE-\HE$ &  & $0.3_{-0.0}^{+0.0}$& $0.5_{-0.0}^{+0.0}$& $-1.6_{-0.0}^{+0.7}$& $0.8_{-0.1}^{+0.0}$ & $0.239\pm0.005$& $0.230\pm0.004$& $0.235\pm0.002$& $2463\pm79$ & \ref{fig:Best_Deep_Euclid}\\
        \noalign{\vskip 1pt}
        EDS & AGN1 & $\IE-\YE$, $\IE-\HE$ & $0.7\leq z\leq4.4$& $0.3_{-0.0}^{+0.0}$& $0.5_{-0.0}^{+0.0}$& $-1.6_{-0.0}^{+0.7}$& $0.8_{-0.1}^{+0.0}$& $0.238\pm0.005$& $0.988\pm0.001$& $0.384\pm0.006$& $507\pm10$  & \ref{fig:Best_Deep_Euclid}\\
        \noalign{\vskip 1pt}
        EDS & AGN1 & $u-z$, $\IE-\HE$ &  & $1.2^{+0.0}_{-0.1}$ & $1.1_{-0.1}^{+0.0}$ & $-1.2_{-0.1}^{+01}$ & $1.7_{-0.0}^{+0.1}$ & $0.775\pm0.012$ & $0.915\pm0.019$ & $0.841\pm0.005$ & $1357\pm45$ & \ref{fig:Deep_LSST}\\
        \noalign{\vskip 1pt}
        EDS & AGN1 & $u-z$, $\IE-\HE$ & $z\leq2.1$ & $1.2^{+0.0}_{-0.1}$ & $1.1_{-0.1}^{+0.0}$ & $-1.2_{-0.1}^{+01}$ & $1.7_{-0.0}^{+0.1}$ & $0.820\pm0.013$ & $0.898\pm0.014$ & $0.858\pm0.002$ & $1356\pm45$ & \ref{fig:Deep_LSST}\\
        \noalign{\vskip 1pt}
        EDS & AGN1 & $\rm [3.6]-[4.5]$, $\IE-\YE$ & & $-0.1_{-0.2}^{+0.0}$ & $1.0_{-0.0}^{+0.0}$ & $-2.0_{-0.3}^{+0.0}$ & $0.0_{-0.0}^{+0.1}$ & $0.562\pm0.002$ & $0.624\pm0.008$ & $0.591\pm0.002$ & $1371\pm40$ & \ref{fig:irac_AGN1} \\
        \noalign{\vskip 1pt}
        EDS & AGN1 & $\rm [3.6]-[4.5]$, $\IE-\YE$ & $0.7\leq z\leq2.4$ & $-0.1_{-0.2}^{+0.0}$ & $1.0_{-0.0}^{+0.0}$ & $-2.0_{-0.3}^{+0.0}$ & $0.0_{-0.0}^{+0.1}$ & $0.764\pm0.008$ & $0.982\pm0.002$ & $0.859\pm0.005$ & $718\pm9$ & \ref{fig:irac_AGN1}\\
        \noalign{\vskip 1pt}
        EDS & All-AGN & $g-r$, $\IE-\YE$ &  & $0.3_{-0.0}^{+\infty}$ & $1.7_{-0.0}^{+\infty}$ & $-3.5_{-0.1}^{+0.0}$ & $0.9_{-0.0}^{+0.0}$ & $0.179\pm0.001$ & $0.576\pm0.009$ & $0.272\pm0.001$ & $5331\pm56$ & \ref{fig:Best_Deep_LSST_All} \\
        \noalign{\vskip 1pt}
        EDS & All-AGN & $g-r$, $\IE-\YE$ & $1.7\leq z\leq2.5$ & $0.3_{-0.0}^{+\infty}$ & $1.7_{-0.0}^{+\infty}$ & $-3.5_{-0.1}^{+0.0}$ & $0.9_{-0.0}^{+0.0}$ & $0.715\pm0.001$ & $0.765\pm0.006$ & $0.739\pm0.003$ & $3299\pm32$ & \ref{fig:Best_Deep_LSST_All} \\
        \noalign{\vskip 1pt}
        EWS & AGN1 & $\IE-\YE$, $\IE-\JE$ &  & $0.5_{-0.0}^{+0.1}$ & $0.7_{-0.0}^{+0.1}$ & $-2.1_{-0.5}^{+0.4}$ & $0.9_{-0.1}^{+0.1}$ & $0.347\pm0.004$ & $0.166\pm0.015$ & $0.224\pm0.001$ & $1850\pm120$  & \ref{fig:Best_Wide_Euclid}\\
        \noalign{\vskip 1pt}
        EWS & AGN1 & $\IE-\YE$, $\IE-\JE$ & $0.7\leq z<5$ & $0.5_{-0.0}^{+0.1}$ & $0.7_{-0.0}^{+0.1}$ & $-2.1_{-0.5}^{+0.4}$ & $0.9_{-0.1}^{+0.1}$ & $0.314\pm0.015$ & $0.705\pm0.004$ & $0.432\pm0.015$ & $390\pm20$ & \ref{fig:Best_Wide_Euclid}\\
        \noalign{\vskip 1pt}
        EWS & AGN1 & $u-z$, $\IE-\HE$ &  & $1.2_{-0.0}^{+0.0}$ & $1.1_{-0.0}^{+0.1}$ & $-1.3_{-0.0}^{+0.2}$ & $1.9_{-0.1}^{+0.0}$ & $0.813\pm0.011$ & $0.922\pm0.017$ & $0.861\pm0.004$ & $659\pm16$ & \ref{fig:WAGN1_LSST}\\
        \noalign{\vskip 1pt}
        EWS & AGN1 & $u-z$, $\IE-\HE$ & $z\leq2.1$ & $1.2_{-0.0}^{+0.0}$ & $1.1_{-0.0}^{+0.1}$ & $-1.3_{-0.0}^{+0.2}$ & $1.9_{-0.1}^{+0.0}$ & $0.850\pm0.000$ & $0.900\pm0.011$ & $0.874\pm0.001$ & $658\pm16$ & \ref{fig:WAGN1_LSST}\\
        \noalign{\vskip 1pt}
        EWS & All-AGN & $u-r$, $\IE-\YE$ & & $0.2_{-0.0}^{+0.0}$ & & $-0.9_{-0.1}^{+0.0}$ & $0.8_{-0.0}^{+0.0}$ & $0.310\pm0.001$ & $0.236\pm0.002$ & $0.268\pm0.001$ & $8070\pm23$ & \ref{fig:WAGNAll_LSST}\\
        \noalign{\vskip 1pt}
        EWS & All AGN & $u-r$, $\IE-\YE$ & $1.7\leq z\leq2.3$ & $0.2_{-0.0}^{+0.0}$ & & $-0.9_{-0.1}^{+0.0}$ & $0.8_{-0.0}^{+0.0}$ & $0.609\pm0.004$ & $0.571\pm0.002$ &$0.589\pm0.003$ & $873\pm3$ & \ref{fig:WAGNAll_LSST} \\
        \hline
    \end{tabular}}
    \label{tab:summary}
\tablefoot{The top table shows a summary of the most promising selection criteria derived in this work for different \Euclid surveys and AGN type. We report also the colours used, the redshift cut considered (if present), the parameters of the best selection (i.e. $l_1$, $l_2$, $m$, and $q$, see Eq.~\ref{eq:sel}), the completeness, the purity, the F1 score, the number of sources selected (true and false positives) in $1\textrm{deg}^2$, as derived from the mock catalogues, and the reference figure. Selections that do not include values for the $l_{2}$ parameter do not require the horizontal colour selection.% $m$ and $q$ parameters do not require the diagonal colour selection. 
    Uncertainties are derived by performing a bootstrap analysis} 
\end{table*}

Efficiently identifying AGN in large surveys is a key priority to properly track accreting SMBHs over a large fraction of the history of the Universe. \Euclid will observe millions of AGN, many of which will have spectroscopic information, but for the majority of them we will need to rely only on \Euclid photometry or a restricted set of ancillary data, such as Rubin/LSST or \textit{Spitzer}/IRAC. In this work, we have simulated Euclid-like photometry to forecast the effectiveness of the \Euclid mission in detecting AGN under a variety of selection techniques. A summary of the most promising selection criteria is reported in Table~\ref{tab:summary}, while we report our main results below. 
%\par

First, AGN1 can be identified in EDS (EWS) using only \Euclid filters with a completeness of $0.239\pm0.005$ ($0.347\pm0.004$) and a purity of $0.230\pm0.004$ ($0.166\pm0.015$). The purity in the EDS  greatly improves to $0.988\pm0.001$ if we restrict the sample to $0.7\leq z\leq4.4$. If redshift information is not available, it is preferable to include an additional selection able to separate AGN1 from dwarf galaxies, derived for example from morphological information. 
%\par

Second, the inclusion of optical ancillary filters, such as those from Rubin/LSST, greatly improves the selection of AGN1 in both \Euclid surveys. Of particular importance are the $u$ and $z$ filters, which allow for the selection of AGN1 with a completeness $0.775\pm0.012$ and $0.813\pm0.011$ and a purity $0.915\pm0.019$ and $0.922\pm0.017$ in the EDS and EWS, respectively. Using such criteria we expect to select $7\times 10^{4}$ AGN1 candidates in the EDS and $10^{7}$ in the EWS. However, the efficiency of these colour selections decrease at $z\simeq2.1$. 

Third, in the EDS the combination of \Euclid filters and the \textit{Spitzer}/IRAC 3.6- and 4.5-$\micron$ filters is a valid alternative to select AGN1. This is valid at least in the redshift range $z=0.7$--2.4, where the proposed selection corresponds to a completeness $C=0.764\pm0.008$ and $P=0.982\pm0.002$.

Finally, the selection of all AGN, including optically obscured AGN and composite systems, is challenging, both with \Euclid filters and with \Euclid filters supplemented by optical or IRAC bands. The best colour selection yields a purity and completeness of $0.765\pm0.006$ and $0.715\pm0.001$, achieved using Rubin/LSST and \Euclid filters in the EDS at $1.7\leq z\leq2.5$. To improve over such colour selections it will be necessary to include ancillary information, for instance, spectroscopic data, longer wavelength data, or morphological information, or to adopt more complex methods, such as machine learning algorithms, as demonstrated, for example, in \citet{Humphrey2022} for the selection of passive galaxies. 
%\par

%%%%%%%%%%%%%%%%%%%%%%%%%%%%%%%%%%%%%%%%

\begin{acknowledgements}
LB acknowledges financial support from PRIN MIUR $2017 - 20173ML3WW$\char`_$001$. LB and LP acknowledge financial contributions by grants "Premiale 2015 - MITiC". The work of DS was carried out at the Jet Propulsion Laboratory, California Institute of Technology, under a contract with NASA. LB and VA acknowledge support from the INAF Large Grant “AGN and Euclid: a close entanglement”, Ob. Fu. 1.05.23.01.14. LB and IP acknowledge support from INAF under the Large Grant 2022 funding scheme (project "MeerKAT and LOFAR Team up: a Unique Radio Window on Galaxy/AGN co-Evolution"). FS acknowledges partial support from the European Union’s Horizon 2020 research and innovation programme under the Marie Skłodowska-Curie grant agreement No. 860744.
\AckEC
\end{acknowledgements}

\bibliographystyle{aa} % style aa.bst
\bibliography{main} % your references Yourfile.bib

%%%%%%%%%%%%%%%%% APPENDICES %%%%%%%%%%%%%%%%%%%%%

\begin{appendix}

\section{\spr{} updates}\label{sec:newspritz}
We have applied updates to some of the luminosity functions reported in the original version of \spr\ in order to improve the comparison with additional observational results, without reducing the agreement with observations already tested. We report such updates in this Appendix.
%\par

Firstly, in the original \spr{} the redshift evolution of the SB-AGN population corresponded to the one presented by \citet{Gruppioni2013}, in which both the luminosity and the number density at the knee of the modified Schechter function continuously increases with redshift, as $(1+z)^{k}$ with $k>1$. However, this representation produces an overestimation of the galaxy stellar mass function (GSMF) of AGN when compared to the values estimated by \citet{Bongiorno2016}. To overcome this issue, we performed a new fit to the number density at the knee, using a new functional form:
\begin{equation}
\Phi^{*} =
  \begin{cases}
  \Phi_{0}^{*}\,(1+z)^{k_{\Phi}}, & \quad \text{if } z\leq z_{\Phi};\\
  \text{constant}, & \quad \text{if } z>z_{\Phi}.
  \end{cases}
\end{equation}
The best non-linear least-squares fit \citep{Scipy} corresponds to $\Phi_{0}^{*} [10^{-3} \rm{Mpc}^{-3}]=0.0181\pm0.0074$, $k_{\Phi}=2.15\pm0.88$, and $z_{\Phi}=1.34\pm0.45$.
%\par

\begin{figure}
    \centering
    \includegraphics[width=\linewidth, trim={0 70 0 50},clip,keepaspectratio]{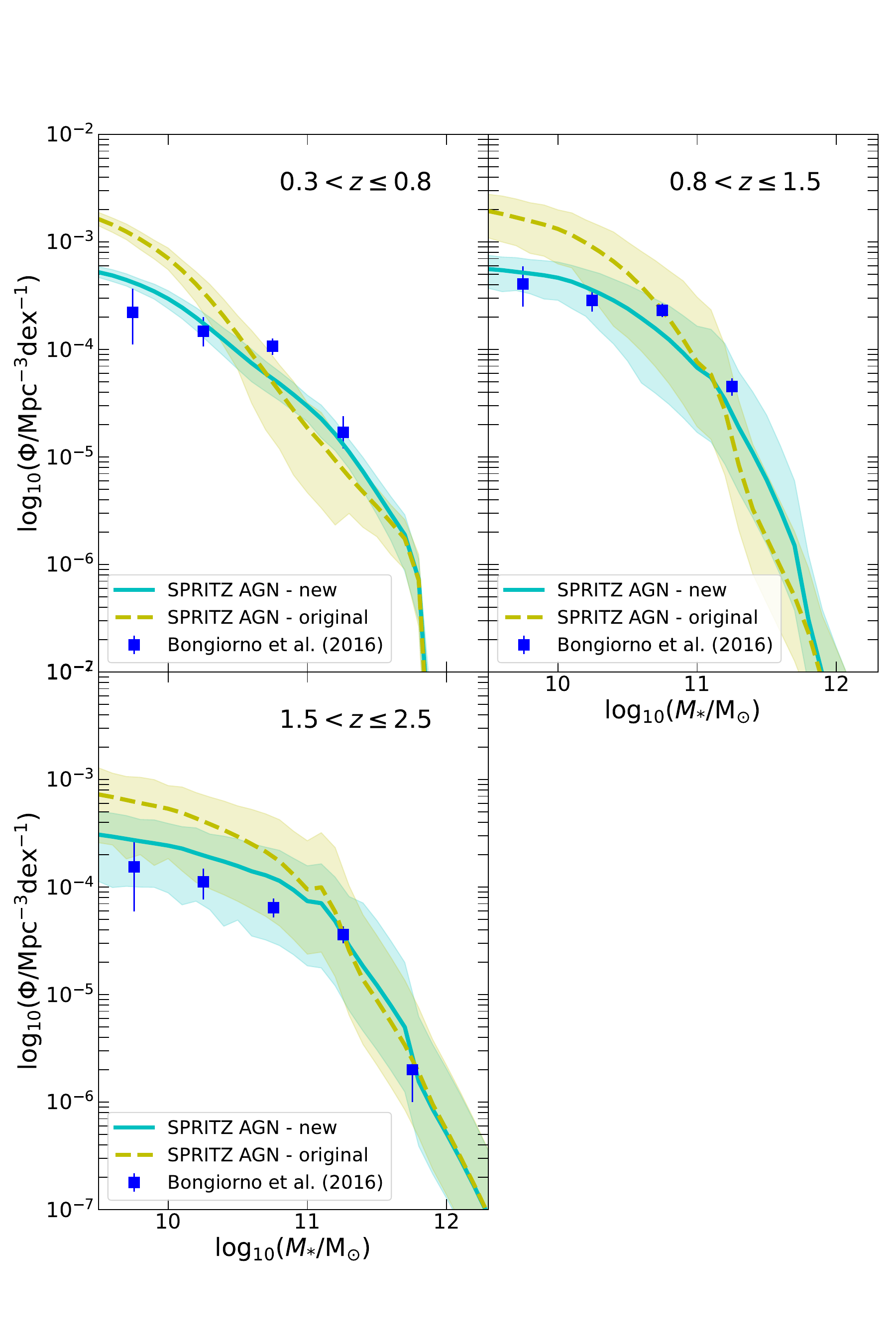}
    \caption{AGN GSMF observed by \citet[][blue squares]{Bongiorno2016} compared with the original (dashed yellow lines) and updated (solid cyan lines) \spr{} AGN GSMFs. Shaded regions show the propagated uncertainties of the starting AGN infrared luminosity functions.}
    \label{fig:AGN_GSMF}
\end{figure}

Secondly, looking again at the observed AGN GSMF by \citet{Bongiorno2016}, the SF-AGN population produces an overestimation at $z<1.5$ at the low-mass end of the function (i.e. $M_{*}<10^{10.5}\si{\solarmass}$). In this case, we modified the faint-end slope of the IR luminosity function (a modified Schechter function) of this population, which was not constrained by the original \hers observations, from $\alpha=1.2$ to $\alpha =1.0$. At the same time, it was necessary to decrease the number density at the knee by 0.5 dex. These changes increase the agreement with the AGN GSMF, while keeping agreement with other observed distributions, such as (but not limited to) the IR luminosity function. A comparison between the observed AGN GSMF, the original, and the new \spr\ AGN GSMF is reported in Fig.~\ref{fig:AGN_GSMF}. 

\begin{figure}[hbpt!]
    \centering
    \includegraphics[width=\linewidth,keepaspectratio]{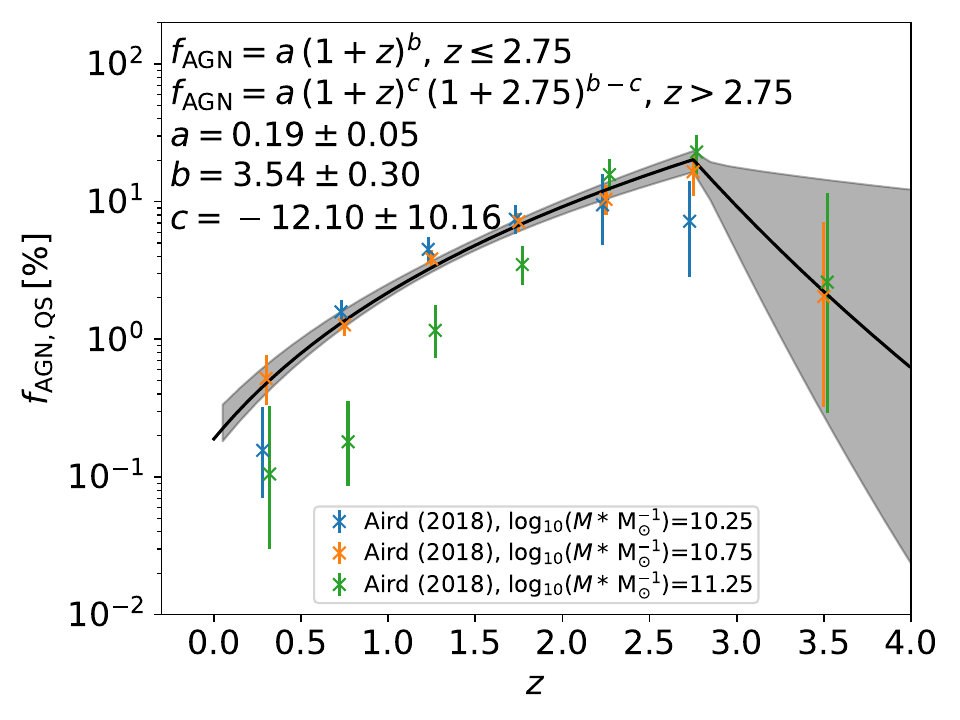}
    \includegraphics[width=\linewidth,keepaspectratio]{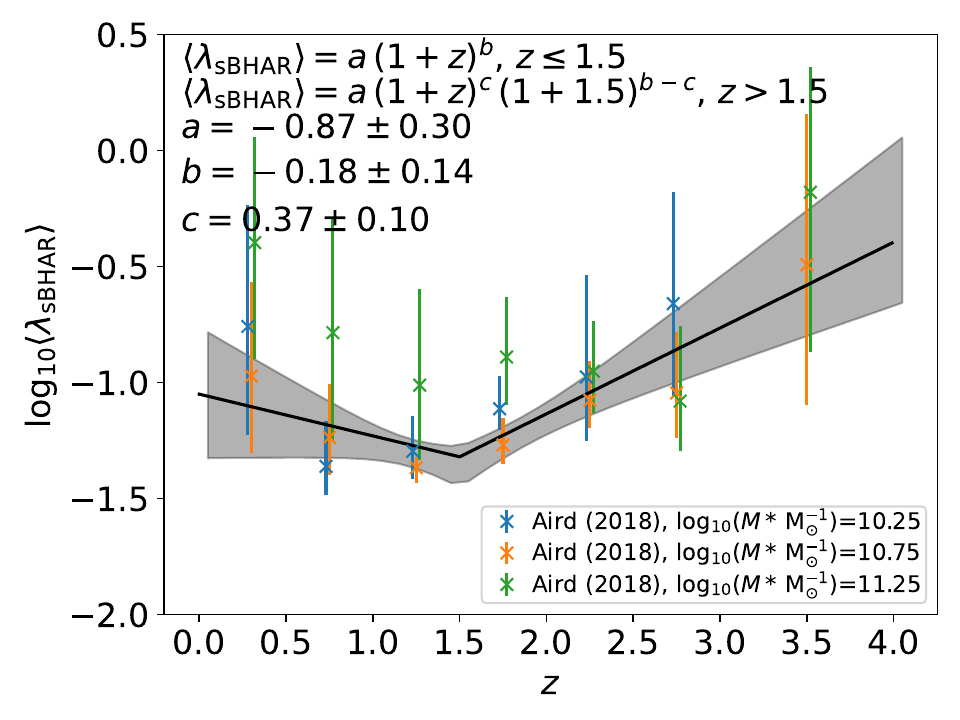}
    \caption{Observed relations used to include Ell-AGN in \spr{}. \textit{Top:} fraction of AGN among quiescent galaxies as a function of redshift. \textit{Bottom:} average black-hole accretion rate as a function of redshift. Data points are slightly shifted horizontally for clarity.}
    \label{fig:fAGN_QS}
\end{figure}

Lastly, in the original \spr, AGN are only present inside SF galaxies. We decided to overcome this limitation by including AGN in quiescent galaxies (QS), which we call Ell-AGN, following the observational results of \citet{Aird2018}. As reported in Fig.~\ref{fig:fAGN_QS}, we parameterise their observed fraction of AGN in quiescent galaxies as a function of redshift as
\begin{equation}
f_{\rm AGN}=
\begin{cases}
    a\,(1+z)^{b}, & \quad \text{if } z\leq2.75,\\
    a\,(1+z)^{c}\,(1+2.75)^{b-c}, & \quad \text{if } z>2.75,
\end{cases}
\end{equation}
with $a=0.19\,\pm\,0.05$, $b=3.54\,\pm\,0.30$, and $c=-12.10\pm10.16$. Using this fraction, we extracted a random number of elliptical galaxies from \spr\ to which we assign AGN. We then assigned each a black-hole accretion rate, following again a redshift-dependent function that well describes the observed average values by \citet{Aird2018},
\begin{equation}
\logten (\langle \lambda_{\rm sBHAR} \rangle)=
\begin{cases}
    a\,(1+z)^{b}, & \quad \text{if } z\leq1.5,\\
    a\,(1+z)^{c}\,(1+1.5)^{b-c}, & \quad \text{if } z>1.5,
\end{cases}
\end{equation}
with $a=-0.87\,\pm\,0.30$, $b=-0.18\,\pm\,0.14$, and $c=0.37\,\pm\,0.10$. We then derived the X-ray luminosity between 0.5 and 2 keV following the definition of the black hole accretion rate presented in the same paper. However, given the small fraction of Ell-AGN, they have a negligible impact on global statistics, such as the GSMF mentioned above.

%%%%%%%%%%%%%%%%%%
\section{AGN colours in \spr}\label{sec:colsec_spritz}
In this section we show that the colours of AGN in \spr{} are consistent with colour selection criteria available in the literature \citep[e.g.][]{Stern2005,Donley2012}. The criteria considered here are based on near-IR colours, which are redwards of the \Euclid wavelength coverage and therefore cannot be used to identify AGN using \Euclid data alone. \citet{Bisigello2021} also showed that galaxies and AGN in \spr{} are correctly placed in one of the AGN diagnostic diagrams based on optical nebular emission lines \citep{BPT1981}. 
%\par

\begin{figure}[hbpt!]
    \centering
    \includegraphics[width=\linewidth,keepaspectratio]{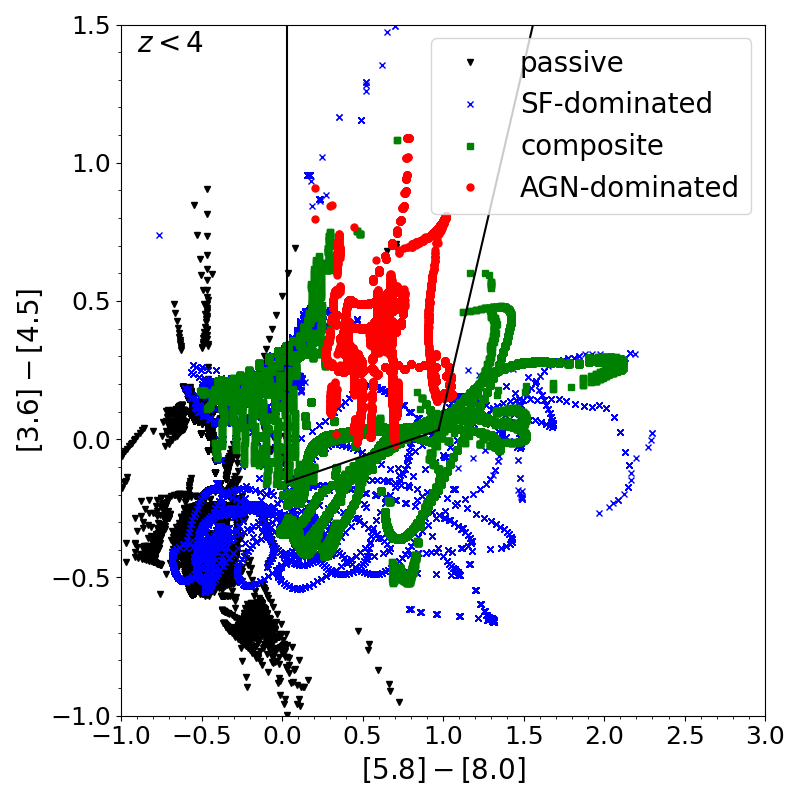}
    \caption{AGN colour selection from \citet{Stern2005} applied to galaxies in \spr{} at $z<4$, without including any photometric uncertainties. Data points indicate objects dominated by AGN activity (red circles), dominated by star formation (blue crosses), composite systems (green squares), and passive galaxies (black triangles).}
    \label{fig:Stern}
\end{figure}

In Fig.~\ref{fig:Stern} we show the \textit{Spitzer}/IRAC $[3.6]-[4.5]$ and $[5.8]-[8.0]$ colours for galaxies in the \Euclid mock catalogue of the EDS at $z<4$, together with the colour selection identified by \citet{Stern2005}, valid over the same redshift range. For this comparison we consider the true colours, without any photometric uncertainties. A total of 99.3$\%$ of the AGN-dominated systems are included in this colour selection, with a small fraction of contaminants, that is, 7$\%$ of objects without AGN activity. On the other hand, only 34$\%$ of the composite systems are selected. This is consistent with other results in the literature using samples of AGN observed at different wavelengths \citep[e.g.][]{Eckart2010,Stern2012}. 
%\par

\begin{figure}[hbpt!]
    \centering
    \includegraphics[width=\linewidth,keepaspectratio]{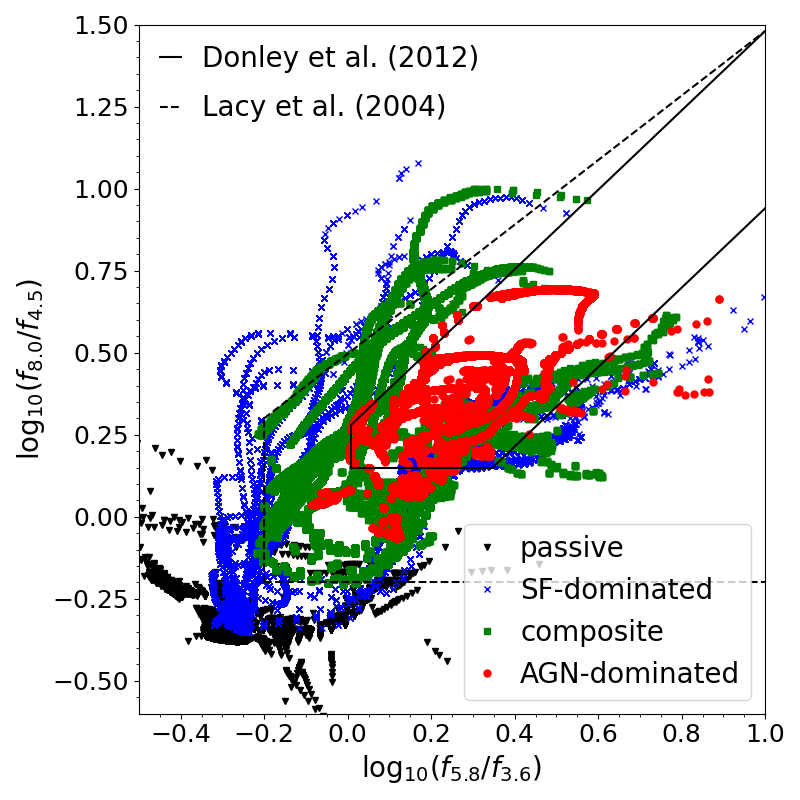}
    \caption{AGN colour selection from \citet[][solid black line]{Donley2012} and \citet[][dashed black line]{Lacy2004} applied to galaxies in \spr{}, without including any photometric uncertainties. Data points indicate objects dominated by AGN activity (red circles), dominated by star formation (blue crosses), composite systems (green squares), and passive galaxies (black triangles).}
    \label{fig:Donley}
\end{figure}

We tested the colours of AGN in the \Euclid mocks using other two-colour selections \citep{Lacy2004,Donley2012} based on \textit{Spitzer}/IRAC fluxes. As shown in Fig.~\ref{fig:Donley}, AGN in the \Euclid mock catalogue occupy the expected colour space. On the one hand, 87$\%$ of AGN-dominated systems are inside the \citet{Donley2012} colour criterion, which is fine-tuned to select objects showing a power-law component in the near-IR, with only 2$\%$ of non-AGN objects falling inside this selection. On the other hand, the less conservative criterion from \citet{Lacy2004} selects all AGN-dominated systems and 96$\%$ of composite galaxies present in the \Euclid mock catalogue. We therefore conclude that the near-IR colours of AGN in \spr{} are consistent with observed AGN. 

%%%%%%%%%%%%%%%%%%%%%%%%%%%%%%%%%%%%%%%%
\section{Selection effects of the colour cuts}\label{sec:Selection}
In this work we tested different colour selections to separate AGN1, or all AGN types, from other galaxy populations. The main \Euclid mock catalogues contain all objects that have a detection, that is, ${\rm S/N}>3$ at the depth of the EWS or of the EDS, in at least one \Euclid filter. It is, however, difficult to obtain any information on the nature of the sources with a detection in a single band. We therefore included in the analysis of each colour selection only the sub-sample of galaxies with ${\rm S/N}>3$ in all the filters included in each colour. In this appendix we analysed the observational biases introduced by the inclusion of this additional S/N threshold with respect to the original sample that requires a detection in a single band. 
%\par

\begin{figure}[hbpt!]
    \centering
    \includegraphics[width=\linewidth,keepaspectratio]{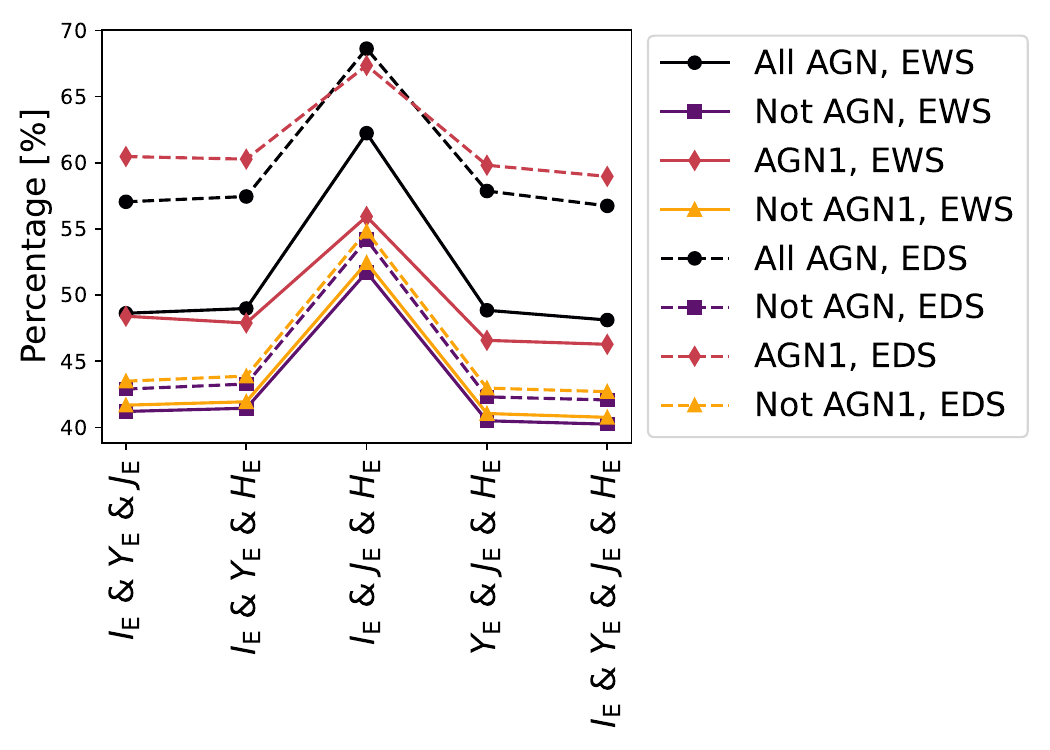}
    \caption{Percentage of AGN1, all AGN, non-AGN1, or non-AGN selected by assuming a ${\rm S/N}>3$ in three or four \Euclid filters with respect to the complete \spr\ \Euclid catalogue. Solid and dashed lines show the percentage in the EWS and EDS, respectively.}
    \label{fig:fraction}
\end{figure}

In Fig.~\ref{fig:fraction} we focus on the colour criteria using only \Euclid filters. Imposing ${\rm S/N}>3$ in three or four \Euclid filters selects an almost constant percentage of galaxies, with a minimum when the detection in all filters is required and a maximum when the detection in \YE\ is not required. Percentages range from 66 to 73$\%$ for AGN1 in the EWS, from 79 to 84\%  for AGN1 in the EDS, from 55 to 64\% for all AGN in the EWS, and from 63 to 71\% for all AGN in the EDS.

\begin{figure}[hbpt!]
    \centering
    \includegraphics[width=\linewidth,keepaspectratio]{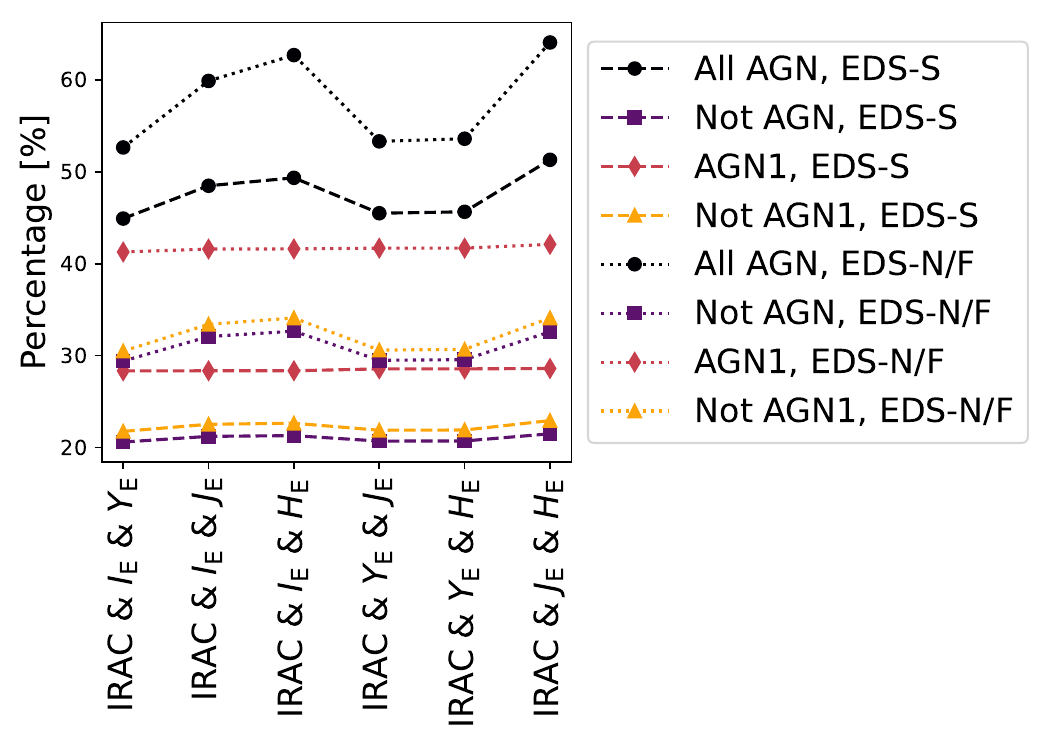}
    \caption{Percentage of AGN1, all AGN, non-AGN1, or non-AGN selected by assuming a ${\rm S/N}>3$ in two \Euclid filters and the 3.6- and 4.5-$\micron$ IRAC filters with respect to the complete \spr\ \Euclid catalogue. Dashed line show the percentage in the EDS-S, while the dotted line shows the one in the EDS-F and in the EDS-N.}
    \label{fig:fraction_IRAC}
\end{figure}

In Fig.~\ref{fig:fraction_IRAC} we report the fraction of objects detected (i.e. ${\rm S/N}>3$) in the 3.6- and 4.5-$\micron$ IRAC filters and in two \Euclid ones. We considered both the depths in the EDF-S and the depths in the EDF-F and EDF-N.

\begin{figure*}[hbpt!]
    \centering
    \includegraphics[width=\linewidth,keepaspectratio]{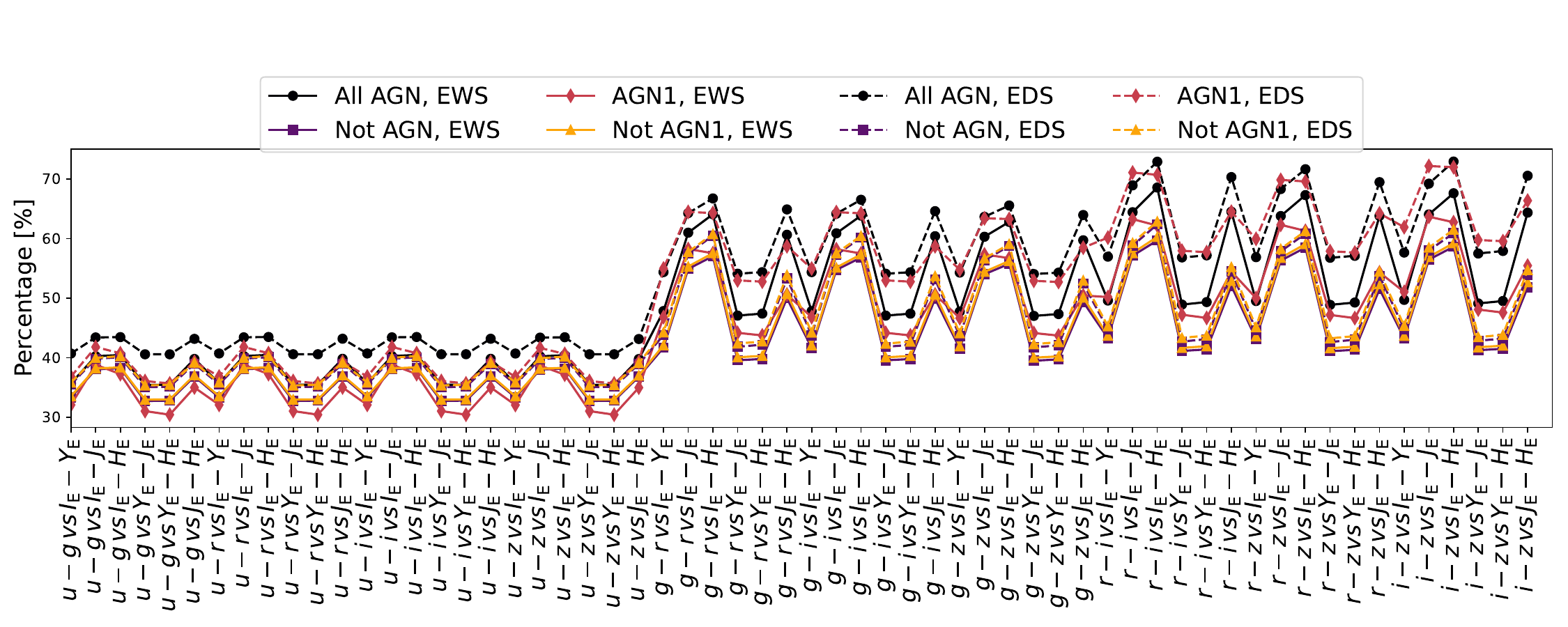}
    \caption{Same as Fig.~\ref{fig:fraction}, but considering Rubin/LSST and \Euclid filters.}
    \label{fig:fraction_LSST}
\end{figure*}

We consider the selections using \Euclid and Rubin/LSST filters in Fig.~\ref{fig:fraction_LSST}. Here we see that the depth of the $u$ filter does not match the depth of the other filters in the EDS for objects that are not AGN1. Therefore, by imposing a detection in the $u$ filter we select around 60$\%$ of AGN1, but less than 40$\%$ of other objects. This may be one of the reasons behind the large F1 score obtained for AGN1 in the EDS. At the same time we see smaller variations depending on the \Euclid filters included in the different colour criteria, with local minima and maxima when we impose a detection in the \YE\ and \JE\ filters or in the \IE\ and \HE\ filters, respectively.

%%%%%%%%%%%%%%%%%%%%%%%%%%
\section{F1 scores for all colour combinations}\label{sec:tables}
In this appendix we provide tables with the F1 scores derived for all colour combinations, either involving only \Euclid filters (Tables~\ref{tab:Deep_Euclid} and \ref{tab:Wide_Euclid}) or both \Euclid and Rubin/LSST filters (Tables~\ref{tab:Deep_Euclid_LSST} and \ref{tab:Wide_Euclid_LSST}). In all tables we report the F1 scores derived when selecting AGN1 and, in parentheses, the values derived when analysing all AGN. We highlight in bold the maximum F1-values derived.

\begin{table*}[hbpt!]
    \centering
    \caption{Best F1-values for different colour-colour selections using \Euclid filters for the EDS. }
    \label{tab:Deep_Euclid}
    \begin{tabular}{cccccc}
		\hline\hline
        \noalign{\vskip 1pt}
		Colours & $\IE-\YE$ & $\IE-\JE$  & $\IE-\HE$ & $\YE-\JE$ & $\YE-\HE$\\
		\hline
        \noalign{\vskip 1pt}
%        \rule{0pt}{\dimexpr.7\normalbaselineskip+1mm}
        $\IE-\JE$ & 0.193(0.121) & \dots & \dots & \dots & \dots\\
        $\IE-\HE$ & \textbf{0.235}(0.121) & 0.180(0.115) & \dots & \dots & \dots\\
        $\YE-\JE$ & 0.128(0.121) & 0.140(0.121) & 0.171(0.123)  & \dots & \dots \\
        $\YE-\HE$ & 0.232(0.121) & 0.142(0.123) & 0.168(0.121) & 0.039(\textbf{0.124}) & \dots\\
        $\JE-\HE$ & 0.189(0.123) & 0.180(0.116) & 0.155(0.116)& 0.053(0.124) & 0.053(0.124)\\\end{tabular}
    \tablefoot{The main values are for selection of AGN1, while values in parentheses correspond to the selection of all AGN types. Bold-faced values are the best ones among all selections.}
\end{table*}

\begin{table*}[hbpt!]
    \centering
    \caption{Same as Table~\ref{tab:Deep_Euclid}, but for the EWS.}
    \label{tab:Wide_Euclid}
    \begin{tabular}{cccccc}
		\hline\hline
        \noalign{\vskip 1pt}
		Colours & $\IE-\YE$ & $\IE-\JE$  & $\IE-\HE$ & $\YE-\JE$ & $\YE-\HE$\\
		\hline
        \noalign{\vskip 1pt}
        $\IE-\JE$ & \textbf{0.224}(0.153) & \dots & \dots & \dots & \dots\\
        $\IE-\HE$ & 0.220(0.153) & 0.190(0.155) & \dots & \dots & \dots\\
        $\YE-\JE$ & 0.224(0.153) & 0.175(0.154) & 0.183(0.155)  & \dots & \dots \\
        $\YE-\HE$ & 0.220(0.153) & 0.177(0.155) & 0.181(0.153) & 0.051(0.156) & \dots\\
        $\JE-\HE$ & 0.208(0.155) & 0.190(0.156) & 0.169(0.156) & 0.056(0.156) & 0.056(0.156)\\           
    \end{tabular}
\end{table*}

\begin{table*}[hbpt!]
	\centering
	\caption{Same as Table~\ref{tab:Deep_Euclid}, but considering colour-colour selections using \Euclid and Rubin/LSST and two IRAC filters in the EDS.}
	\label{tab:Deep_Euclid_LSST}
	\begin{tabular}{ccccccc} 
		\hline\hline
        \noalign{\vskip 1pt}
		Colours & $\IE-\YE$ & $\IE-\JE$ & $\IE-\HE$ & $\YE-\JE$ & $\YE-\HE$ & $\JE-\HE$\\
		\hline
        \noalign{\vskip 1pt}
		$u-g$ & 0.392(0.236) & 0.190(0.190) & 0.336(0.196) & 0.021(0.226) & 0.050(0.224) & 0.056(0.145)\\
		$u-r$ & 0.646(0.231) & 0.480(0.194) & 0.600(0.197) & 0.033(0.190) & 0.076(0.190) & 0.079(0.171)\\
        $u-i$ & 0.757(0.193) & 0.756(0.164) & 0.787(0.167) & 0.086(0.139) & 0.127(0.137) & 0.113(0.130)\\ 
        $u-z$ & 0.792(0.195) & 0.817(0.173) & \textbf{0.841}(0.174) & 0.239(0.161) & 0.280(0.161) & 0.204(0.146)\\
		$g-r$ & 0.618(\textbf{0.272}) & 0.649(0.226) & 0.640(0.214) & 0.063(0.170) & 0.083(0.168) & 0.072(0.170) \\
		$g-i$ & 0.503(0.212) & 0.666(0.214) & 0.698(0.204) & 0.152(0.149) & 0.172(0.155) & 0.106(0.140)\\
		$g-z$ & 0.430(0.159) & 0.620(0.165) & 0.669(0.155) & 0.270(0.151) & 0.333(0.143) & 0.176(0.127)\\
		$r-i$ & 0.225(0.160) & 0.336(0.167) & 0.330(0.150) & 0.238(0.157) & 0.220(0.127) & 0.087(0.118)\\
		$r-z$ & 0.171(0.138) & 0.291(0.154) & 0.328(0.156) & 0.248(0.139) & 0.278(0.134) & 0.145(0.121)\\
		$i-z$ & 0.168(0.118) & 0.152(0.117) & 0.195(0.112) & 0.107(0.122) & 0.126(0.122) & 0.106(0.120)\\
		\hline
        \noalign{\vskip 1pt}
        $\rm [3.6]-[4.5]$ & 0.445(0.207) & 0.418(0.214)& \textbf{0.591}(0.219)& 0.250(0.208)& 0.281(0.208)& 0.248(\textbf{0.222})\\
        \hline
	\end{tabular}
\end{table*}

\begin{table*}[hbpt!]
	\centering
	\caption{Same as Table~\ref{tab:Deep_Euclid}, but considering colour-colour selections using Euclid and Rubin/LSST filters for the EWS.}
	\label{tab:Wide_Euclid_LSST}
	\begin{tabular}{ccccccc} 
		\hline\hline
        \noalign{\vskip 1pt}
		Colours & $\IE-\YE$ & $\IE-\JE$ & $\IE-\HE$ & $\YE-\JE$ & $\YE-\HE$ & $\JE-\HE$\\
		\hline
        \noalign{\vskip 1pt}
        $u-g$ & 0.397(\textbf{0.292}) & 0.223(0.254) & 0.331(0.259) & 0.029(0.288) & 0.064(0.287) & 0.056(0.265)\\
        $u-r$ & 0.654(0.268) & 0.558(0.251)& 0.619(0.254)& 0.051(0.232)& 0.102(0.235)& 0.084(0.230)\\
        $u-i$ & 0.764(0.223)& 0.790(0.215)& 0.805(0.214)& 0.111(0.202)& 0.171(0.203)& 0.125(0.201)\\
        $u-z$ & 0.810(0.215)& 0.851(0.213)& \textbf{0.861}(0.211)& 0.289(0.217)& 0.315(0.216)& 0.231(0.213)\\
        $g-r$ &  0.631(0.233)& 0.686(0.232)& 0.675(0.226)& 0.085(0.165)& 0.112(0.164)& 0.085(0.183)\\
        $g-i$ &  0.550(0.176)& 0.726(0.206)& 0.752(0.199)& 0.180(0.168)& 0.218(0.168)& 0.122(0.167)\\
        $g-z$ & 0.501(0.166) & 0.684(0.170)& 0.736(0.165)& 0.316(0.159)& 0.366(0.159)& 0.195(0.159)\\
        $r-i$ & 0.297(0.153)& 0.417(0.150)& 0.439(0.152)& 0.292(0.157)& 0.284(0.157)& 0.113(0.158)\\
        $r-z$ &  0.283(0.151)& 0.357(0.149)& 0.407(0.151)& 0.302(0.156)& 0.343(0.156)& 0.171(0.157)\\
        $i-z$ &  0.272(0.151)& 0.207(0.148)& 0.230(0.150)& 0.158(0.154)& 0.176(0.155)& 0.130(0.156)\\
		\hline
	\end{tabular}
\end{table*}

%%%%%%%%%%%%%%%%%%%%%%%%%%%%%%%%%%%%%%%%%%%%%%%%%%
\end{appendix}

\end{document}